\documentclass[10pt,journal,compsoc]{IEEEtran}
\usepackage{ifpdf, flushend,subfigure}
\ifCLASSINFOpdf
  \usepackage[pdftex]{graphicx}
  \graphicspath{{../pdf/}{../jpeg/}}
  \DeclareGraphicsExtensions{.pdf,.jpeg,.png}
\else
  \usepackage[dvips]{graphicx}
  \graphicspath{{../eps/}}
  \DeclareGraphicsExtensions{.eps}
\fi
\usepackage{epstopdf}
\usepackage{multirow}
\usepackage{float}
\usepackage[cmex10]{amsmath}
\usepackage {amssymb}

\usepackage{graphicx}
\usepackage{array}
\usepackage{mdwmath}
\usepackage{mdwtab}
\usepackage{eqparbox}
\usepackage{nomencl}
\usepackage{cite}
\usepackage{algorithm}
\usepackage{algorithmic}
\usepackage{makeidx}
\usepackage{ifthen}
\usepackage{subfigure}
\usepackage{caption}

\usepackage{pdflscape}
\newcommand{\argmax}{\operatornamewithlimits{argmax}}
\newcommand{\beq}{\begin{equation}}
\newcommand{\eeq}{\end{equation}}
\newcommand{\beqn}{\begin{eqnarray}}
\newcommand{\eeqn}{\end{eqnarray}}
\newcommand{\beqno}{\begin{eqnarray*}}
\newcommand{\eeqno}{\end{eqnarray*}}
\newcommand{\bma}{\begin{displaymath}}
\newcommand{\ema}{\end{displaymath}}
\newcommand{\bnu}{\begin{enumerate}}
\newcommand{\enu}{\end{enumerate}}
\newcommand{\bce}{\begin{center}}
\newcommand{\ece}{\end{center}}
\newcommand{\btb}{\begin{tabular}}
\newcommand{\etb}{\end{tabular}}
\newcommand*{\qedb}{\hfill\ensuremath{\square}}%

\newtheorem{theorem}{Theorem}[section]

\newtheorem{proposition}[theorem]{Proposition}

\newtheorem{definition}[theorem]{Definition}

\begin{document}
\title{Price-based Resource Allocation for Edge Computing: A Market Equilibrium Approach}

\author{Duong~Tung~Nguyen,~\IEEEmembership{Student Member,~IEEE,}
        Long~Bao~Le,~\IEEEmembership{Senior Member,~IEEE,}
        and~Vijay~Bhargava,~\IEEEmembership{Life~Fellow,~IEEE} }

\IEEEtitleabstractindextext{%
\begin{abstract}
The emerging edge computing paradigm
promises to deliver superior user experience and enable
a wide range of Internet of Things (IoT) applications.
In this work, we propose a new market-based framework for efficiently allocating 
resources of heterogeneous capacity-limited 
edge nodes (EN) to multiple competing services at the network edge. 
By properly pricing 
the geographically distributed  
ENs, 
the proposed framework 
generates 
a market equilibrium (ME) solution that
 not only maximizes 
the  edge 
computing resource utilization but also allocates optimal
 (i.e., utility-maximizing)
 resource bundles to the services given their budget constraints. 
When the utility of a service is defined as the maximum revenue that the service can achieve from its resource allotment, 
the equilibrium can be computed centrally by solving the Eisenberg-Gale (EG) convex program.
 drawn from the	
economics literature. 
We further show that the equilibrium allocation is Pareto-optimal and satisfies desired fairness properties including sharing incentive, proportionality, and envy-freeness.
Also, two distributed algorithms are introduced, which efficiently converge to an ME.
	When each service aims to maximize its net profit (i.e., revenue minus cost) instead of the revenue, we derive a novel convex optimization problem and rigorously prove that its solution is exactly an ME. 
	Extensive numerical results are presented to validate the effectiveness of the proposed techniques. 
\end{abstract}

}

\maketitle


\section{Introduction}

The last decade has witnessed
an explosion of data traffic over the communication network 
attributed to
the rapidly growing cloud computing and pervasive mobile devices.
This trend is expected to continue for the foreseeable future  with 
 a whole new generation of applications
including  4K/8K UHD video, 
hologram, 
interactive mobile gaming, 
tactile Internet,  virtual/augmented reality (VR/AR), 
mission-critical communication, smart homes, 
and a variety of IoT applications \cite{mchi16}.
As the cloud infrastructure and number of devices will continue to expand at an accelerated rate, a tremendous burden will be put on the network. 
Thus, 
it is imperative 
for network operators to develop innovative solutions to meet the soaring traffic demand and accommodate diverse requirements of various services and use cases in 
the next generation communication network. 

Thanks to the economy of scale and supercomputing capability advantages, cloud computing 
will likely continue to play a prominent role in the future computing landscape.
However, cloud data centers (DC) are often geographically distant from the end-user, which induces 
enormous network traffic, along with significant communication delay and jitter. 
Hence, despite the immense power and potential, 
 cloud computing alone is 
facing growing limitations in satisfying the stringent requirements 
in terms of latency, reliability, security, mobility, and 
localization
of many new systems and applications 
(e.g., 
embedded artificial intelligence,
manufacture automation, 
  5G wireless systems) 
	\cite{mchi16}. 
To this end, edge computing (EC) \cite{msat17}, also known as fog computing (FC) \cite{mchi16},
has emerged as a new computing paradigm that complements the cloud to enable the implementation of
innovative services right at the network edge. 
%
%

EC forms a virtualized platform that 
 distributes computing, storage, 
control, and networking services closer to end-users to smarten the edge network. 
The size of an EN is flexible ranging from smartphones, PCs, 
smart access points (AP), base stations (BS) to edge clouds  \cite{wshi16}. 
For example, a smartphone is the edge between wearable devices and the cloud, a home gateway is the edge between smart appliances and the cloud, a cloudlet, a telecom central office, a micro DC 
is the edge between mobile devices and cloud core network. 
Indeed, the distributed EC infrastructure
encompasses any computing, storage, and networking nodes along the path between end devices and cloud DCs, not just exclusively nodes located at the customer edge \cite{wshi16}. 
By providing elastic resources and intelligence at the edge of the network,
EC offers many remarkable capabilities, including local data processing and analytics, distributed caching, 
location awareness, resource pooling and scaling, enhanced privacy and security, and 
reliable connectivity. 
%

These capabilities combined with the shorter communication distance allow operators to efficiently handle both downstream and upstream data between the cloud and the customer edge, which translates into drastic
network traffic reduction
and significant 
user experience improvement.
%
For instance, 
with edge caching, location-awareness, and real-time data processing and analysis, not only can service providers serve user content requests locally,
but also can adaptively optimize video coding and resolution according to the user device information 
and the varying wireless channel conditions. 
Also, it is envisioned that  
most of data produced by IoT sensors will be processed at the edge and only  important information and metadata will be sent to the cloud for further analytics. 
Additionally, EC is the key enabler for  ultra-reliable low-latency applications such as AR/VR, cognitive assistance, autonomous driving, industrial automation, remote robotics, 
and healthcare. 
A myriad of benefits and other use cases (e.g., computation offloading, caching, advertising, smart homes/grids/cities) of EC 
 can be found in \cite{wshi16,msat17,mchi16}.

Today, EC is still in the developing stages 
and presents many new challenges,
such as network architecture design, programming models and abstracts, IoT support, service placement, resource provisioning and management, security and privacy,  incentive design, 
and reliability and scalability of edge devices
\cite{wshi16,msat17,mchi16}. 
To unlock the huge potential of this new technology, it requires 
significant collaborative efforts between various entities in the ecosystem.
In this work, we focus on the EC resource
allocation problem.
Unlike cloud computing, where computational capacity of large DCs is virtually unlimited and network delay is 
high, 
EC is characterized by relatively low network latency but considerable processing delay due to the limited computing power of ENs. Also, there are a massive number of distributed computing nodes compared to a small number of large DCs.
%
Moreover, ENs may come with different sizes (e.g., number of computing units) and configurations (e.g., computing speed)
ranging from a smartphone to an edge cloud with tens/hundreds of servers. These nodes are dispersed in numerous locations 
with varying network and service delay towards end-users. 

On the other hand, different services  
may have 
different requirements and 
properties. 
Some services can only be handled by 
ENs satisfying 
certain criteria.
Additionally, different services
may be given different priorities.
%
%
While every service not only wants to obtain 
as much 
resource as possible but also prefers to be served by its closest ENs with 
low response time,
the capacities of ENs are limited. 
Also, due to the diverse preferences of the services towards the ENs, some nodes can be under-demanded while other nodes are over-demanded. 
%
%
Thus, a fundamental problem  is: 
 \textit{given a set of geographically distributed heterogeneous ENs, how can we efficiently allocate their limited computing resources to competing services with different desires and characteristics, 
considering service priority and fairness?} 
%
%
This work introduces 
a novel market-based solution framework which aims	not only to maximize the resource utilization of the ENs but also to make every service happy with the allocation decision. 

  The basic idea behind our approach is to assign different prices to resources of different ENs. In particular, highly
sought-after resources are priced high while prices of under-demanded resources are low. We assume that each service has a certain budget for resource procurement. The budget can be virtual or real money. Indeed, budget is used to capture service priority/differentiation. It can also be interpreted as the market power of each service.
 Given the resource prices, each service  buys the favorite resource bundle that it can afford. 
When all the resources are fully allocated, the resulting prices and allocation form a \textit{market equilibrium} (ME).
%
If there is
only one EN, an ME can be found easily by adjusting the
price gradually until demand equals supply or locating the
intersection of the demand and supply curves. However,
when there are multiple heterogeneous ENs and multiple
services with diverse objectives and different buying power,
the problem becomes challenging.
We consider two distinct market models in this work.

In the first
model, the money does not have intrinsic value to the
services. Given resource prices, each service aims to maximize its
revenue from the allocated resources, without caring about
how much it has to pay as long as the total payment does not
exceed its budget. 
This model arises in many real-world scenarios.
For example, in 5G networks, the Mobile Edge Computing (MEC) servers of a Telco are shared among different network slices, each of which runs a separate service  (e.g., voice, video streaming, AR/VR, connected vehicles, sensing) and serves a group of customers who pay for the service. The Telco can allot different budgets to the slices depending on their importance and/or potential revenue generation (e.g., the total fee paid by the users/subscribers of each slice). 
%

Similarly, an application provider (e.g., Uber, Pokemon Go) 
 or a sensor network may own a number of ENs in a city and need to allocate the edge resources to handle requests of different groups of users/sensors. The budget can be decided based on criteria such as the populations of users/sensors in different areas and/or payment levels (subscription fees) of different groups of users.
Another example is that 
a university (or other organizations) can grant different virtual budgets to different departments or research labs so that they can fairly share the edge servers on the campus.
The first model may also emerge in the setting of cloud federation at the edge where several companies (i.e., services) pool their resources together and each of them 
contributes a fixed portion of resource of every EN.
Here, the budgets are proportional to the initial contributions of the  companies. Instead of resource pooling, these companies may agree upfront on
their individual budgets, and then 
buy/rent a given set of ENs together. 


In these scenarios, it is important to consider both fairness and efficiency. Thus, 
conventional schemes such as social welfare maximization, maxmin fairness, and auction models may not be suitable.  In particular, a welfare
maximization allocation often gives most of the resources to users who have high marginal utilities  while 
users with low marginal utilities receive a very small amount of resources, even nothing. Similarly, in auction models, the set of losers are not allocated any resource. Hence, these solutions can be unfair to some users. On the other
hands, a maxmin fairness solution often allocates too many resources to users with low marginal utilities, hence, it
may not be efficient. 

To strive the balance between fairness and efficiency, we advocate the General Equilibrium Theory \cite{karr54}, with  a specific focus on the Fisher market model \cite{wbra00}, as an effective solution concept for this problem.  
Specifically, the first model can be cast as a Fisher market in which services act as buyers as ENs act as different goods in the market. 
For the linear additive utility function
 as considered in this work, given resource prices, 
a service may have an infinite set of optimal resource bundles,
 which renders difficulty in designing distributed algorithms. 
 We suggest several methods to overcome this challenge.
Moreover, we show that the obtained allocation is Pareto-optimal, which means there is no other allocation that would make some service better off without making someone else worse off \cite{eco}. In other words, there is no strictly ``better'' allocation.  Thus, a Pareto-optimal allocation is efficient.

 We furthermore link the ME to the fair division literature \cite{hmou04} and prove that the 
allocation satisfies remarkable fairness properties including envy-freeness, 
 sharing-incentive, and proportionality, which provides strong incentives for the services to participate in the proposed scheme. 
Indeed, these properties were rarely investigated explicitly in the ME literature. 
\textit{Envy-freeness}  means that every service prefers its allocation to the allocation of any other service. 
In an envy-free allocation, every service feels that its share is at least as good as the share of any other service, and thus no service feels envy. 
%
\textit{Sharing-incentive} is another well-known fairness concept. It  ensures that services get better utilities than what they would get in the \textit{proportional sharing} scheme that gives each service an amount of resource from every EN proportional to its budget. 
Note that proportional sharing is an intuitive way to share resources fairly in terms of quantity. 
For the federation setting, sharing-incentive implies that every service gets better off by pooling their resources (or money) together.   
Finally, it is natural for a service to expect to obtain a utility of at least  $b/B$ of the maximum utility that it can achieve by getting all the resources, where $b$ is the payment of the service and $B$ is the total payment of all the services. 
The \textit{proportionality} property guarantees that the utility of  every service at the ME is at least proportional to its payment/budget. Thus, it makes every service feel fair in terms of the achieved utility.

In the second model, the money does have intrinsic value to the services. The services not only want to maximize their revenues but also want to minimize their payments. In particular,
each service aims to maximize the sum of its remaining budget (i.e., surplus) and the revenue from the procured resources, which is equivalent to maximizing the net profit (i.e., revenue minus cost).
This model is prevalent in practice.
For example, several service providers (SP), each of which has a certain budget, may compete 
for the available resources of an edge infrastructure provider (e.g., a Telco, a broker).
The SPs only pay for their allocated resources and can take back their remaining budgets.
Obviously, a SP will only buy a computing unit if the potential gain from that unit outweighs the cost. 
It is natural for the SPs to maximize their net profits in this case. 
%
%
The traditional Fisher market model does not capture this setting since the utility functions of the services depend on the resource prices.

It is worth mentioning that, conventionally, the optimal dual variables associated with the supply demand constraints (i.e., the capacity constraints of the ENs) are often interpreted as the resource prices \cite{boyd} and common approaches such as  network utility maximization (NUM) \cite{dpal06} can be used to compute an ME.  However, these approaches do not work for our models that take budget into consideration.
Indeed, the main difficulty in computing an ME in both models stems from the budget constraints which contain both the dual variables (i.e., prices) and primal variables (i.e., allocation). In the second model, the prices also appear in the objective functions of the services.  Therefore, the ME computation problem becomes challenging.
Note that the pair of equilibrium prices and equilibrium allocation has to not only clear the market but also simultaneously maximize the utility of every service (as elaborated later in Section 4). 

Fortunately, for a wide class of utility functions, the ME in the first model can be found by solving a simple Eisenberg-Gale (EG) convex program \cite{EG,EG1,AGT}. However, the EG program does not capture the ME in the second model. 
%
Interesting, by reverse-engineering the structure of the primal and dual programs in the first model, 
we can rigorously construct a novel convex optimization problem whose solution is an ME of the second model. 
Our main contributions include:
%
%
\begin{itemize}

\item \textit{Modeling}. We formulate a new market-based EC resource allocation framework 
and advocate the General Equilibrium theory as an effective solution method for the proposed problem.

\item \textit{Centralized solution}. The unique ME in the first model can be determined by the EG program.
We also prove 
some salient fairness features of the ME.
 
\item \textit{Decentralized algorithms}. We introduce several distributed algorithms that 
 efficiently overcome the difficulty raised by the non-unique demand functions of the services and  converge to the ME. 

\item \textit{Extended Fisher market.} 
We systematically derive a new convex optimization problem whose optimal solution is an exact ME in the extended Fisher market model where buyers value the money.

\item \textit{Performance Evaluation.} Simulations are conducted to illustrate the efficacy of the proposed techniques. 

\end{itemize}


The rest of the report is organized as follows. Section \ref{related} describes related work. The system model and problem formulation are given in Section \ref{model} and Section  \ref{formu1}, respectively. 
The centralized solution using the EG program is analyzed in Section \ref{EG}. Then, we introduce several distributed algorithms in Section \ref{dist}. The market model in which buyers aim to maximize their net profits is studied in Section \ref{formu2}.
Simulation results are shown in Section \ref{sim} followed by conclusions and discussion of future work in Section \ref{con}.

\vspace{-0.1in}

\section{Related Work}
\label{related}


The potential benefits and many technical aspects of EC have been studied extensively in the recent literature.
First, the hybrid edge/fog-cloud system can be leveraged to improve the performance of emerging applications such as  cloud gaming and healthcare \cite{lgu17,ylin17}. 
A. Mukherjee {\em et. al.} \cite{amuk17} present a power and latency aware cloudlet selection strategy for computation offloading in a multi-cloudlet environment. 
The tradeoff between power consumption and service delay in a fog-cloud system is investigated in \cite{rden16} where the authors formulate a workload allocation problem to minimize the 
system energy cost 
under latency constraints.
 %
A latency aware workload offloading scheme in a cloudlet network is formulated in \cite{xsun17} to minimize the average 
 response time for mobile users.

 In \cite{mjia17}, M. Jia {\em et. al.} explore the joint optimization of cloudlet placement and user-to-cloudlet assignment to minimize service latency while considering load balancing.
A unified service placement and request dispatching framework is presented in \cite{lyan16} to evaluate the tradeoffs between the user access delay and service cost.
Stackelberg game and matching theory are employed in \cite{hzha17} to study the
joint optimization among 
data service operators (DSO), data service subscribers (DSS), and a set of ENs in a three-tier edge network where the DSOs can obtain computing resources from different ENs to serve their DSSs.
Another major line of research has recently focused on the 
 joint allocation of communication and computational resources for task offloading in 
the Mobile Edge Computing (MEC) environment \cite{ssar15,xche154,xlyu17}. 
MEC allows mobile devices to offload computational tasks to resource-rich servers located near or at cellular BSs, which could potentially  reduce the devices' energy consumption and task execution delay.   
However, these benefits could be  jeopardized if multiple 
users 
 offload their tasks to MEC servers simultaneously. In this case, a user may not only suffer severe interference but also receive a very small amount of EC resource, 
which would consequently reduce data rate, increase transmission delay, and cause high task execution time on the servers. 
Hence, offloading decision, allocation and scheduling of radio resources, and computational resources should be jointly considered in an integrated framework. 

Instead of optimizing the overall system performance from a single network operator's point of view,
we study the EC resource allocation problem from the 
game theory and market design perspectives \cite{AGT}. 
Specifically, we exploit the General Equilibrium  \cite{karr54},  a Nobel prize-winning theory, to construct an efficient market-based resource allocation framework. 
%
Although this concept was proposed more than 100 years ago  \cite{wbra00}, 
only until 1954, the existence of an ME was 
proved 
under mild conditions in the seminal work of Arrow and Debreu \cite{karr54}. 
However, their proof of existence  based  on fixed-point theorem is non-constructive and does not 
give an algorithm to compute an equilibrium
\cite{AGT}.
Recently, theoretical computer scientists have expressed great interests in understanding algorithmic aspects of the General Equilibrium concept. Various efficient algorithms and complexity analysis for ME computation have been accomplished over the past decade \cite{ndev08,vvaz11,nche16,AGT,xche17,jgag15}. 
	Note that although the existence result has been established, there is no general technique for computing an ME. 

%
%

Our proposed models are inspired by the Fisher market \cite{wbra00} which is a special case of the exchange market model in the General Equilibrium theory. 
An \textit{exchange market} model
 consists of a set of economic agents trading different types of divisible goods. Each agent has an initial endowment of goods
and  a utility function 
representing her preferences for the different bundles of goods. Given the goods' prices, 
every agent sells the initial endowment, and then uses the revenue to buy the best bundle of goods they can afford \cite{AGT,karr54}.
The goal of the market is to find the equilibrium prices and allocations 
that maximize every agent's utility
respecting the budget constraint, and the market clears. 
 In the Fisher market model, every agent comes to the market with an initial endowment of money only and wants to buy goods available in the market. 
We cast the EC resource allocation problem as a Fisher market. We not only show appealing fairness properties of the equilibrium allocation, but also introduce efficient distributed algorithms to find an ME.
More importantly, we systematically devise a new and simple convex program to capture the market in which money has intrinsic value to the buyers, which is beyond the scope of 
the Fisher and exchange market models.

Note that 
 there is a rich literature on cloud resource allocation and pricing  \cite{nluo17}. In  \cite{hxu13,atoo15}, the authors propose
different profit maximization frameworks for cloud providers.
References \cite{lmas152,mhad17,ipet15} study how to efficiently share resource and profit among cloud providers in a cloud federation. Several resource procurement mechanisms are introduced in \cite{apra14} to assist a cloud user to select suitable cloud vendors in a multi-cloud market. 
In \cite{dard13}, the interaction between a cloud provider and multiple services is modeled as a generalized Nash game. This model is extended to a multi-cloud multi-service environment in \cite{dard17}. A single-cloud multi-service resource provision and pricing problem with flat, on-demand, and on-spot VM instances is formulated in \cite{vcar18} as a Stackelberg  game, which not only maximizes the revenue of the cloud provider but also minimizes costs of the services.

Auction theory has been widely used to study cloud resource allocation \cite{lmas151,schi17,xwan15}.
A typical system consists of one or several clouds and multiple users. First, the users submit bids, which include their desired resource bundles in terms of VM types and quantities as well as the price that they are willing to pay, to an auctioneer. Then, the auctioneer solves a winner determination problem to identify accepted bids. Finally, the auctioneer calculates the payment that each winner needs to pay to ensure truthfulness. In auction, the common objectives are to maximize the social welfare or maximize the profit of the cloud provider. 
Additionally, only winners receive cloud resources.  Furthermore, most of existing auction models do not consider elastic user demands. For example, previous works often assume that cloud users are single-minded, who are interested in a specific bundle only  and have zero value for other bundles.

Different from the existing works on cloud economics and resource allocation in general, our design objective is to find a fair and efficient way to allocate resources from multiple nodes (e.g., ENs) to budget-constrained agents (i.e., services), which makes every agent happy with her resource allotment and ensures high edge resource utilization. 
The proposed model also captures practical aspects, for example, a service request can be served at different ENs and service demands can be defined flexibly rather than fixed bundles as in auction models.

\section{System Model}
\label{model}



An EC environment is depicted in Fig.~\ref{fig:sys}. Besides  local execution and remote processing at cloud DCs, data and requests from  end-devices (e.g., smartphones,  set-top-boxes, sensors) can be handled by the EC platform.  
Note that some data and computing need to be done in the local to keep data privacy.
A request typically first goes to a Point of Aggregation (PoA) (e.g., switches/routers, BSs, APs), then it will be routed to an EN for processing.  
Indeed, enterprises, factories, organizations (e.g., hospitals, universities, museums), commercial buildings (shopping malls, hotels, airports), and other third parties (e.g., sensor networks) can also outsource their services and computation to the intelligent edge network. Furthermore, service/content/application providers like Google, Netflix, and Facebook 
can proactively install their content and services onto ENs to serve better  their customers.
In the EC environment, various sources (e.g., smartphones, PCs, servers in a lab, underutilized small/medium
data centers in schools/hospitals/malls/enterprises, BSs,
telecom central offices) can act as ENs.



We consider a system encompassing 
various services and a set of geographically distributed ENs with different configurations and limited computing capacities. Each service has a budget for resource procurement
 and wants to offload as many requests as possible to the edge network. The value of an EN to a service is measured in terms of the maximum revenue that it can generate by using the EN's resource. 
An EN may have different values to different services. 
Since some ENs (e.g., ones with powerful servers) 
can be over-demanded while some others are under-demanded, it is desirable to harmonize the interests of the services so that each service is happy with its  allotment while ensuring high resource utilization. 
%
%
An intuitive 
solution is to assign  prices to ENs and let each service choose its favorite resource bundle. 
%
We assume that there is a platform 
lying between the services and the ENs. 
 Based on the information collected from  the ENs  (e.g., computing capacity) and the services (e.g., budgets, preferences), the platform computes an ME solution including resource prices and allocation, which not only maximizes the satisfaction of every service but also fully allocates the ENs' resources.

In the first model, each service seeks solely to maximize its revenue under the budget constraint, without concerning about the money surplus after purchasing resources.
This can be the case where the services and ENs belong to the same entity, and each service is assigned a virtual budget representing the service's priority. 
For instance, a Telco can give different budgets to different network slices, each of which runs a service (e.g., voice, video streaming, AR/VR, connected vehicles).
 %
In the second model, the remaining money does have intrinsic value to the services. In this case, each service aims to maximize the sum of its remaining budget and the  revenue from the procured resources. For example, this can be the case where services and ENs are owned by different entities, and each SP (e.g., Google, Facebook, enterprises) 
has a certain budget for leasing resources from an infrastructure provider (e.g., a Telco).
 For simplicity, we assume that the values of ENs to the services are fixed. 
Our model can be extended to capture time-varying valuation in a multi-period model by considering each pair of an EN and a time slot as an independent EN. 

\section{Problem Formulation}
\label{formu1}



\subsection{EC Resource Allocation Problem}

Let $\mathcal{M}$, $\mathcal{N}$, M, and N be the sets of ENs and services, and the numbers of ENs and services, respectively.  
%
Denote $i$ as the service index and $j$ as the EN index. We assume that each EN $j$ has 
$c_j$ homogeneous computing units (e.g., servers) 
\cite{hzha17}. If an EN has several types of computing units, 
we can always divide the EN into several clusters, each of which contains only homogeneous units. Then, each cluster can be considered as a separate EN.  While the computing units in each EN are homogeneous, different ENs can have different types of computing units. 
%
%
Let $x_{i,j}$ be the number of computing units of EN $j$ allocated to service $i$. The vector of resources allocated to service $i$ is $x_i = \big( x_{i,1}, x_{i,2}, \ldots, x_{i,M} \big)$. Finally, define $B_i$ as the budget of service $i$.

Our goal is to compute an ME including an equilibrium price vector $p = (p_1, p_2, . . . , p_M)$, where $p_j$ is price of EN $j$, 
and a resource allocation matrix $\mathcal{X}$,
in which the element at the $i$th row and $j$th column is $x_{i,j}$. The utility $U_i(x_i,p)$ of service $i$ is defined as a function   of the amount of resources $x_i$ that it receives and the resource prices $p$. 
The capacity constraint of ENs renders: $\sum_{i=1}^N x_{i,j} \leq c_j, ~\forall j \in \mathcal{M} $.
Without loss of generality, we normalize the capacity of every EN to be 1 (i.e., $c_j = 1, ~\forall j$) and scale 
related parameters (e.g., price, resource allocation) accordingly. 
This normalization is just to simplify expressions and equations.  Hence, we have:
 $\sum_{i=1}^N x_{i,j} \leq 1, ~ \forall j, ~~ x_{i,j} \geq 0,~~ \forall  j.$

Each service is a player in our market game. Given a price vector $p$, service  $i$ aims to maximize its utility $U_i(x_i,p)$ subject to the budget constraint $\sum_j x_{i,j} p_j \leq B_i$.

%

\begin{definition}
\label{MEdef} 
An ME solution ($p^*$,$X^*$) needs to satisfy the two conditions:
\end{definition}

\begin{itemize}

\item \textit{Condition 1}: Given the equilibrium resource price vector $p^* = (p_1^*, p_2^*, . . . , p_M^*)$, every service $i$  receives its optimal resource bundle $x_i^*$, i.e., we have
\beqn
x_i^* =  (x_{i,1}^*, \ldots, x_{i,M}^*) \in \argmax_{x_i \geq 0; \sum_j p_j^* x_{i,j} \leq B_i} U_i(x_i,p^*) 
\eeqn 

\item \textit{Condition 2}:  All the resources are fully allocated,  i.e. , we have:  $\sum_{i} x_{i,j} = 1, ~~\forall j$.
%

\end{itemize}   

The first condition can be interpreted as the \textit{user satisfaction condition} while the second condition is often called the \textit{market clearing condition} in Economics \cite{eco}.  The first condition ensures that the equilibrium allocation $x_{i}^*$ maximizes the utility of service $i$ at the equilibrium prices $p^*$ considering the user budget constraint. The second condition maximizes the resource utilization of the ENs. It also means the ENs' resources are fully sold in the market, which consequently maximizes the profit of every EN since the equilibrium prices are non-negative. The services are players competing for the limited EC resources, while the platform tries to satisfy the market clearing condition. Prices are used to coordinate the market.

Let $u_i(x_i)$ be the gain/profit/revenue of service $i$ can achieve from the procured resources. We consider two models.  In the first model (basic model), every service $i$ wants to maximize $U_i(x_i,p) =  u_i(x_i)$ and does not care about how much it has to pay as long as the total payment is under its budget. 
%
Here, utility of a service is its revenue.
In the second model, instead of revenue, the services aim to maximize their net profits (i.e., revenue minus cost).
The service utility in this model is $U_i(x_i,p) =  u_i(x_i) - \sum_j p_j x_{i,j},~\forall i$.
We focus on the first model throughout the report. The second model is examined in Section  \ref{formu2}.

\vspace{-0.1in}
\subsection{Service Utility Model}


In practice, the services may use different criteria to define $u_i(x_i)$. 
Our framework takes $u_i(x_i)$ as an input to compute an ME solution. How each service evaluates the ENs is not the focus of this work.  While the proposed model is generic, we consider linear functions for the ease of exploring the framework.
Extensions to more general functions will be discussed throughout the work. Let $a_{i,j}$ be the gain of service $i$ from one unit of resource of EN $j$. Then, we have: $u_i(x_i) = \sum_j a_{i,j} x_{i,j}, ~\forall i$.

In the following, we present an example of how $a_{i,j}$ can be computed. 
 We consider only delay-sensitive services, which are also the main target application of EC. For simplicity, we assume that the transmission bandwidth is  sufficiently large and the data size of a request is small (e.g., Apple Siri, Google Voice Search,  Google Maps, AR, and Translation).
Hence, the data transmission delay (i.e., size/bandwidth) is assumed to be negligible and  we consider only propagation delay and processing delay \cite{zliu15,dard13}. 

The total delay of a request of service $i$ from the time a user sends the request to the time she receives a response includes the round-trip delay $d_i^{\sf UE-PoA}$ between the user and a PoA of the service, the round-trip network delay 
$d_{i,j}^{\sf n}$ between the PoA and an EN $i$ hosting the service, and the processing delay at the EN $d_{i,j}^{\sf p}$. Note that an EN can be located in the same place with a PoA (e.g., a BS). 
In reality, $d_i^{\sf UE-PoA}$ is quite small, 
and we assume it is fixed similar to \cite{xsun17}. 
In other words, we study the system only \textit{from the aggregation level to the EC platform}. 
 For simplicity, we assume that each service is located at one PoA (e.g., an IoT gateway, a BS, a building). If a service has several PoAs, we need to take sum over all the PoAs to get the total number of requests 
of the service 
handled by the EC platform. 
%
Denote $T_i^{\sf max}$ as the maximum tolerable delay of service $i$, 
we have
\beqn
d_{i,j}^{\sf p} + d_{i,j}^{\sf n} \leq T_i^{\sf max}, \quad \forall i,~j.
\eeqn
Obviously,  the maximum number of requests  $\lambda_{i,j}^{\sf max}$ that EN $j$ can process is zero if $d_{i,j}^{\sf n} \geq T_i^{\sf max}$.

We model the processing delay at ENs using the widely used M/G/1 queues and assume that the workload is evenly shared among computing units \cite{ltan17,hzha17,dard13,zliu15}. 
The average response time $d_{i,j}^{\sf p}$ of EN $j$ for processing service $i$ can be computed as follows: 
\beqn
\label{queuemodel}
d_{i,j}^{\sf p} = \frac{1}{ \mu_{i,j} - \frac{\lambda_{i,j}}{x_{i,j}}  }, \quad \forall i,~j
\eeqn
where $\mu_{i,j}$ be the service rate of one computing unit of EN $j$ for handling  service $i$, 
and $\lambda_{i,j}$ is the request arrival rate (i.e., number of requests per time unit) of service $i$ to EN $j$. For queue stability, we have $\frac{\lambda_{i,j}}{x_{i,j}} < ~\mu_{i,j},~ \forall i,~j.$
Otherwise, the queuing delay will be infinite as requests accumulated.

From (\ref{queuemodel}), we have 
\beqn
\frac{1}{ \mu_{i,j} - \frac{\lambda_{i,j}}{x_{i,j}}  } \leq T_i^{\sf max} - d_{i,j}^{\sf n} \\ \nonumber
\Rightarrow \lambda_{i,j} \leq x_{i,j} \Big( \mu_{i,j} - \frac{1}{T_i^{\sf max} - d_{i,j}} \Big) .
\eeqn

Therefore, if $d_{i,j}^{\sf n} < T_i^{\sf max}$, the maximum number of requests that service $i$ can process at EN $j$ is
\beqn
 \lambda_{i,j}^{\sf max} &=& \max \Big\{  x_{i,j} \Big( \mu_{i,j} - \frac{1}{T_i^{\sf max} - d_{i,j}}\Big),~0 \Big\}\\ \nonumber
 &=& x_{i,j} q_{i,j} , \quad \forall i,~j
\eeqn
where $q_{i,j} =  \max \Big\{ \Big( \mu_{i,j} - \frac{1}{T_i^{\sf max} - d_{i,j}}\Big),~0 \Big\}$. 
%
%
%
Define a successful request as the request whose total delay is smaller or equal to the maximum delay tolerance. 
Let $r_i$ be the benefit of successfully serving one request of service $i$ \cite{hzha17}. Then, given $x_{i,j}$ computing units, the revenue of service $i$ is 
\beqn
\label{a_uti}
u_{i,j}(x_{i,j}) = r_i q_{i,j} x_{i,j} = a_{i,j} x_{i,j}, \quad \forall i,~j
\eeqn
with $a_{i,j} = r_i q_{i,j}$. 
Thus, we have 
\beqn
\label{utifunc}
u_i(x_i) = \sum_{j = 1}^M u_{i,j} = \sum_{j=1}^M a_{i,j} x_{i,j}, \quad \forall i
\eeqn
in which $a_{i,j}$ can be computed beforehand. Note that we implicitly assume the request pool of a service is unlimited. We will discuss later how some assumptions can be relaxed.

\begin{definition}
\label{homo} 
A function $u(.)$ is \textit{homogeneous of degree} $d$, where $d$ is a constant, if $u(\alpha x) = \alpha^d u(x),~\forall ~\alpha > 0$ \cite{AGT}. 
\end{definition}
From  (\ref{utifunc}), it is easy to verify that 
$u_i(x_i)$ is a linear function that is homogeneous of degree 1. 

	\textbf{Remark:} The value of an EN to a service can be defined flexibly. For example, a service may give higher values to  ENs in a populated area or ENs with high reliability. A suitable weight can be added to $a_{i,j}$. 
	In the proposed model, each service informs the platform its budget and how much it values different ENs. Based on these information, the platform computes suitable resource allocation satisfying given design objectives. How each service utilizes its allocated resources in the operation stage is not the focus of this work.  
The key concern of our work is how to harmonize the interests of different services that may have different preferences towards the ENs. Also, we  consider  only delay-sensitive services to illustrate one way to model the service utility function.
 It can be justified by the fact that non-delay-sensitive services can be handled effectively by cloud DCs and the precious edge resources can be reserved for important low-latency services. Nevertheless, our model is generic enough to handle other service types as long as we can define the  utility of a service as a suitable function of its allocated EC resources. 
 Finally, although we consider computing resources only, the proposed framework can apply to a system in which each service evaluates an EN based on a combination of different resource types of the EN, such as computing, storage, and bandwidth.  


\vspace{-.02in}
\section{Centralized Solution}
 \label{EG}

In the first model, each service $i$ aims to maximize $U_i(x_i,p) = u_i(x_i) = \sum_j a_{i,j}x_{i,j}$ subject to the budget constraint $\sum_j p_j x_{i,j} \leq B_i$, $\forall i$.
If $p$ is a price vector, the ratio $a_{i,j}/p_j$ is defined as the \textit{bang-per-buck} of EN $j$ to service $i$, which indicates the utility gained by service $i$ through one unit of money spent on EN $j$ (assuming 0/0 = 0). The  \textit{maximum bang-per-buck (MBB)} of service $i$ over the set of ENs is $\alpha_i = \max_j \{a_{i,j}/p_j\}$ \cite{ndev08}. 
The demand set $D_i(p)$ of service $i$ includes all ENs giving it the MBB value, i.e., 
$D_i(p)  = \{ j :  a_{i,j}/p_j = \alpha_i \},~\forall i$.
Intuitively, to maximize its utility, each service will spend full budget to buy resources from only ENs giving it the MBB. Therefore,   a pair $(X, p)$ is an ME if: i) given prices  $p$, service $i$ will exhaust its budget to buy resources only from ENs in $D_i(p)$; and ii) the market clears at prices $p$.
In the following, we will show that the ME in the first model can be inferred from the optimal solution of a convex optimization problem. Also, we will describe some properties of the equilibrium.
Specifically, for the case of buyers with linear utilities, the ME 
can be found by solving the EG convex program given below \cite{EG,AGT}:
%
%
%
\beqn
\label{EGprogram1}
\vspace{-0.1in} 
\underset{\mathcal{X},u}{\text{maximize}} ~\sum_{i=1}^N B_i \ln u_i
\eeqn
subject to
\vspace{-0.3in} 
\beqn
\label{EQ11}
u_i =\sum_{j=1}^M a_{i,j} x_{i,j}, \quad \forall i  \\
\vspace{-0.5in}
\label{EQ12}
\sum_{i=1}^N x_{i,j} \leq  1, \quad \forall j \\
\label{EQ13}
x_{i,j} \geq 0, \quad \forall i,~j.
\eeqn
 This problem always has an interior feasible solution by simply setting $x_{i,j} = \epsilon > 0$, for all $i$ and $j$, where $\epsilon$ is sufficiently small such that all constraints (\ref{EQ12})-(\ref{EQ13}) are satisfied with strict inequality. 
Hence, Slater’s condition holds and the the Karush--Kuhn--Tucker (KKT) conditions are necessary and sufficient for optimality \cite{boyd}.
Denote $\eta_i$, $p_j$, and $\nu_{i,j}$ as the dual variables associated with constraints (\ref{EQ11}), (\ref{EQ12}), and (\ref{EQ13}), respectively. We have the Lagrangian
\beqn
L(u,X,\eta,p,\nu) = \sum_i B_i \ln u_i + \sum_j p_j ( 1 - \sum_i x_{i,j})   \\ \nonumber
+ \sum_i \eta_i \Big( \sum_j a_{i,j} x_{i,j}  - u_i \Big) + \sum_i \sum_j \nu_{i,j} x_{i,j}.
\eeqn
\vspace{-0.1in}
The KKT conditions give 
\beqn
\label{kkt1}
\frac{\partial L}{\partial u_i} = \frac{B_i}{u_i} -\eta_i = 0,~\forall i \\
\vspace{-0.1in}
\label{kkt2}
\frac{\partial L}{\partial x_{i,j}} = B_i \frac{a_{i,j}}{u_i} - p_j + \nu_{i,j} = 0, \quad \forall i,~j \\
\vspace{-0.2in}
\label{kkt3}
 u_i =\sum_j a_{i,j} x_{i,j} ,~ \forall i; ~p_j (1 - \sum_i x_{i,j}) = 0, \quad \forall j \\
\label{kkt4}
\nu_{i,j} x_{i,j} = 0, ~ \forall i,~j;~ p_j \geq 0,  ~ \forall j; ~ \nu_{i,j} \geq 0, ~ \forall i,~j.
\eeqn
We can infer the following
\beqn
\label{con1}
\forall i,j: \frac{u_i}{B_i} \leq \frac{a_{i,j}}{p_j}\\
\label{con2}
\forall i,j: \text{if} ~x_{i,j} > 0 \Rightarrow \nu_{i,j} = 0 \Rightarrow \frac{u_i}{B_i} = \frac{a_{i,j}}{p_j}\\
\label{con3}
\forall j: p_j > 0 \Rightarrow \sum_i {x_{i,j}} = 1; ~ \sum_i {x_{i,j}} < 1   \Rightarrow p_j = 0.
\eeqn

If $p$ is a price vector, the ratio $a_{i,j}/p_j$ is defined as the \textit{bang-per-buck} of EN $j$ to service $i$, which indicates the utility gained by service $i$ through one unit of money spent on EN $j$ (assuming 0/0 = 0). The  \textit{maximum bang-per-buck (MBB)} of service $i$ over the set of ENs is $\alpha_i = \max_j \{a_{i,j}/p_j\}$ \cite{ndev08}. 
The demand set $D_i(p)$ of service $i$ includes all ENs giving it the MBB value, i.e., 
$D_i(p)  = \{ j :  a_{i,j}/p_j = \alpha_i \},~\forall i$.

The dual variable $p_j$ in the EG program can be interpreted as the price of EN $j$. Hence, conditions (\ref{con1}) and (\ref{con2}) imply that $x_{i,j} > 0 $ if and only if $j \in D_i(p)$, i.e., each service buys resources only from ENs giving it the MBB. This also maximizes $u_i(x_i)$.
The following theorem captures the relationship between the EG program and the ME solution as well as some properties of the equilibrium.

\begin{theorem} 
\label{theorem1}
The optimal solution to the EG convex program (\ref{EGprogram1})-(\ref{EQ13}) is an ME. Specifically, the Lagrangian dual variables corresponding to the ENs' capacity constraints (\ref{EQ12}) are the equilibrium prices. 
At the equilibrium, the resource allocation not only maximizes the utility  but also exhausts the budget of every service.
Furthermore, each service purchases resources only from ENs giving its MBB.
Additionally, the optimal utilities of the services as well as equilibrium prices are unique. 
\end{theorem}
\textbf{Proof}:  Let $X^*$ and $u_i^*$ be the optimal solution to the EG program. Then, $X^*$ and $u_i^*$ need to satisfy the KKT conditions (\ref{kkt1})-(\ref{con3}). Denote $\eta^*$, $p^*$, and $\nu^*$ as the optimal dual variables. 
From (\ref{kkt2}), we have 
\beqn
\label{budeq1}
B_i \frac{a_{i,j}}{u_i^*} = p_j^* - \nu_{i,j}^* , \quad \forall i,~j.
\eeqn
Multiplying both sides of (\ref{budeq1}) by $x_{i,j}^*$ and adding the resulting equalities,  we get
\beqn
\label{budeq2}
\frac{B_i}{u_i^*} \sum_j a_{i,j} x_{i,j}^* = \sum_j (p_j^* - \nu_{i,j}^*) x_{i,j}^* , \quad \forall i,~j. 
\eeqn
Since $\nu_{i,j}^* x_{i,j}^* = 0, \forall i,~j$, and $u_i* =  \sum_j a_{i,j} x_{i,j}^*, ~\forall i$,  equation (\ref{budeq2}) implies $\sum_j p_j^* x_{i,j}^* = B_i, ~\forall i$. Thus, the optimal solution to the EG program  (\ref{EGprogram1})-(\ref{EQ13}) fully exhausts the budget of every service. Furthermore, as shown above, at the optimality, each service buys resources only from ENs giving its MBB value. In other words, the optimal solution to the EG program maximizes the utility of every service subject to the budget constraint because every service uses all of its money to purchase its MBB resources. This can be inferred from (\ref{con1}) and (\ref{con2}).   

We now 
consider the market clearing condition. From (\ref{con3}), we can observe that resources of ENs with positive price $p_j$ are fully allocated. For ENs with zero prices, their resources can be allocated arbitrarily without affecting the optimal utility of service since the price is zero \cite{AGT}. Thus, the market clears. Since ($X^*$,~ $p^*$) satisfies both conditions of an ME, the optimal solution to the EG program is an ME. 
%

Finally, since the objective function (\ref{EGprogram1}) is strictly concave in $u_i$ for all $i$, the optimal utilities are unique. 
The uniqueness of equilibrium prices can be inferred from (\ref{con2}).
\qedb

From (\ref{budeq1}), if $p_j^* = 0, \text{then} ~ \nu_{i,j}^* = 0$ and $a_{i,j} = 0, ~\forall i,~j$, which means an EN has price of zero only when it is not wanted by all services. We can remove this EN from our system. In the following, we consider only the case where $p_j > 0, \forall j$. 
Also, it can be shown that Theorem \ref{theorem1} is not only applied to linear utilities, 
but also true for a wider class of homogeneous  concave utility functions \cite{EG1}. Please refer to \textbf{Appendix D} for more details.
Next, we study the properties of the equilibrium allocation. First, from (\ref{EGprogram1})-(\ref{EQ13}), it can be easily verified that the equilibrium allocation is scale-free. It means that it does not matter if service $i$ reports $a_i = (a_{i,1},\ldots,a_{i,M})$ or $e_i a_i$ for some constant $e_i$, the allocation that it receives is the same. 
Also, if a service divides its budget into two parts and acts as two different services with the same original utility function, then the total allocation it obtains from the new ME is equal to the original equilibrium allocation. 
%
Furthermore, the equilibrium allocation is not only Pareto-optimal but also possesses many appealing fairness properties such as envy-freeness, sharing incentive, and proportionality. 

An allocation is \textit{Pareto-optimal} if there is no other allocation that would make some service better off without making someone else worse off \cite{eco}, which means there is no strictly ``better'' allocation. Hence, a Pareto-optimal allocation is efficient and non-wasteful because the remaining resources (if any) cannot improve utility of any service. 
 \textit{Envy-freeness} means that every service prefers its allocation to the allocation of any other service. When the services have equal budgets, an envy-free allocation $\mathcal{X}$ implies $u_i(x_i) \geq u_i(x_{i'})$ for all $i$ and $i' \in \mathcal{N}$ \cite{hmou04}. 
In an envy-free allocation, every service feels that her share is at least as good as the share of any other service, and thus no service feels envy.
Since the budgets can be different, we need to extend  the classical definition of envy-freeness. 
An allocation $\mathcal{X}$ is envy-free if $ u_i(\frac{x_i}{B_i}) \geq u_i ( \frac{x_{i'}}{B_{i'}}),~\forall i,~i' \in \mathcal{N}$.

Let $\hat{x}$ be the allocation where each service receives
resource  from every EN proportional to its budget, i.e., $\hat{x}_{i,j} = \frac{B_i}{\sum_i' B_{i'}}, \forall i,~j$. 
\textit{Sharing-incentive} property implies $u_i(x_i) \geq u_i(\hat{x}_i),~\forall i.$ 
Indeed, $\hat{x}$ is an intuitive resource-fair allocation that allocates resources  from every EN to each service proportional to the service budget. We can also understand that each service $i$ contributes an amount of $\hat{x}_{i,j}$ to EN $j$ in a resource pool consisting of the ENs.
Sharing-incentive ensures that every service prefers the equilibrium allocation to its initial resource contribution to the pool. This can be interpreted as resource-fairness.

Finally, if $u_i(x_i) \geq \frac{B_i}{\sum_{i'} B_{i'}} u_i (C)$, for all $i$, in which $u_i(C)$ is the utility of service $i$ when it receives all the resources from the market (i.e., $C = (1,...,1), C \in \mathcal{R}^M$), we say that the allocation $\mathcal{X}$ satisfies the \textit{proportionality}  property.
Indeed,  $u_i(C)$ is the maximum utility that every service  $i$ can achieve from the EC resource pool. The proportionality property guarantees that the utility of  every service at the ME is at least proportional to its payment/budget. Thus, this property can be interpreted as utility-fairness.
Obviously, these fairness properties encourage services to participate in the proposed resource allocation scheme. 

\begin{theorem} 
\label{theorem2}
At equilibrium, the allocation is Pareto-optimal and envy-free. It also satisfies the sharing-incentive and proportionality properties. 
\end{theorem}
\textbf{Proof:} 
Since at the equilibrium, every service exhausts its budget and receives its favorite resource bundle, it does not envy with other services. Hence, the equilibrium allocation is envy-free. The Pareto-optimality follows directly from the first-welfare theorem in Economics \cite{eco,AGT}. Indeed, Pareto-optimality can also be inferred from the Nash Bargaining concept \cite{jnas50}. In particular, the problem (\ref{EGprogram1})-(\ref{EQ13}) has the objective in the form of a Nash Social Welfare function with closed, compact, and convex feasible region. Thus, it enjoys all compelling properties of a Nash Bargaining solution such as Pareto efficiency and scale-invariance. For linear utilities, we can prove the properties above directly 
as follows. 

- \textit{Pareto Optimality:} 
We show this by contradiction. Assume allocation $X^*$ is not Pareto-optimal. Then, there exists an allocation $X'$ such that $u_i(x'_i) \geq u_i(x_i^*)$ for all $i$, and $u_i(x'_i) > u_i(x_i^*)$ for some $i$. Note that $u_i(x_i) = \sum_j a_{i,j} x_{i,j}$. Consider any feasible allocation $X'$. 
Recall the MBB of buyer $i$ is $\alpha_i = \max_j \frac{a_{i,j}}{p_j}$. We have
\beqn
\label{PO1}
\sum_j x'_{i,j} p_j \geq \sum_j x'_{i,j} \frac{a_{i,j} }{\alpha_i} \geq \sum_j x_{i,j}^* a_{i,j} \frac{1}{\alpha_i} = \sum_j x_{i,j}^* p_j .
\eeqn
\vspace{-0.1in}
The second inequality is due to 
$u_i(x'_i) \geq u_i(x_i^*),~\forall i$.
Thus
\beqn
\label{PO2}
\sum_j x'_{i,j} p_j \geq B_i, \quad \forall i.
\eeqn
Since $u_i(x'_i) > u_i(x_i^*)$ for some $i$, 
$\sum_j x'_{i,j} p_j \geq B_i$ for some $i$.
Adding both sides of (\ref{PO2})  over all buyers renders
\beqn
\label{PO3}
\sum_i B_i < \sum_i \sum_j x'_{i,j} p_j = \sum_i x_{i,j} \sum_j p_j \leq \sum_j p_j
\eeqn
because $\sum_i x'_{i,j} \leq 1, ~ \forall j$ (i.e., the capacity constraints of ENs).
However, (\ref{PO3}) means the total prices of all the ENs is greater than the total budget of all buyers, which cannot occur. Thus,  the equilibrium allocation $X^*$ is Pareto-optimal. 

- \textit{Envy-freeness}: To prove that  $X^*$ is envy-free, 
we need to show: $B_{i'} u_i(x_i^*) \geq B_i u_i(x_{i'}^*), ~\forall i,~i' \in \mathcal{N}$. Let $b_{i,j}$ be the total money that service $i$ spends on EN $j$.
We have 
\beqn
\label{envy}
B_{i'} u_i(x_i^*) &=& B_{i'} \sum_j a_{i,j} x_{i,j}^* = B_{i'} \sum_j a_{i,j} \frac{b_{i,j}^*}{p_j} \\ \nonumber
&=& B_{i'} \sum_j \frac{a_{i,j}}{p_j} b_{i,j}^* = B_i' \alpha_i \sum_j b_{i,j}^* \\ \nonumber
&=& B_i' \alpha_i B_i  = B_i \alpha_i \sum_j b_{i',j}^* \\ \nonumber
\vspace{-0.1in}
&\geq& B_i \sum_j \frac{a_{i,j}}{p_j} b_{i',j}^* = B_i \sum_j a_{i,j} \frac{b_{i',j}^*}{p_j}\\ \nonumber
&=& B_i \sum_j a_{i,j} x_{i',j}^* = B_i u_i(x_{i'}^*), ~\forall i,~j. 
\eeqn
Note that the equalities in the second line of (\ref{envy}) can be inferred from the fact that each buyer only buys resources from ENs in its demand set $D_i$ while the first inequality in the fourth line holds because $\alpha_i \geq \frac{a_{i,j}}{p_j},~\forall i,~j$.

- \textit{Proportionality:} From Theorem \ref{theorem1}, $\sum_i x_{i,j}^* = 1,~\forall j$. Thus, for linear utilities and the envy-free property, we have
\beqn
u_i(C) = u_i \big(\sum_i x_i^* \big) 
 = u_i \big(x_i^* \big) + \sum_{i' \neq i} u_i \big( x_{i'}^* \big) \\ \nonumber
\leq  u_i \big(x_i^* \big) + \sum_{i' \neq i} \frac{B_{i'}}{B_i} u_i \big(x_i^* \big) = \frac{\sum_{i'} B_{i'}}{B_i} u_i \big(x_i^* \big).
\eeqn 
Hence, $u_i(x_i^*) \geq \frac{B_i}{\sum_{i'} B_{i'}} u_i (C), \forall i$. 

- \textit{Sharing-incentive:} At the ME ($X^*, p^*$), no service spends more than its budget. We have 
\beqn
\sum_i \sum_j x_{i,j}^* p_j^* \leq \sum_i B_i \Rightarrow \sum_j p_j^* \sum_i x_{i,j}^* \leq \sum_i B_i 
\eeqn
Thus, $\sum_j p_j^* \leq \sum_i B_i$. Consequently, resource bundle $\hat{x}_i$ costs service $i$: $\sum_j \hat{x}_{i,j} p_j^* = \sum_j \frac{B_i}{\sum_{i'} B_{i'}} p_j^* \leq B_i, \forall i.$ So, service $i$ can afford to buy bundle $\hat{x}_i$ at prices $p^*$. 
However, out of all feasible bundles that are affordable to service $i$, its favorite one is $x_i^*$.
It means $u_i(x_i^*) \geq u_i(\hat{x}_i), ~\forall i.$
%
\qedb
\vspace{-0.1in}

\section{Decentralized Solution}
\label{dist}

A common approach for implementing distributed algorithm is to let the platform iteratively compute prices  of the ENs  and broadcast the updated prices to the services. Then, each service finds its optimal demand bundle and sends the updated demand to the platform. 
This price-based strategy can be implemented in a tatonnment style or using the dual decomposition method  \cite{dpal06}. 
Unfortunately, linear utilities may result in non-unique optimal demand bundles because multiple ENs may give the same MBB to a buyer. 
Hence, the algorithm cannot terminate without aggregated demand coordination from the platform.
Consider an example with two services and three ENs. 
The system parameters are:  $B_1$ = \$1, $B_2$ = \$4, $a_1 = (1, 10, 4)$, and $a_2 = (4, 8, 8)$.

\begin{figure}[ht]
	\centering
		\subfigure[With coordination]{
		  \includegraphics[width=0.22\textwidth,height=0.10\textheight]{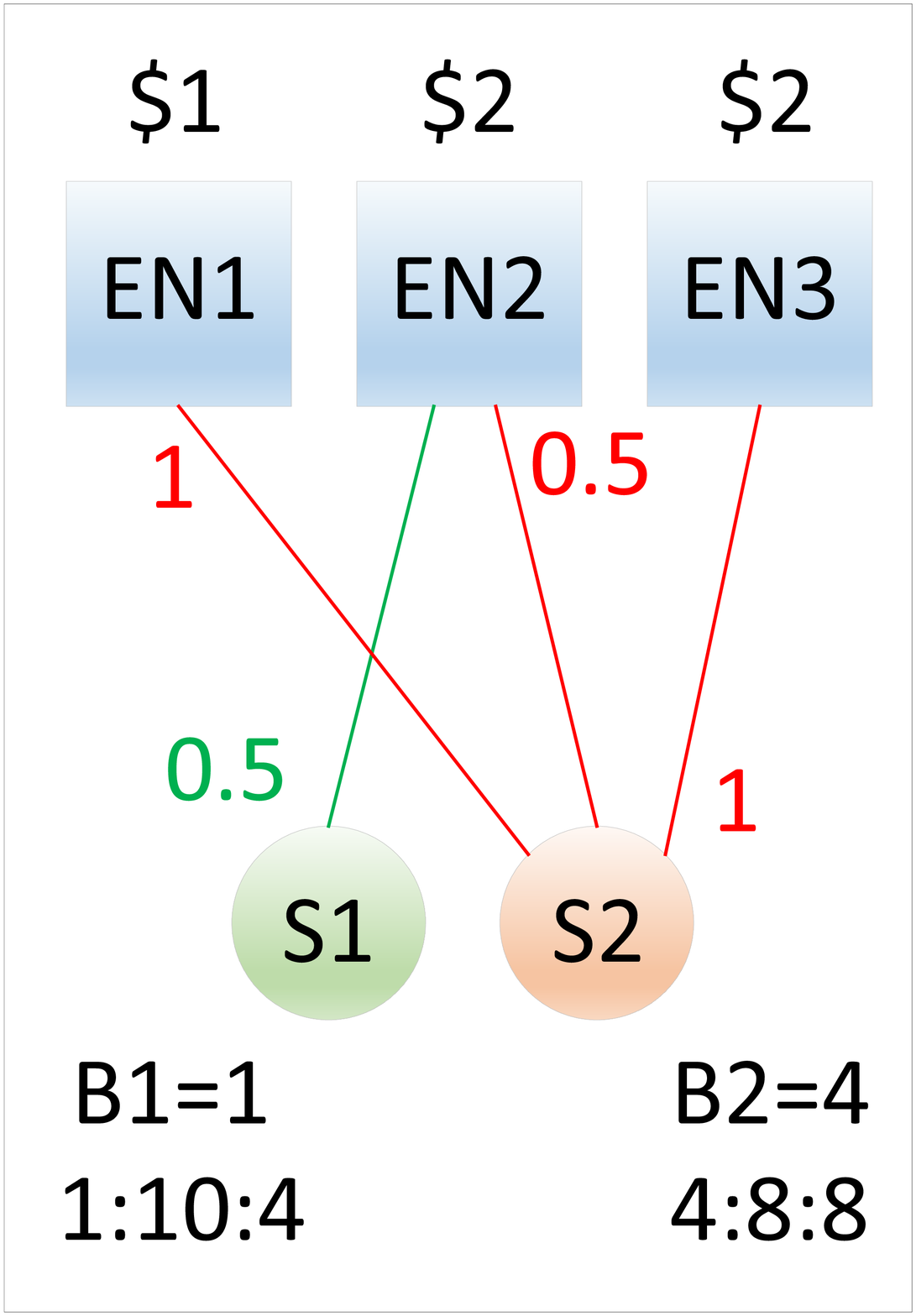}
	    \label{fig:fail1}
	} \hspace*{-0.5em}
		 \subfigure[Without coordination]{
	     \includegraphics[width=0.22\textwidth,height=0.10\textheight]{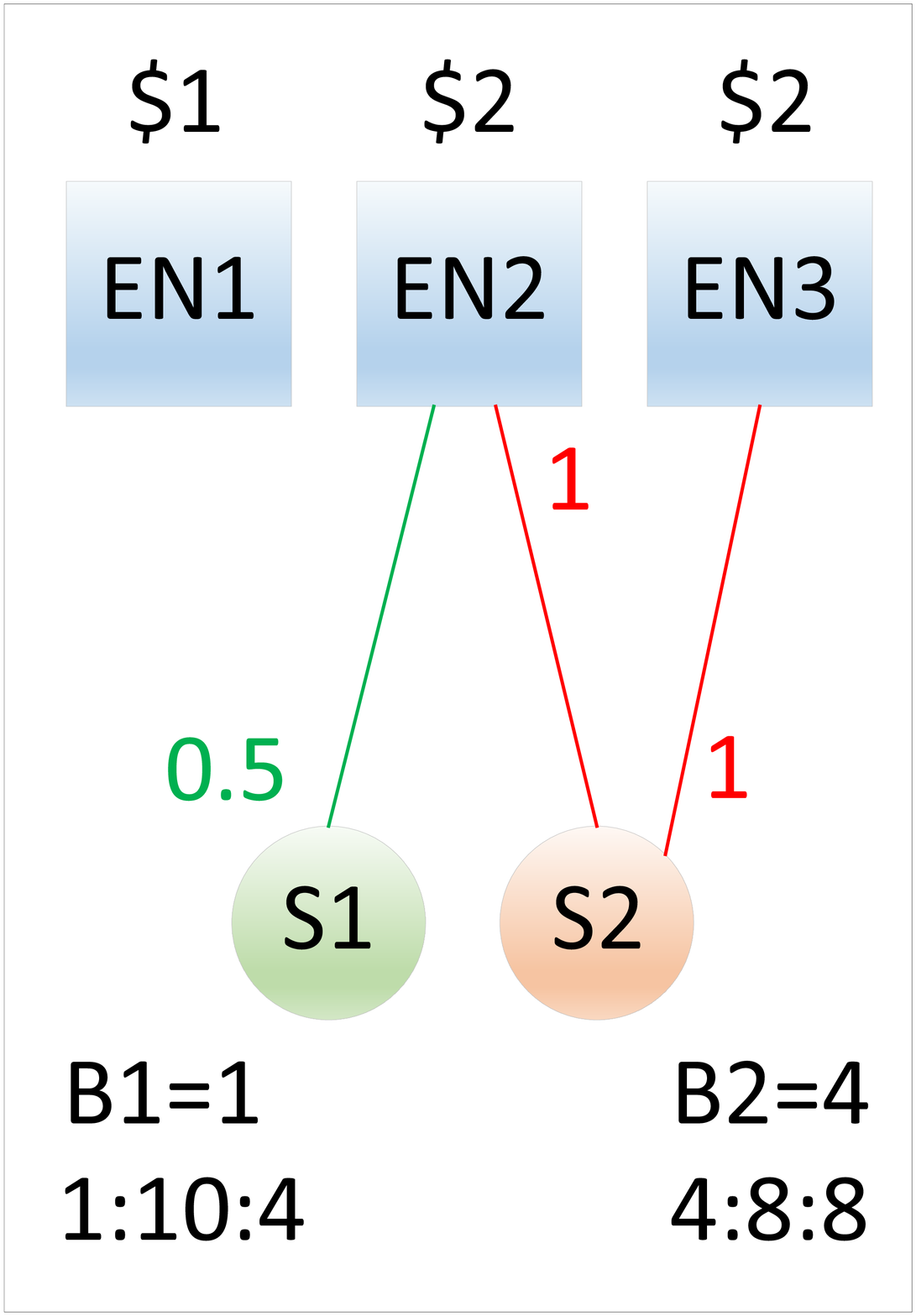}
	     \label{fig:fail2}
	}
	\caption{Market equilibrium with linear utilities}
\end{figure}

Fig.~\ref{fig:fail1} presents the ME from the centralized EG program. The value associated with each edge between a service and an EN indicates the amount of resource that the service buys from the EN. For example, in Fig. ~\ref{fig:fail1}, we have: $x_{1,1} = 0, ~x_{1,2} = 0.5$, and $x_{1,3} = 0.$ The equilibrium price vector is $p = (1, 2, 2)$. The demand sets 
are: $D_1 = \{2\}$ and $D_2 = \{1, 2, 3\}$. Given the equilibrium prices, the set of optimal (i.e., utility-maximizing) resource bundles of service 2 is infinite. 
Hence, even if a distributed algorithm reaches the exact equilibrium prices at some iteration, it may not stop
 since the total  demand reported by the buyers may not equal to the total supply. For instance, in Fig.~\ref{fig:fail2}, although the platform announces the exact equilibrium prices, service 2 may choose to buy all resources from EN2 and EN3.
Then, the algorithm may never terminate. In the following, we present two distributed algorithms 
 to find the ME. 



\vspace{-0.18in}

\subsection{Dual Decomposition with Function Approximation}
\label{dualapp}

Using Lagrangian relaxation \cite{boyd,dpal06}, we can decompose the EG convex program into 
sub-problems, each of which can be solved by a service. 
We observe that the EG program (\ref{EGprogram1})-(\ref{EQ13}) can be written equivalently as follows. 
\beqn
\label{EGprogram11}
\underset{\mathcal{X}}{\text{maximize}} && \sum_{i=1}^N B_i \ln u_i(x_i) \\ \nonumber
\vspace{-0.2in} 
\label{EQ18}
\text{subject to} ~ && \sum_{i=1}^N x_{i,j} \leq  1, ~ \forall j; \quad  x_{i,j} \geq 0, ~ \forall i,~j. 
\eeqn
Relaxing the coupling constraints, the partial Lagrangian is 
\beqn
L(X,p) &=&  \sum_i B_i \ln u_i(x_i) + \sum_j p_j \big( 1 - \sum_i x_{i,j} \big) \\ \nonumber
&=& \sum_i \Big( B_i \ln u_i(x_i) - \sum_j p_j x_{i,j} \Big) + \sum_j p_j. 
\eeqn 
Thus, given a price vector $p$, each service solves 
\beqn
\label{subprob}
\underset{x_i \geq 0}{\text{maximize}} ~ B_i \ln u_i(x_i) - \sum_j p_j x_{i,j}.
\eeqn
%
To overcome the difficulty raised by the non-uniqueness of the optimal demand of the services with linear utilities, we propose to approximate the linear utility function by a Constant Elasticity of Substitution (CES) function, which is widely used in Economics and Computer Science \cite{eco,AGT}.
A CES function has the following form: $u_i^{\sf CES} (x_i) = \Big( \sum_{j=1}^M (a_{i,j} x_{i,j} )^\rho \Big)^{\frac{1}{\rho}}, ~ \rho < 1,~ \rho \neq 0.$
 %
Indeed, the linear utility function is a special case of the CES function family as $\rho \rightarrow 1$. We can approximate the original linear utility function by a CES function where $\rho = 1 - \epsilon$ with $\epsilon$ is arbitrarily small. As $\epsilon \rightarrow 0$, $u_i^{\sf CES} \rightarrow u_i$.
Clearly, 
a CES function is strictly concave and homogeneous \cite{eco}. Hence, the EG program and Theorem \ref{theorem1} also apply to CES functions \cite{EG1,AGT}. Additionally, we can observe that maximizing a CES function above is equivalent to maximizing  $u_i(x_i) = \sum_j (a_{i,j} x_{i,j} )^\rho$. 
 Since a CES function is strictly concave,  the optimal demand bundle of a service is unique. 
Consider the following optimization problem
\beqn
\label{prob1}
\underset{x_{i} \geq 0}{\text{maximize}} ~u_i(x_i) \quad
\text{subject to} ~  \sum_j p_j x_{i,j} \leq B_i. 
\eeqn

\begin{proposition} Given a positive price vector $p$ and a CES approximation function, each service $i$ can either solve Problem (\ref{subprob}) or Problem (\ref{prob1}). Both the problems have the same closed form solution as follows: 
\beqn
\label{cesdemand}
x_{i,j} = \Big( \frac{a_{i,j}^{\rho}}{p_j}\Big)^{\frac{1}{1-\rho}} \frac{B_i}{ \sum_{j=1}^M \Big( \frac{a_{i,j}}{p_j} \Big)^{\frac{\rho}{1-\rho}}}.
\eeqn
\end{proposition}
\textbf{Proof:} Refer to \textbf{Appendix A}.\qedb

Thus, based on the dual decomposition method where each service solves the sub-problem (\ref{subprob}), we have the following distributed algorithm with CES function approximation (\textbf{Algorithm 1}). 
 With a sufficiently small step size, it is guaranteed to terminate and converge to an (approximate) global optimal solution \cite{boyd,dpal06}. Our simulation results confirm that \textbf{Algorithm 1} produces a solution arbitrarily close to the optimal one from the centralized EG program. 
%

	\begin{algorithm}[H]
\footnotesize
\caption{\textsc{Function Approximation Algorithm}}
\label{CESalg}
\begin{algorithmic}[1]

\STATE Initialization: iteration t = 0, set initial prices of ENs $p(0) = p_0$, and set step size $\alpha(0)$ and tolerance $\gamma$ to be small. 

\REPEAT 
  \STATE At iteration $t$, the platform broadcasts prices p(t) to the buyers. 
	
  \STATE Each buyer computes its optimal demand $x_i(t)$ 
	using (\ref{cesdemand}) and sends it to the platform. 

  \STATE The platform updates 
 the prices  \\ 
$p_j(t+1) = \max \Big\{ p_j(t) + \alpha(t) \big( 1 - \sum_{i=1}^N x_{i,j}(t) \big), 0 \Big\}, \quad \forall j$

	
\UNTIL {$\big| p_j(t+1) - p_j(t) \big| < \gamma, ~\forall j$, or the number of iterations $t$ is too large.}

\STATE Output: equilibrium prices $p^*$ and optimal allocation $X^*$.

\end{algorithmic}
\end{algorithm}




\subsection{Proportional Response Dynamics Strategy}
\label{propdyn}

In this section, we present the Proportional Response Dynamics (\textbf{PropDyn}) algorithm  proposed by the P2P community.
This distributed algorithm is very simple to implement and has been proved to converge to an ME \cite{fwu07}. 
Basically, in every iteration $t$, each service updates its bids 
proportional to the utilities it receives from the previous iteration. 
Specifically, $b_{i,j}(t) = B_i \frac{u_{i,j}(t-1)}{u_i(t-1)}, ~ \forall i,~ j,~ t.$
Since the ENs' capacities are normalized, the price of an EN equals to the total bids sent to it, i.e., $p_j(t) = \sum_{i} b_{i,j}(t)$. By bidding $b_{i,j}(t-1)$ to  EN $j$, 
service $i$ obtains an amount of resource $x_{i,j}(t-1) = b_{i,j}(t-1)/p_j$, and gains a utility $u_{i,j}(t-1) = a_{i,j} x_{i,j}(t-1)$. 
Finally, $u_i(t-1) = \sum_j u_{i,j}(t-1)$ is the total utility of service $i$ 
at iteration $t-1$. 
%
The salient feature of this algorithm is that it can be implemented efficiently in a distributed manner. In particular, each EN only needs to know the total bid that it receives to compute the price while each buyer only needs to know its own 
information and learns its utilities achieved in the previous iteration to compute its new bids.
The algorithm terminates when the price deviation of every EN is 
sufficiently small \cite{fwu07}. The major difference between this novel algorithm and traditional distributed algorithms is that in each iteration, every service computes its new bids as mentioned above instead of its optimal demand bundle. 

	\begin{algorithm}[H]
\footnotesize
\caption{\textsc{Best Response Dynamics Algorithm \cite{mfel09}}}
\label{CESalg}
\begin{algorithmic}[1]


\STATE 	 Sort ENs according to the decreasing order of  $\frac{a_{i,j}}{b_{-i,j}}$. \\

   Output a sorted list $L_i = \{i_1,~i_2,\ldots,~i_M\}. $
	
\STATE  Find the largest k such that ~\\ ~\\

$ \frac{\sqrt{a_{i,i_k} b_{-i,i_k}}} {\sum_{j=1}^k \sqrt{a_{i,i_j} b_{-i,i_j}}} \Big(B_i + \sum_{j=1}^k b_{-i,i_j} \Big) - b_{-i,i_k} \geq 0$ ~\\ ~\\

\STATE Set $b_{i_l} = 0 ~\text{for} ~l > k$, and for $1 \leq l \leq k$, set ~\\ ~\\
 
$b_{i_l} = \frac{\sqrt{a_{i,i_l} b_{-i,i_l}} }{\sum_{j=1}^k \sqrt{a_{i,i_j} b_{-i,i_j}} } \Big(B_i + \sum_{j=1}^k b_{-i,i_j}  \Big)- b_{-i,i_l}$ ~\\

	


\end{algorithmic}
\end{algorithm}

To illustrate the effectiveness of the \textbf{PropDyn} mechanism as well as the ME concept, we compare it with the \textit{Proportional Sharing Best Response (BR)} mechanism (\textbf{PropBR}) proposed in \cite{mfel09}, which aims to find a Nash Equilibrium (NE).  In a non-cooperative game, a NE is a stable state of a system where no player can gain by a unilateral change of strategy if the strategies of the others are fixed \cite{AGT}. 
Both \cite{fwu07} and \cite{mfel09} study a  proportional sharing system where the resource of every node is shared proportionally to the services according to their bids. Specifically, we have $x_{i,j} = \frac{b_{i,j}}{b_{i,j} + b_{-i,j}},~ \forall i,~j$, where $b_{-i,j}$ is the total bid of all the services except $i$.
In both mechanisms, the actions of the 
services are the bids ($b_{i,j}$) 
submitted to the ENs.
However, instead of updating its bids following the rule in \textbf{PropDyn},  each service in the \textbf{PropBR} mechanism tries to selfishly maximize its utility given strategies taken by other services  \cite{mfel09}. 

\textbf{Algorithm 2} is the BR algorithm that buyer $i$ will execute given the total bid $b_{-i,j}$ of other buyers. 
The whole algorithm is implemented in rounds. In each round, each buyer in turns runs \textbf{Algorithm 2} and updates its bid vector $b_i$ 
to the platform. The platform broadcasts new bids to all buyers in the system. A round completes when all buyers have updated their bids. Obviously, whenever this BR dynamics strategy converges, it converges to an NE. As mentioned in \cite{mfel09}, the algorithm normally converges after a few rounds. 

Interestingly, our simulation shows that buyers do not gain significantly by playing BR. Indeed, most of buyers achieve lower utilities in the \textbf{PropBR} scheme compared to  the  \textbf{PropDyn} scheme.
 Furthermore, to play BR dynamics, each buyer has to know 
total bids of others and the actual capacity of every EN \cite{mfel09}. In \textbf{PropDyn}, buyers only need to know their own information.
Therefore, in a proportional sharing system, buyers  may not have incentives to play BR.
\section{Net Profit Maximization }
\label{formu2}

Different from the \textbf{basic model}, in the second model, the services try to optimize their net profits (i.e., revenue minus cost) instead of revenue. Specifically, the net profit of service $i$ is $v_i(x_i) =  \sum_j (a_{i,j} - p_j) x_{i,j},~\forall i.$ 
Given prices $p$, the objective of service  $i$ is to maximize $U_i(x_i,p) = v_i(x_i)$ subject to: $\sum_j x_{i,j} p_j \leq B_i,~\forall i$ and $x_{i,j} \geq 0,~ \forall i,~j$.
%
%
Indeed, maximizing the net profit $v_i(x_i)$ is equivalent to maximizing $\sum_j (a_{i,j} - p_j) x_{i,j} + B_i = \sum_j a_{i,j} x_{i,j} + s_i $, where $s_i = B_i - \sum_j p_j x_{i,j}$ is the surplus money of service $i$ after purchasing $x_i$.
Inspired by the EG program for the \textbf{basic model}, we would like to construct a similar convex program to capture the ME in this new model. 

Note that without budget consideration, this game-theoretic problem can be solved efficiently by writing down a social welfare maximization problem (i.e., maximizing sum of utilities of all the services), then use the dual decomposition method \cite{dpal06} to decompose it into sub-problems, each of which is solved by one service.  Each sub-problem is exactly a net profit maximization problem of a service. Unfortunately, this strategy fails when we consider budget since the social welfare maximization problem cannot be decomposed due to the coupling budget constraints. 


Our derivation of the new convex optimization problem is based on reverse-engineering the \textbf{basic model}.
\begin{proposition} 
The equilibrium prices in the \textbf{basic model} can be found by solving the following convex problem.
\beqn
\label{dual1}
\underset{p,\eta}{\text{minimize}} ~&&\sum_{j=1}^M p_j - \sum_{i=1}^N B_i \ln (\eta_i) \\ \nonumber
\label{dual1EQ1}
\text{subject to} ~ &&p_j \geq a_{i,j} \eta_i, ~\forall i,~j; 
~ p_j \geq 0,~\forall j.
\eeqn
\end{proposition}
\textbf{Proof:}
 We can obtain this convex problem by using Lagrangian and Fenchel conjugate function \cite{boyd} to construct the dual problem of the original EG program. Indeed, $\eta_i$ and $p_j$ are the dual variables associated with (\ref{EQ11}) and (\ref{EQ12}).  See our \textbf{Appendix B} for the full proof. \qedb

Clearly, to maximize $v_i(x_i) =  \sum_j (a_{i,j} - p_j) x_{i,j}$, service $i$ will never buy resource from EN $j$ if $a_{i,j} < p_j$. In other words, service $i$ would only buy resources from ENs in the set $A_i = \big\{j: \frac{p_j}{a_{i,j}} \leq 1\big\}.$
From (\ref{dual1}), we have $ \eta_i \leq \frac{p_j}{a_{i,j}}, ~\forall i$. 
From these observations, we conjecture that the following prorgram captures the equilibrium prices in our second market model (i.e., net profit maximization).
\beqn
\label{dual2}
\underset{p,\eta}{\text{minimize}}~~ \sum_{j=1}^M p_j - \sum_{i=1}^N B_i \ln (\eta_i)
\eeqn
\vspace{-0.1in}
subject to
\beqn
\label{dual2EQ1}
p_j \geq a_{i,j} \eta_i, ~\forall i,~j;~ \eta_i \leq 1,~ \forall i;~\eta_i \geq 0, ~\forall i;~p_j \geq 0, ~\forall j. \nonumber
\eeqn
\vspace{-0.2in}


\begin{theorem} The solution of the following convex program is exactly an ME of the new market model.
\beqn
\label{EGprogram2}
  \underset{\mathcal{X},u,s}{ \text{maximize}} ~ &&\sum_{i=1}^N \Big( B_i \ln u_i - s_i \Big) \\ \nonumber
\vspace{-0.2in}
\label{EQ21}
\text{subject to}  && u_i \leq \sum_{j=1}^M a_{i,j} x_{i,j} + s_i,\forall i \\ \nonumber
\vspace{-0.1in}
&& \sum_{i=1}^N x_{i,j} \leq 1, \forall j;x_{i,j} \geq 0, \forall i,j;~ s_i \geq 0, ~\forall i.
\eeqn
At the equilibrium, the total of money spent and surplus money of every service equals to its budget. 
Additionally, the optimal utility of every service is unique and greater or equal to its budget. For any buyer who has surplus money, her utility equals her budget.
\end{theorem}
\textbf{Proof:}
See our Appendix C . \qedb

%
%


The convex problem (\ref{EGprogram2}) is indeed the dual program of problem (\ref{dual2}).
We can interpret problem (\ref{EGprogram2}) as follows. First, the utility of a service is the sum of its revenue and its surplus money. 
The first part of the objective function is the weighted sum of logarithmic utilities of the services similar to that of the EG program. However, since the surplus money does not contribute (i.e., not visible) to the market, we should subtract this amount from the aggregated utility function, i.e., the objective function. Finally, similar to the EG program, although budget constraints are not included in  (\ref{EGprogram2}), the optimal solution satisfies these constraints. It is worth noting that, somewhat surprisingly, although our reverse-engineering approach is specialized for linear revenue functions only,   the convex program (\ref{EGprogram2}) works  also for a wider class of homogeneous concave revenue functions. 
Interested readers can find more details in Appendix E.

%
%

\section{Numerical Results}
\label{sim}
\subsection{Simulation Settings}

\begin{figure}[ht!]
	\centering
		\includegraphics[width=0.35\textwidth,height=0.10\textheight]{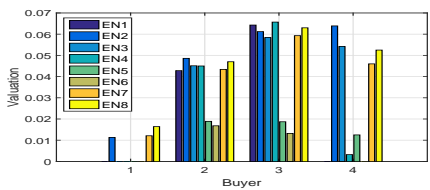}
			\caption{Valuations of ENs to the buyers}
	\label{fig:value} \vspace{-0.2in}
\end{figure}

We consider a square area with dimensions of 10km x 10km. The locations of ENs and services are generated randomly in the area. We generate a total of 100 ENs and 1000 locations. We assume that each service is located at one location. For the sake of clarity in analysis, in the \textbf{base case}, we consider a small system with 8 ENs and 4 services (i.e., M = 8 and N = 4), which are selected randomly in the set of 100 ENs and 1000 services. The network delay between a service and an EN is assumed to be proportional to the distance between them. The maximum tolerable delay of the services follows a uniform distribution over the interval [15,~25]. The service rate $\mu_{i,j}$ is generated randomly from 80 to 240 requests per time unit. The service price is from 2 to 3 per 100000 requests. The number of computing units in the ENs ranges from 10 to 20. From these parameters, we can compute $a_{i,j}$ 
of the services as in (\ref{a_uti}). 
The net profit maximization model is considered in Section \ref{sim2}. 
In the \textbf{ base case}, we assume that the services have equal budget. 
Fig.~\ref{fig:value} depicts the valuations of the ENs to the buyers in the \textbf{base case}. The \textbf{base case} is used in all the simulations unless mentioned otherwise. 

\subsection{Performance Comparison}

In the first model captured by the EG program, the absolute value of the budget only affects the equilibrium prices by a scaling factor (e.g., all the prices increase twice as the budget of every service is double) and does not affect the allocation and utilities of the services. 
The budget is normalized such that the total budget of all services is one.
 The prices act as a means to allocate resources only.
\begin{figure}[ht]
		\subfigure[Same budget]{
		  \includegraphics[width=0.245\textwidth,height=0.10\textheight]{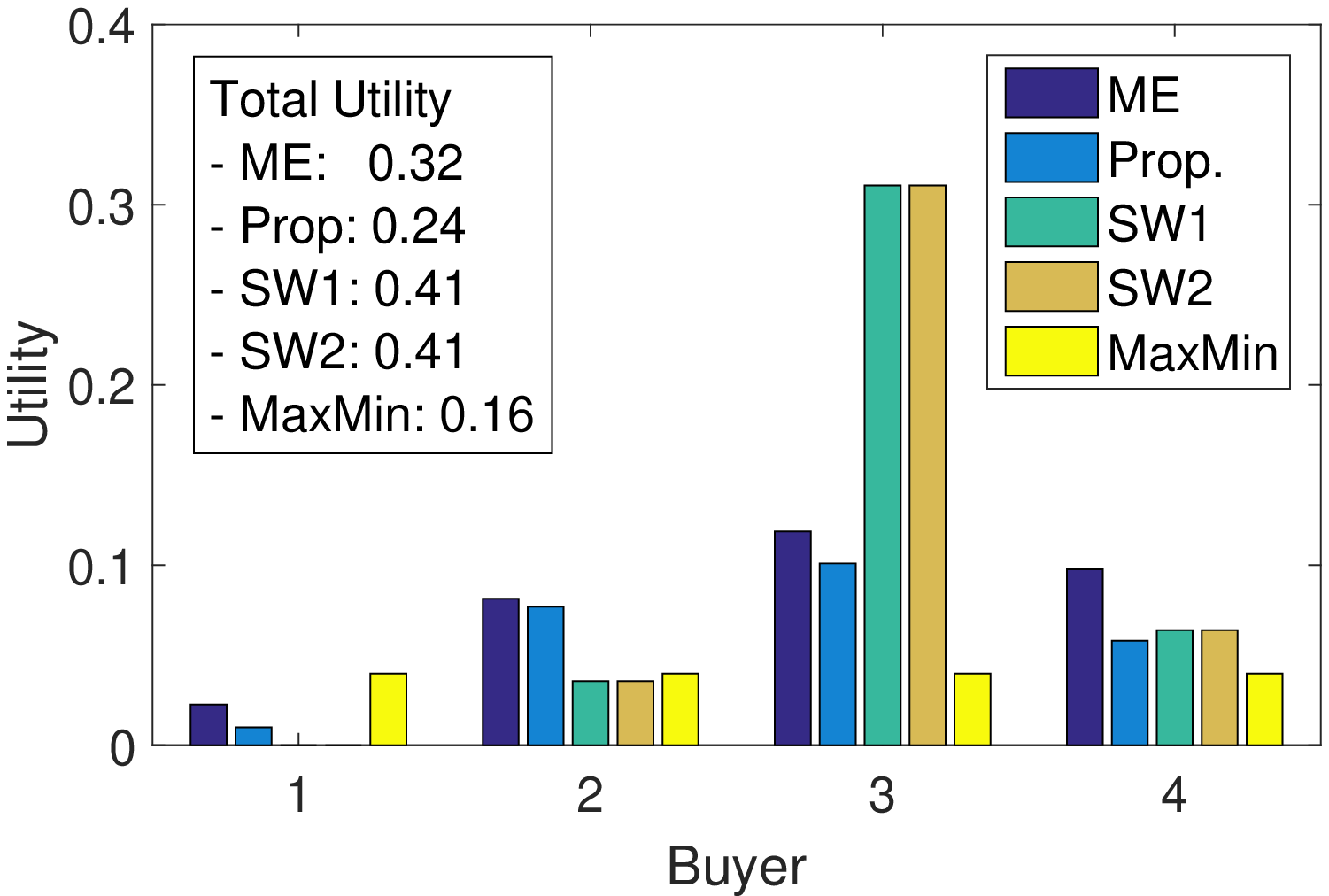}
	    \label{fig:compa1}
	} \hspace*{-1.9em}
		 \subfigure[Budget ratio (4/4/1/4)]{
	     \includegraphics[width=0.245\textwidth,height=0.10\textheight]{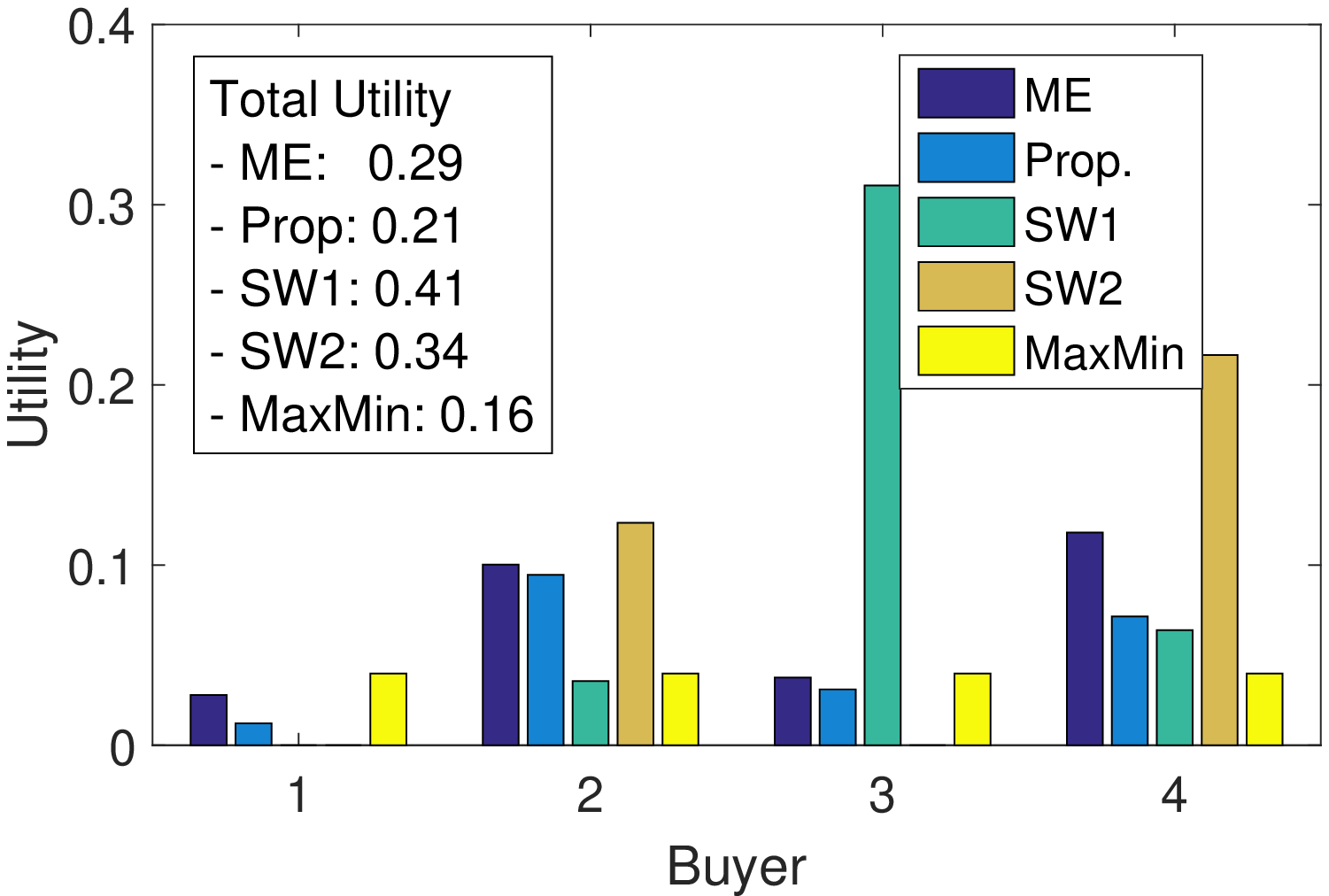}
	     \label{fig:compa2}
	}\vspace{-0.2cm}
	\caption{Performance comparison}
\end{figure}

We consider five schemes, including: the proposed \textit{ME}, the proportional sharing (\textit{Prop.}), the social welfare maximization with equal weights (\textit{SW1}), social welfare maximization with different weights (\textit{SW2}), and the maxmin fairness (\textit{maxmin}) schemes. In the proportional sharing, 
each buyer $i$ receives $\frac{B_i}{\sum_i B_i}$ portion of resource of every EN. 
In the social welfare maximization schemes, budget is not considered, and  the objective is to maximize $\sum_i w_i u_i(x_i)$ subject to the capacity constraints of the ENs. $w_i$ is the weighting factor of service $i$. In \textit{SW1}, all weights are equal. In \textit{SW2}, the weight of each service is its budget. Finally, without budget consideration, the \textit{maxmin} scheme aims to maximize $\min_i u_i(x_i)$ under ENs' capacity constraints.

\begin{figure}[ht]
		\subfigure[Same budget]{
		  \includegraphics[width=0.245\textwidth,height=0.10\textheight]{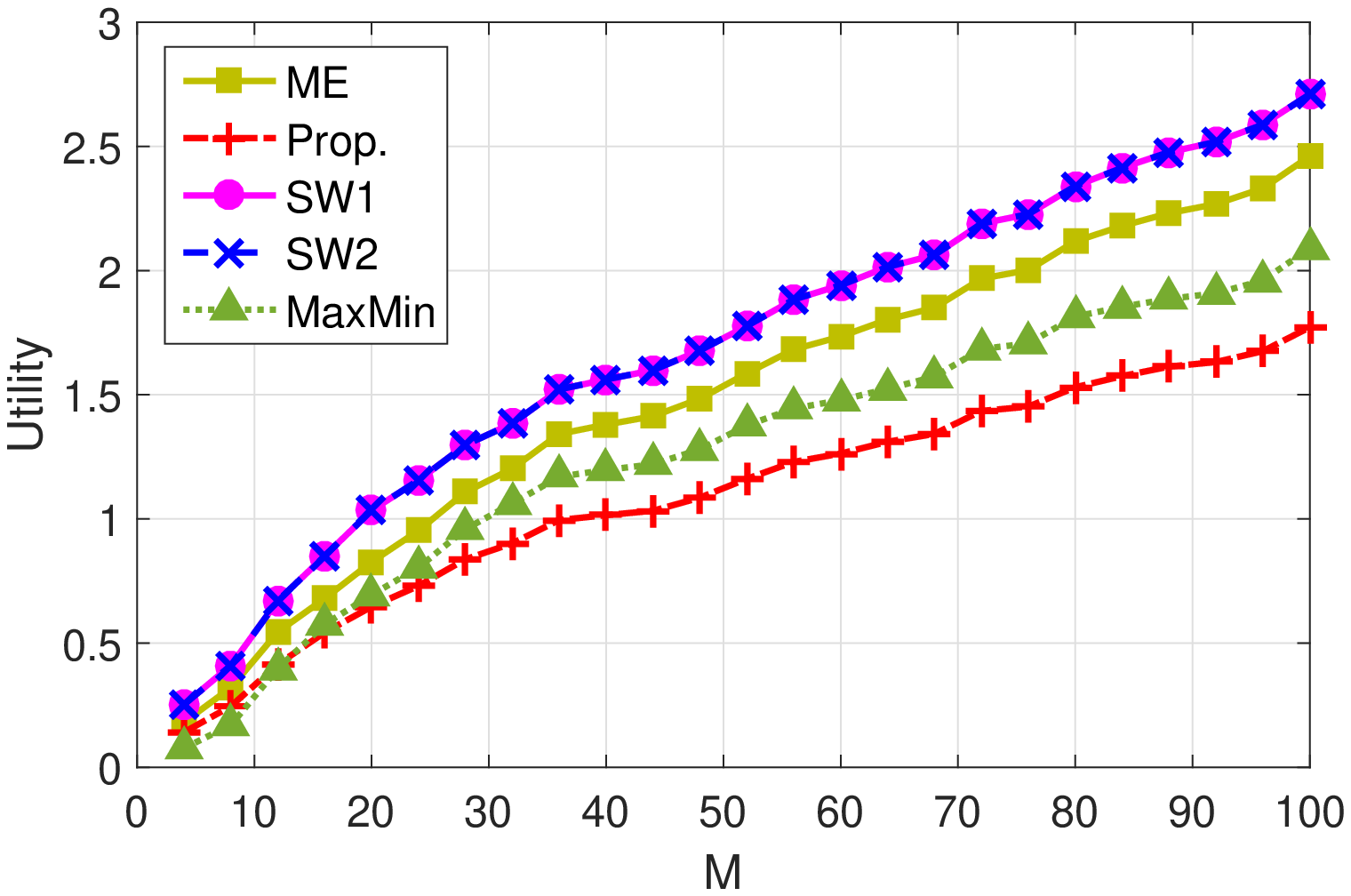}
	    \label{fig:compa3}
	} \hspace*{-1.9em}
		 \subfigure[Budget ratio (4/4/1/4)]{
	     \includegraphics[width=0.245\textwidth,height=0.10\textheight]{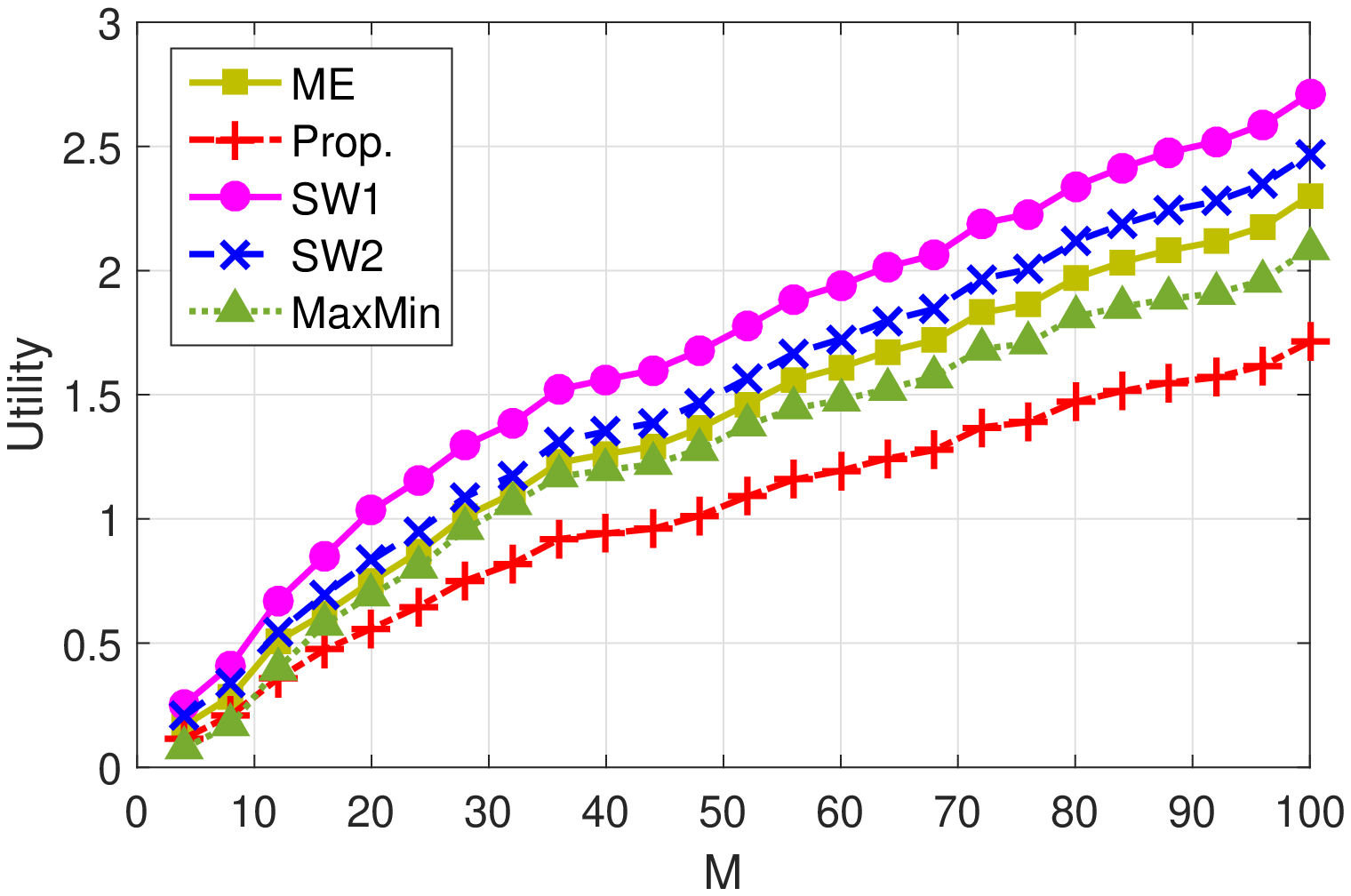}
	     \label{fig:compa4}
	}\vspace{-0.2cm}
	\caption{Utility efficiency comparison (N = 4)}
\end{figure}
\vspace{-0.2in}

\begin{figure}[ht]
		\subfigure[Same budget]{
		  \includegraphics[width=0.245\textwidth,height=0.10\textheight]{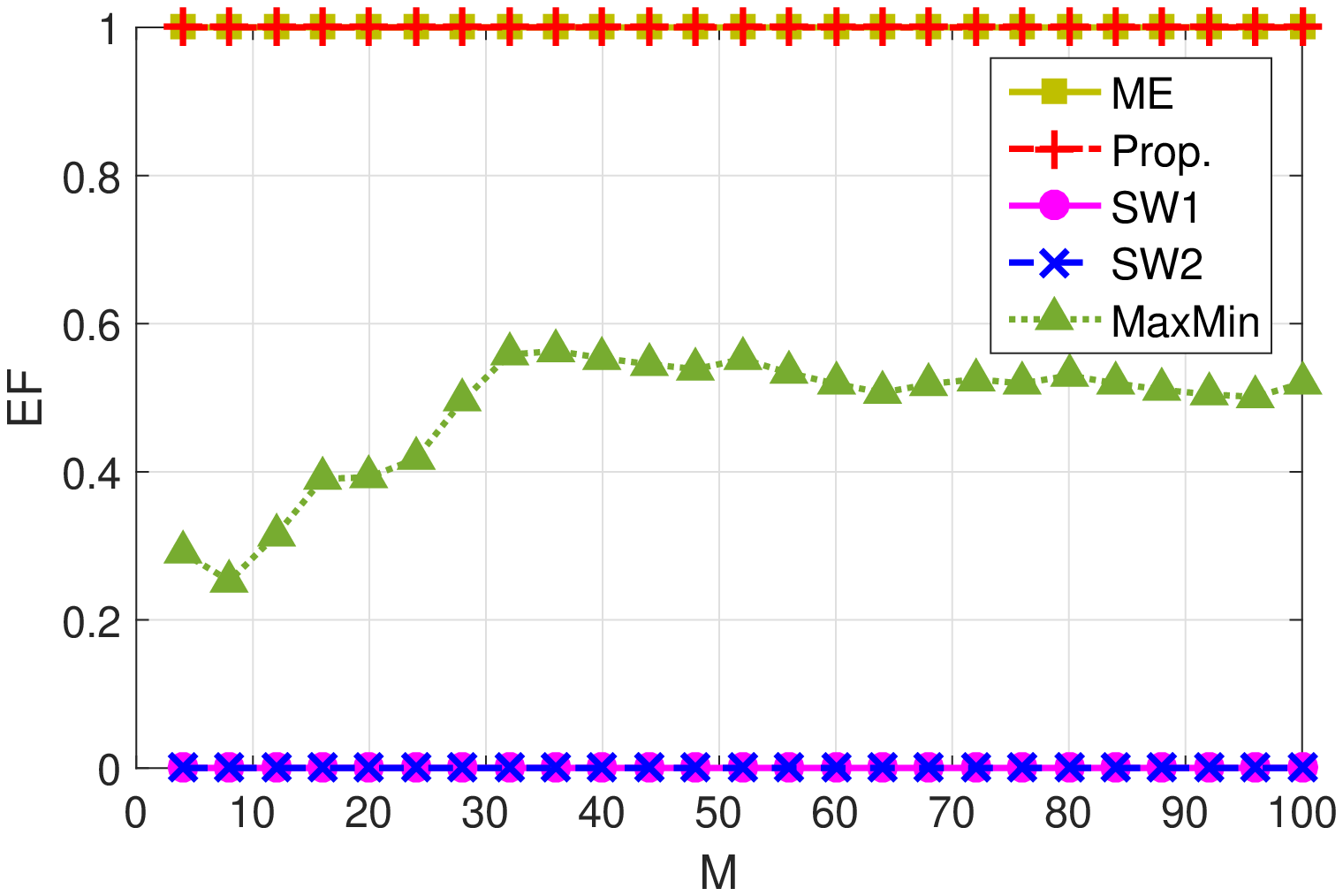}  
	    \label{fig:compa5}
	} \hspace*{-1.9em}
		 \subfigure[Budget ratio (4/4/1/4)]{
	     \includegraphics[width=0.245\textwidth,height=0.10\textheight]{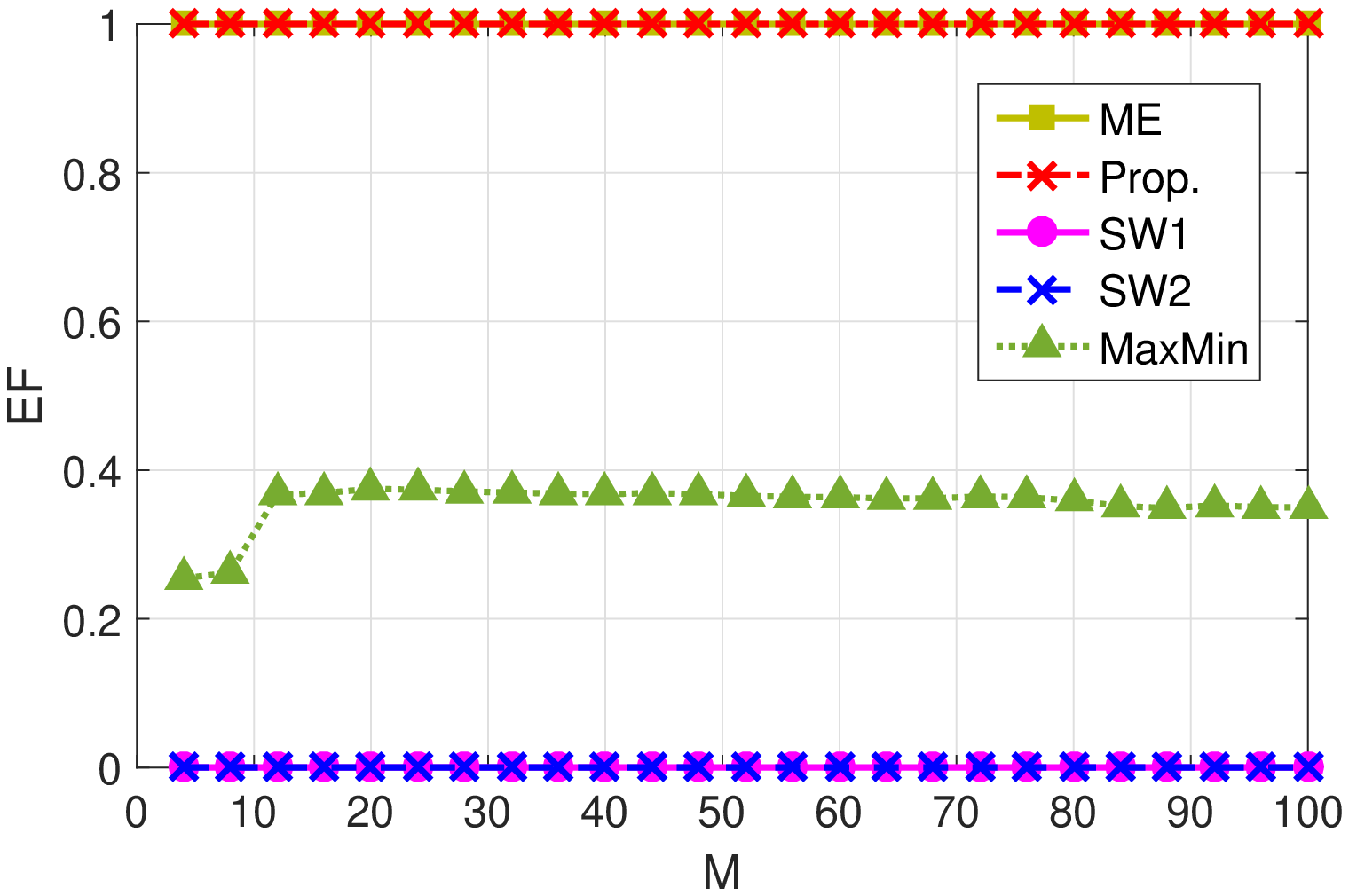}
	     \label{fig:compa6}
	}\vspace{-0.2cm}
	\caption{Envy-freeness comparison (N = 4)}
\end{figure}

\begin{figure*}[ht]
		\subfigure[ME scheme]{
		  \includegraphics[width=0.35\linewidth,height=0.11\textheight]{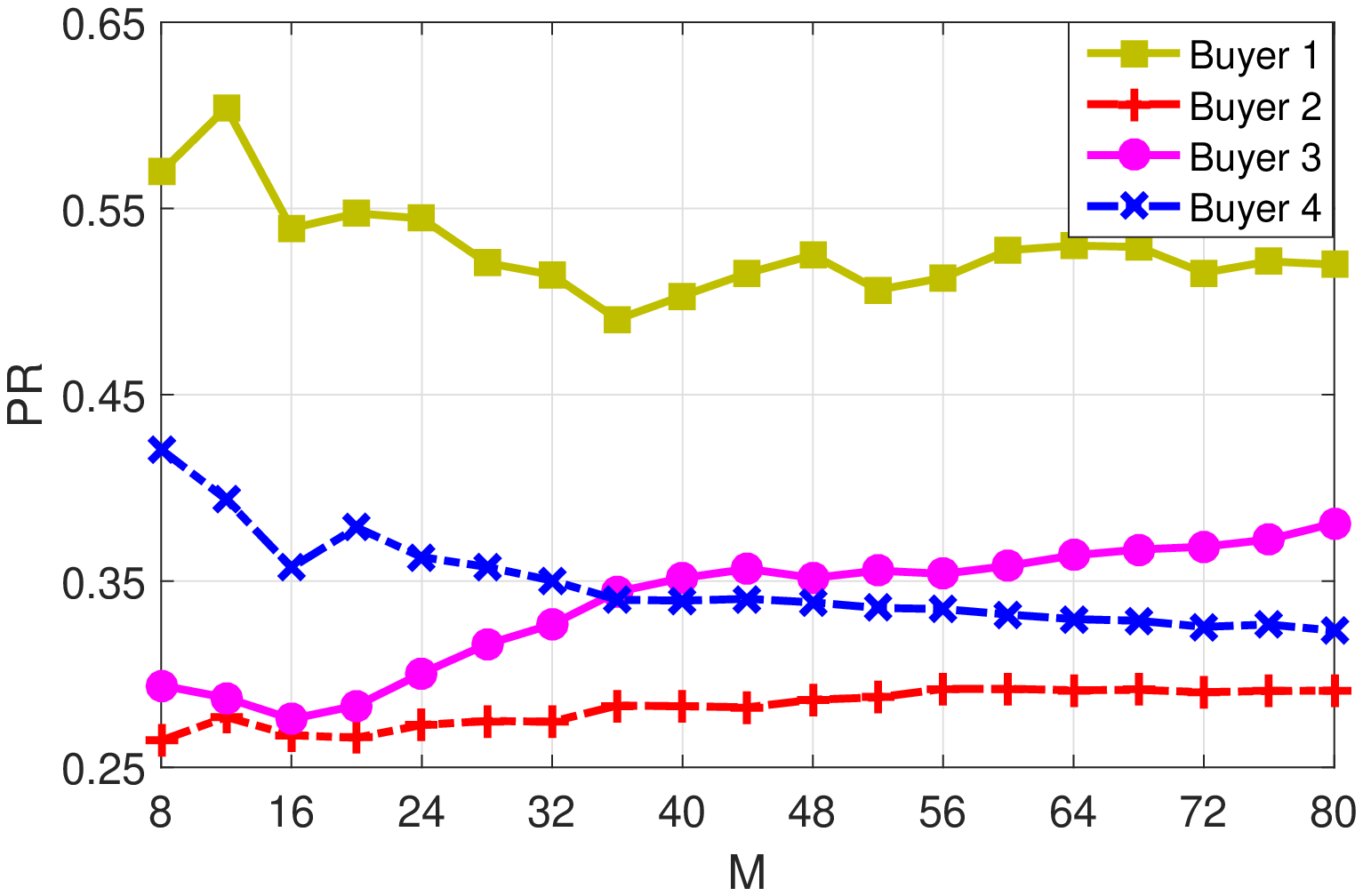}
	    \label{fig:pr1}
	} \hspace*{-1.9em}
	  \subfigure[SW scheme]{
	     \includegraphics[width=0.35\linewidth,height=0.11\textheight]{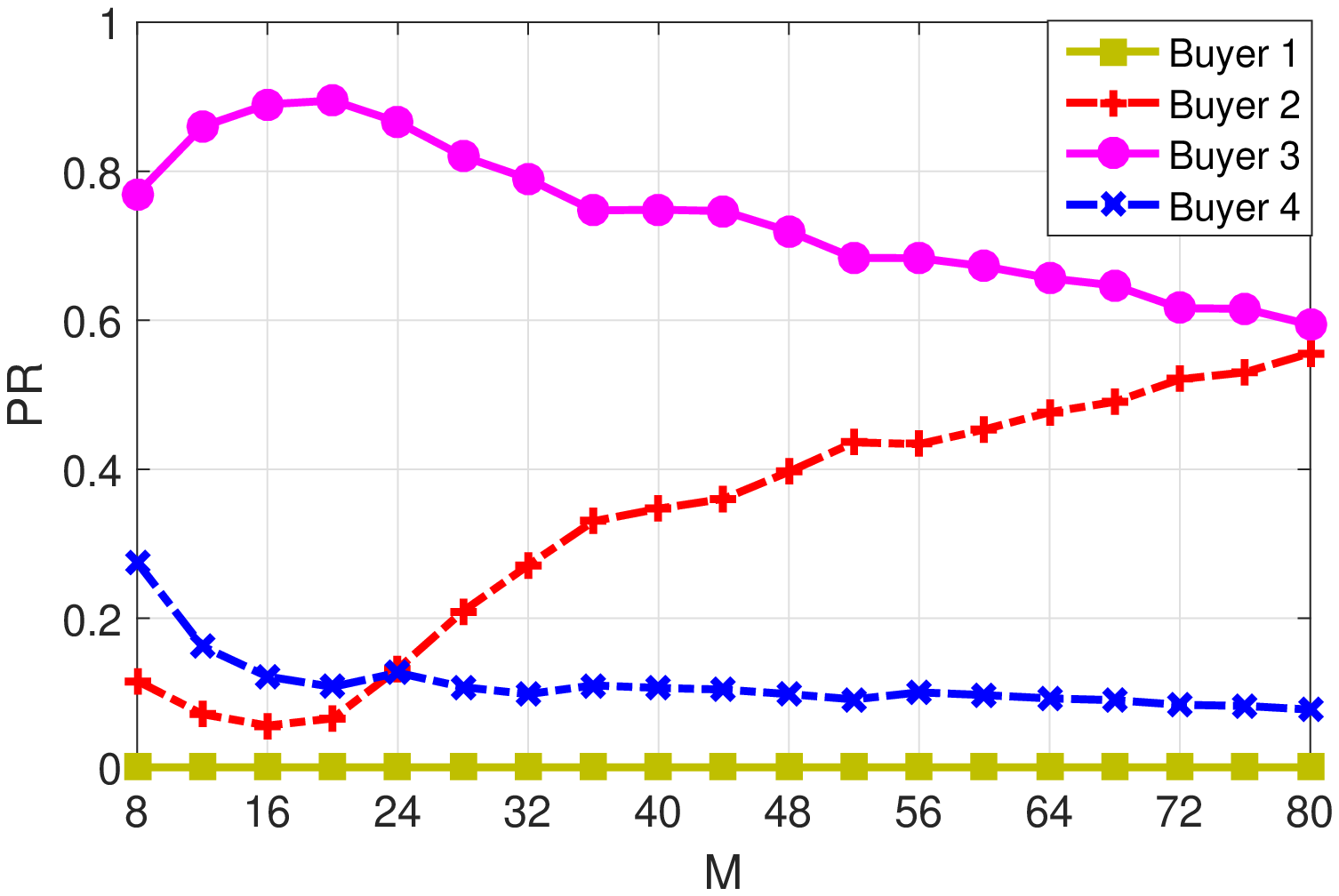}
	     \label{fig:pr2}
	} \hspace*{-1.9em} 
		 \subfigure[MM scheme]{
	     \includegraphics[width=0.35\linewidth,height=0.11\textheight]{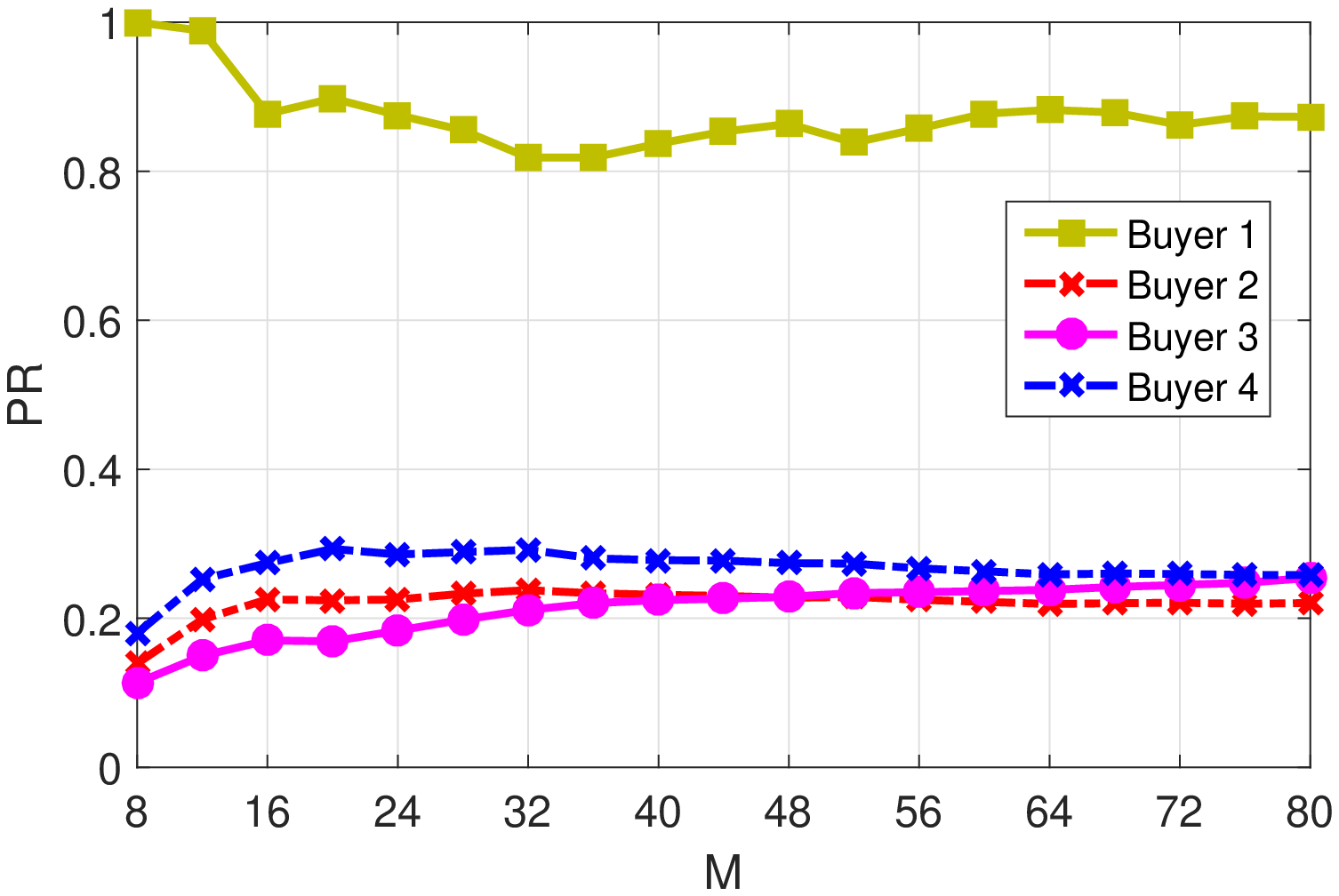}
	     \label{fig:pr3}
	} 
	\caption{Proportionality ratio comparison (N = 4, same budget)}
\end{figure*}
 \vspace{-0.1in}

\begin{figure*}[ht]
		\subfigure[Budget ratio 0.67/1.33/1/1]{
		  \includegraphics[width=0.35\linewidth,height=0.10\textheight]{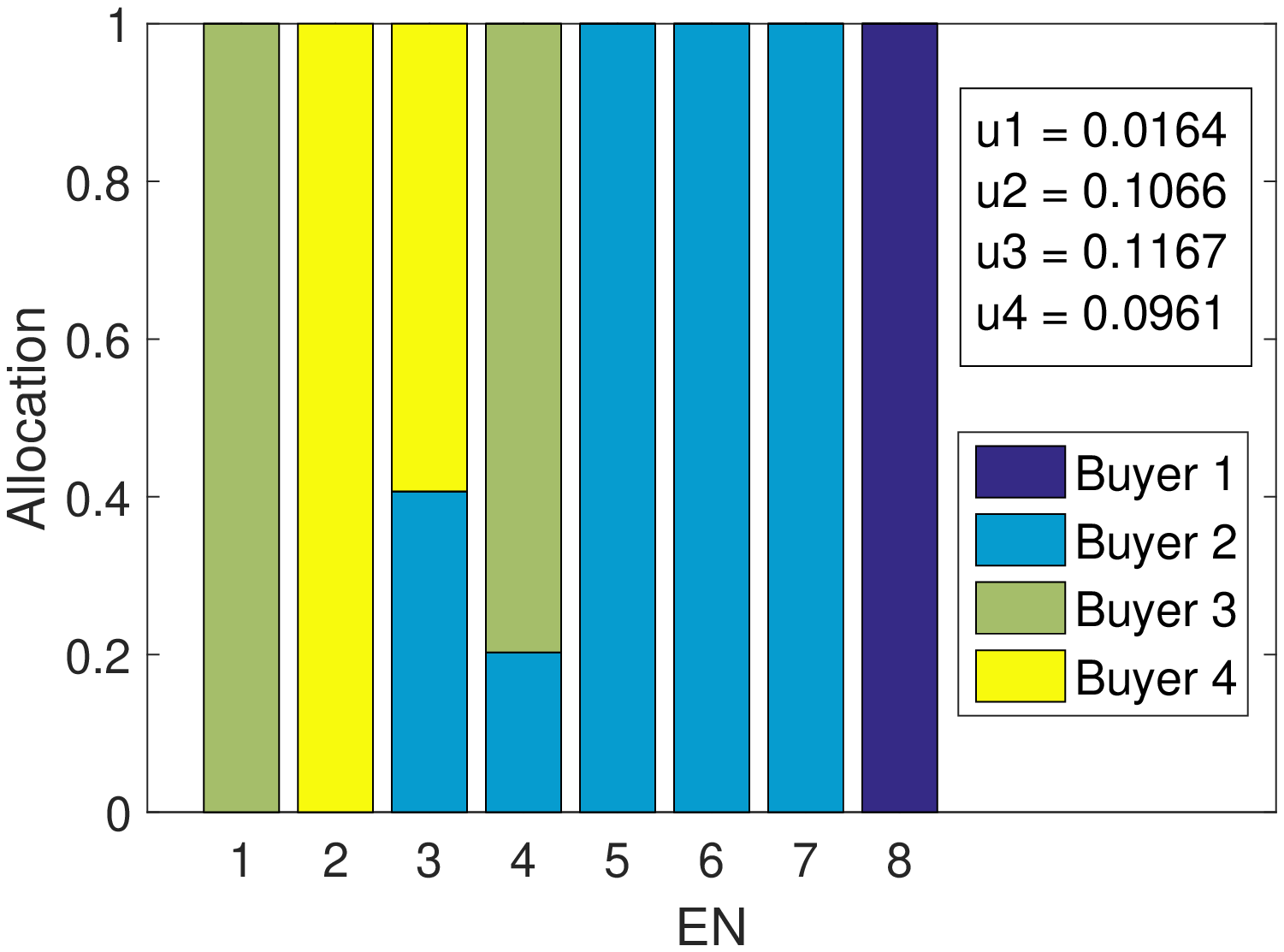}
	    \label{fig:rhalfbar}
	} \hspace*{-1.9em}
	  \subfigure[Budget ratio 1/1/1/1]{
	     \includegraphics[width=0.35\linewidth,height=0.10\textheight]{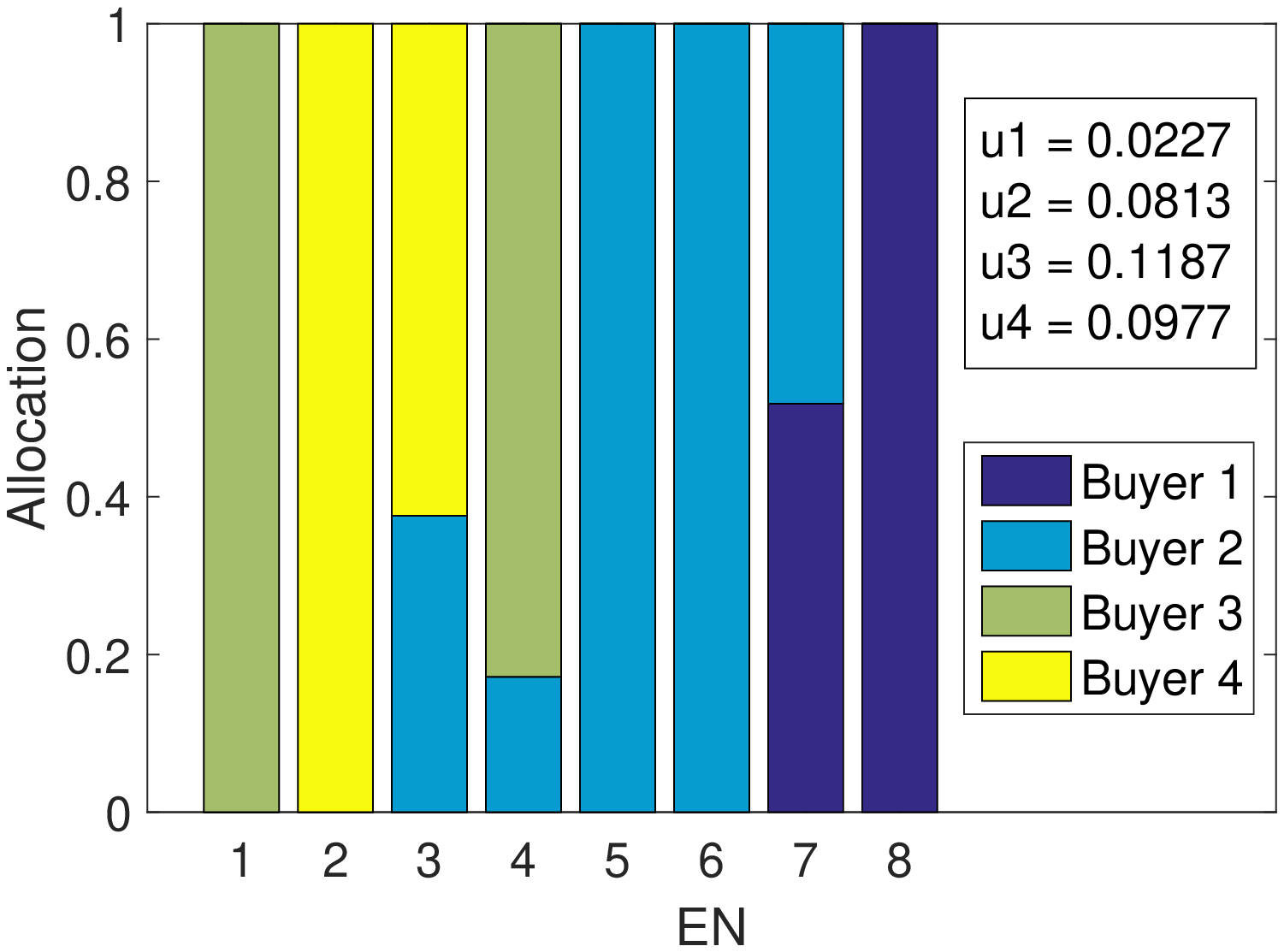}
	     \label{fig:r1bar}
	} \hspace*{-1.9em} 
		 \subfigure[Budget ratio 1.33/0.67/1/1]{
	     \includegraphics[width=0.35\linewidth,height=0.10\textheight]{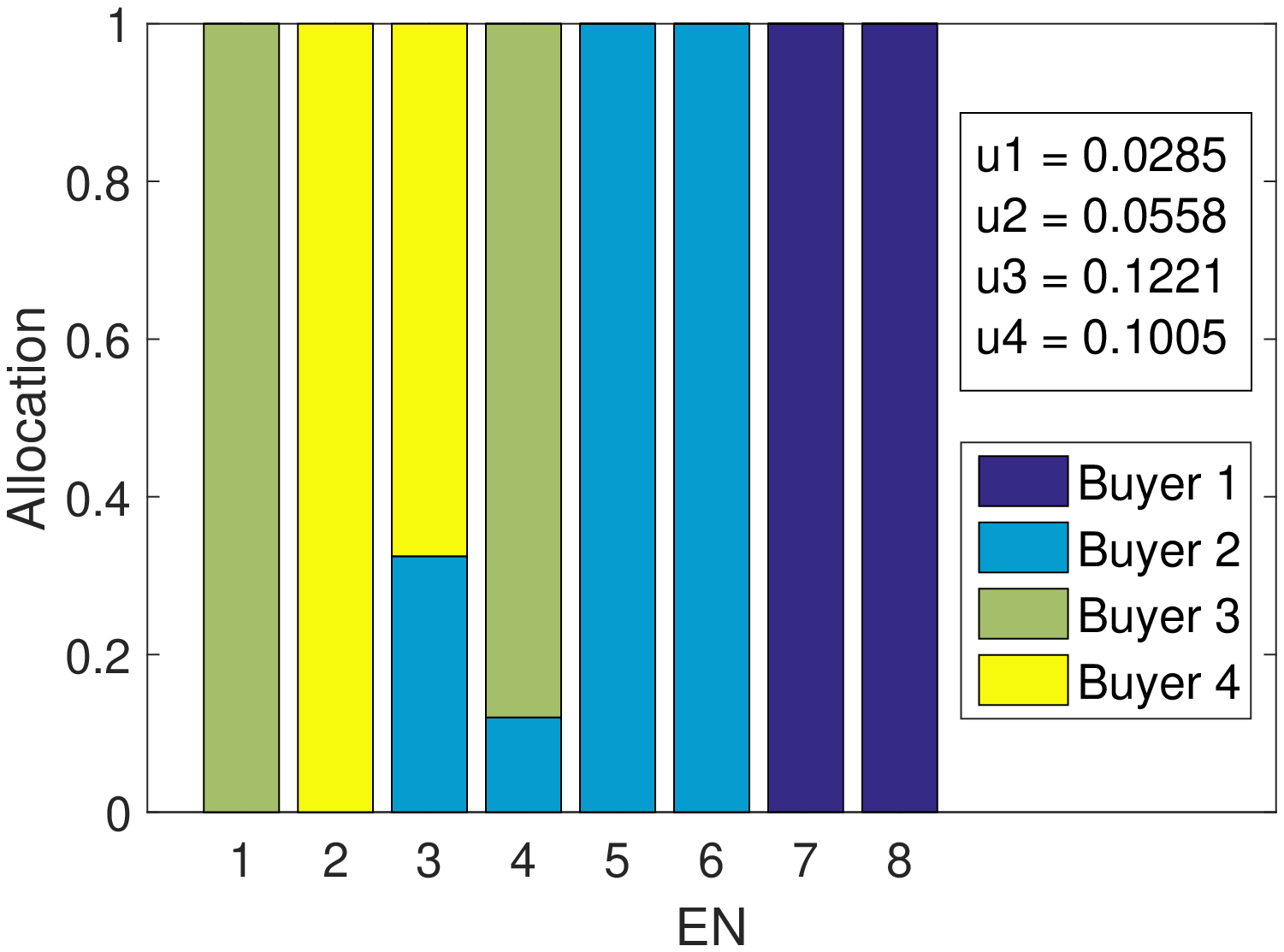}
	     \label{fig:r2bar}
	} 
		
	\caption{Impact of budget ratio on the equilibrium allocation}
\end{figure*}
  \vspace{-0.1in}

Figs.~\ref{fig:compa1}--\ref{fig:compa6} present performance comparison among these schemes under both equal budget and different budget settings. We can observe that the ME scheme balances well the tradeoff between system efficiency and fairness. 
First, the \textit{ME} scheme considerably outperforms the \textit{Prop.} scheme, which confirms the sharing-incentive property of the ME solution. The \textit{maxmin} scheme produces a fair allocation among the buyers but the total utility of the buyers is much  lower compared to other schemes. Noticeably, although the total utility is largest, both schemes \textit{SW1} and
 \textit{SW2} produce undesirable allocations since some buyers (e.g., buyer 1) are not allocated anything and have zero utility in these schemes. 
  %
Also, as can be seen from Figs.~\ref{fig:compa5},~\ref{fig:compa6} which compare envy-freeness indices ($EF = \min_{i,j} \frac{u_i(x_i)/B_i}{u_i(x_j)/B_j}$ \cite{mfel09}) among the different schemes. An allocation is envy-free if EF equals to one. We can observe that our proposed ME scheme significantly outperforms the social welfare maximization and \textit{maxmin} 
schemes in terms of envy-free fairness.
Finally, Figs.~ {fig:pr1}--{fig:pr3} depict the proportionality ratio (PR) of different schemes. When there are 4 buyers with equal budget, every buyer naturally expects to obtain a PR of at least 1/4. It can be observed that our proposed ME scheme satisfies the proportionality property as shown in Theorem 5.2.

\subsection{Sensitivity Analysis}

\begin{figure}[ht]
	\centering
		\includegraphics[width=0.35\textwidth,height=0.12\textheight]{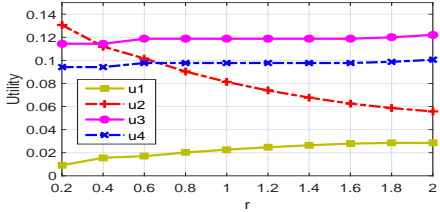} \vspace{-0.2cm}
			\caption{Impact of budget ratio on the buyers' utilities}
	\label{fig:r12uti}\vspace{-0.2cm}
\end{figure}

First, we examine the impact of budget on the equilibrium allocation by varying the budget ratio among the buyers. 
Figs.~ \ref{fig:rhalfbar}-\ref{fig:r2bar} show impact of budget on the equilibrium allocation
as we vary the budget ratio between services 1 and 2.
We observe that buyer 1 is allocated more resources as her budget increases, which also increases her utility. 
The allocation and utility of buyer 2 decrease as her budget decreases. 
Fig.~\ref{fig:r12uti} further supports this observation where $r$ is the budget ratio between  services 1 and 2.
Hence, we can conclude that the proposed algorithm is \textit{effective to capture service priority} in terms of budget in the allocation decision. 
%

\begin{figure}[ht]
	\centering
		\includegraphics[width=0.35\textwidth,height=0.10\textheight]{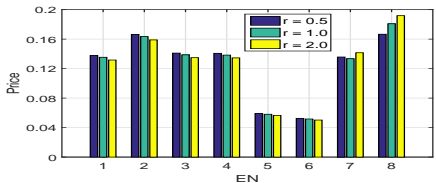} \vspace{-0.2cm}
			\caption{Impact of budget ratio on the equilibrium prices}
	\label{fig:r12price} 
\end{figure}

Fig.~\ref{fig:r12price} shows the dependence of the equilibrium prices on the budget ratio of the buyers. For example, since only EN2, EN7, and EN8 can satisfy the delay requirement of service 1 as seen in Fig.~\ref{fig:value}, the prices of EN7 and EN8 change considerably as budget of service 1 varies. Also, because EN5 and EN6 are less valuable to the buyers, their equilibrium prices are significantly lower than the prices of other ENs while the prices of EN2 and EN8 are highest because they have high values to all the buyers. These observations imply the proposed method is \textit{effective in pricing}. 

%

\begin{figure}[ht!]
	\centering
		\includegraphics[width=0.35\textwidth,height=0.10\textheight]{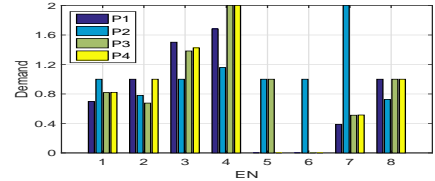} \vspace{-0.2cm}
			\caption{Impact of resource prices on total demand}
	\label{fig:demand}
\end{figure}

In Fig.~\ref{fig:demand}, we change the equilibrium prices by a small amount to obtain different price vectors $P1, P2, P3, P4$  such that demand of each buyer at these prices is unique. 
Price vector $P1$ ($P2$) is obtained by increasing (decreasing) every price in the equilibrium price vector in the \textbf{base case} by 0.01. Price vectors $P3$ ($P4$) is generated by adding a random number between 0 and 0.0005 (0.0003). We can observe that even under small price variation, the market is  not cleared. Some ENs are over-demanded while some others are under-demanded. It means \textit{proper equilibrium pricing is important to clear the market}.

\begin{figure}[ht]
		\subfigure[M = 8, varying N]{
		  \includegraphics[width=0.245\textwidth,height=0.10\textheight]{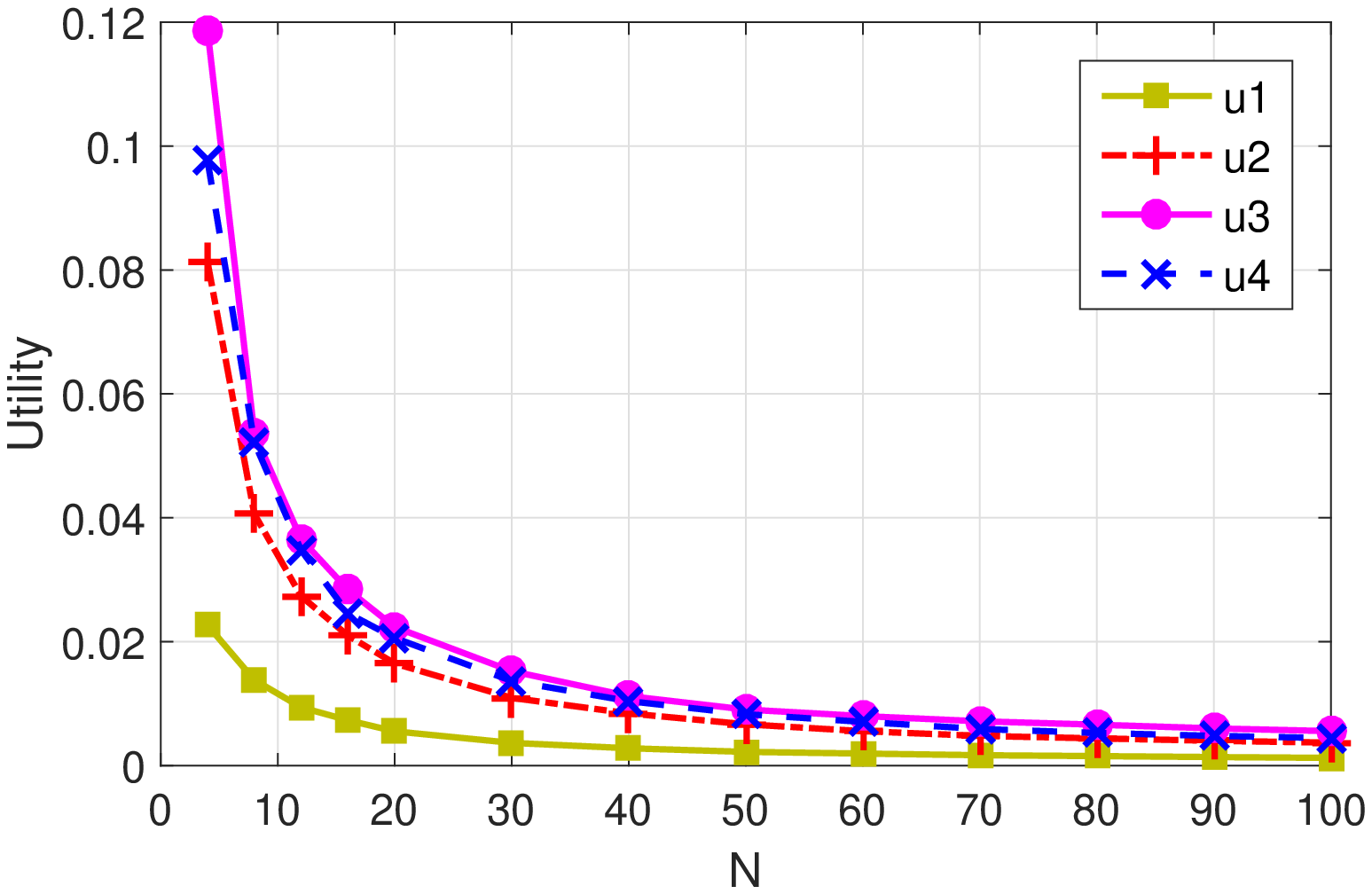}
	    \label{fig:varyNu}
	}  \hspace*{-1.9em}
		 \subfigure[N = 4, varying M]{
	     \includegraphics[width=0.245\textwidth,height=0.10\textheight]{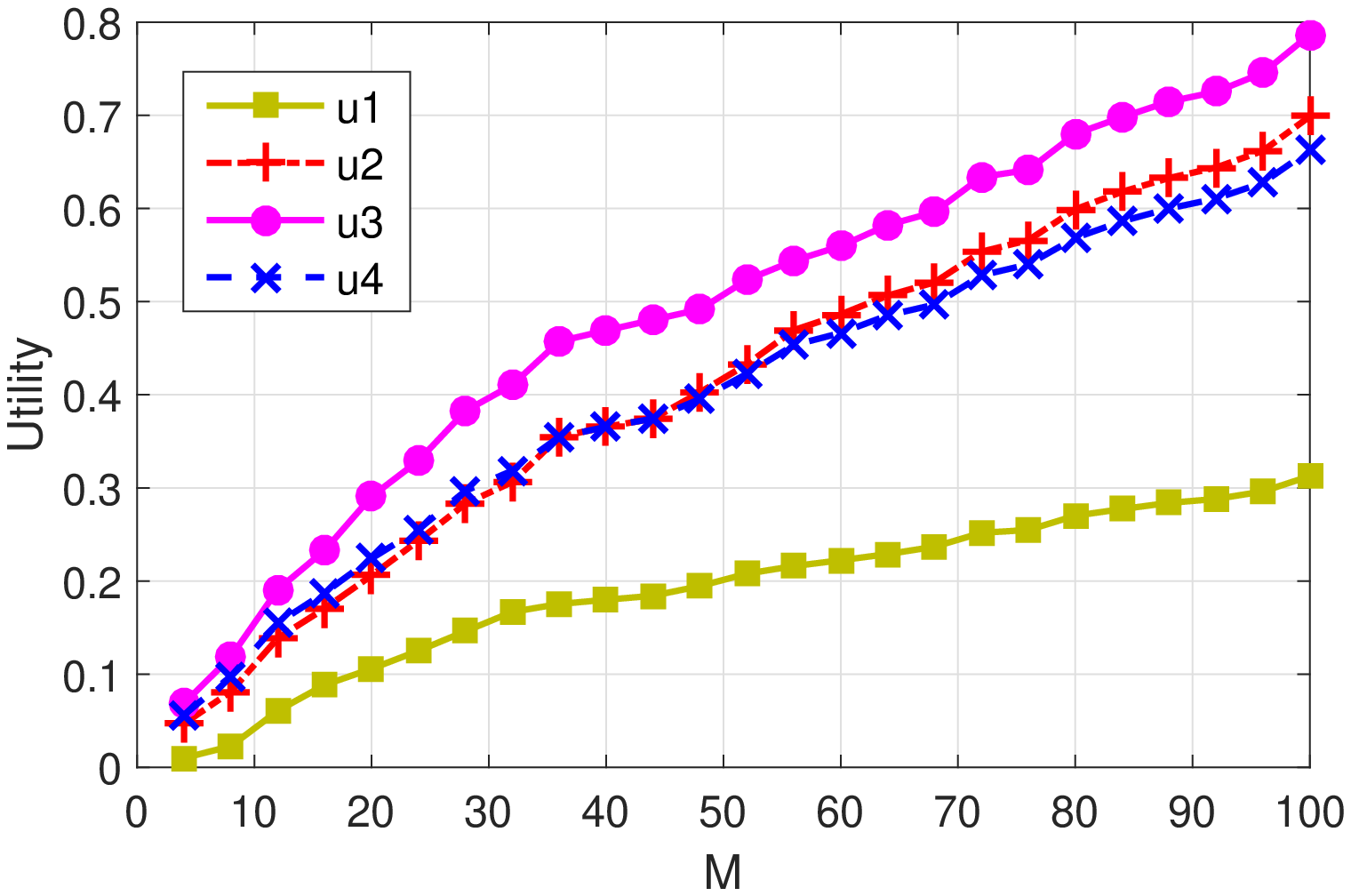}
	     \label{fig:varyMu}
	}\vspace{-0.2cm}
	\caption{Impact of M and N on the players' utilities}
\end{figure}

The impact of the number of players (i.e., number of ENs  and number of services) on the ME is illustrated in Figs.~\ref{fig:varyNu}, \ref{fig:varyMu}. The buyers have the same budget in this case. We show the utilities of buyers 1,~2,~3, and~4 in these figures. As expected, as the number of buyers increases, the utility of individual buyer decreases since the same set of ENs has to be shared among more services. On the other hand, the service utility increases significantly as the number of ENs increases. 
\vspace{-0.1in}

\vspace{-0.2cm}
\subsection{Analysis of Distributed Algorithms}

\subsubsection{Proportional Dynamics Allocation}

\begin{figure}[ht]
		\subfigure[Price trace]{
		  \includegraphics[width=0.245\textwidth,height=0.10\textheight]{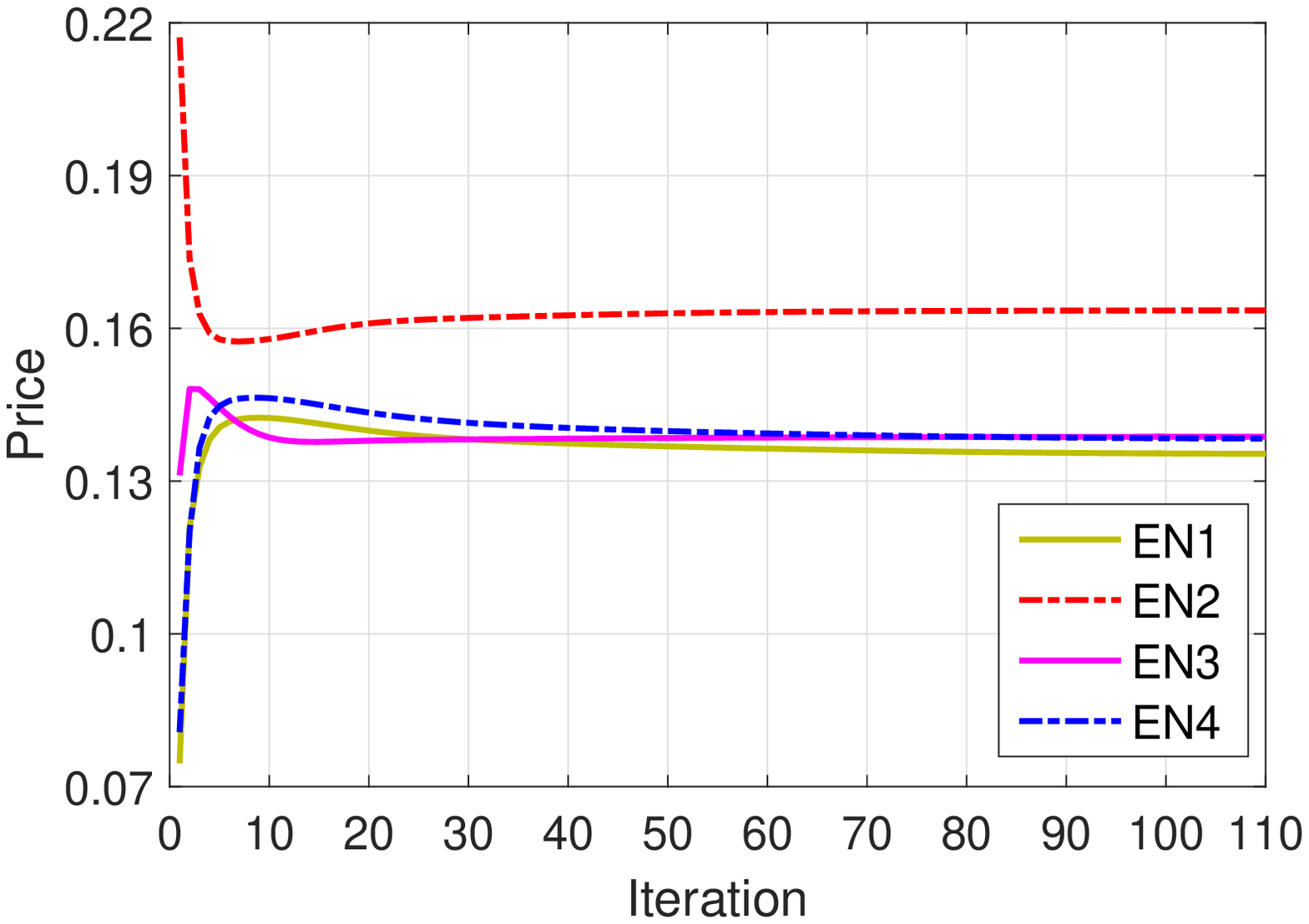}
	    \label{fig:PDptrace}
	}  \hspace*{-1.9em}
		 \subfigure[Buyer 1's bids]{
	     \includegraphics[width=0.245\textwidth,height=0.10\textheight]{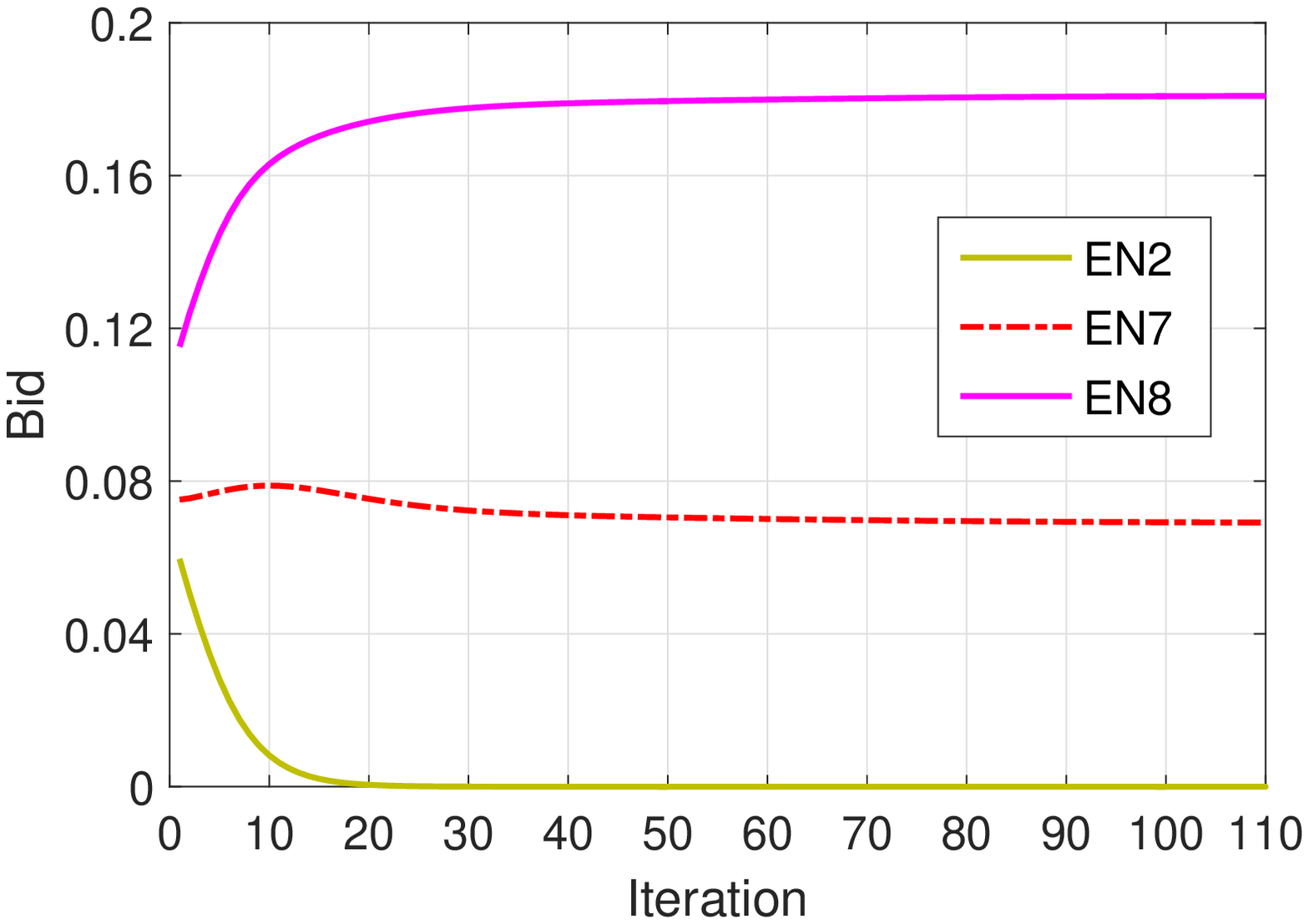}
	     \label{fig:PDbtrace}
	}\vspace{-0.2cm}
	\caption{Convergence of EN prices and bids}
\end{figure}

\begin{figure}[ht]
	\centering
		\subfigure[Tolerance = $10^{-4}$]
		{
		  \includegraphics[width=0.245\textwidth,height=0.10\textheight]{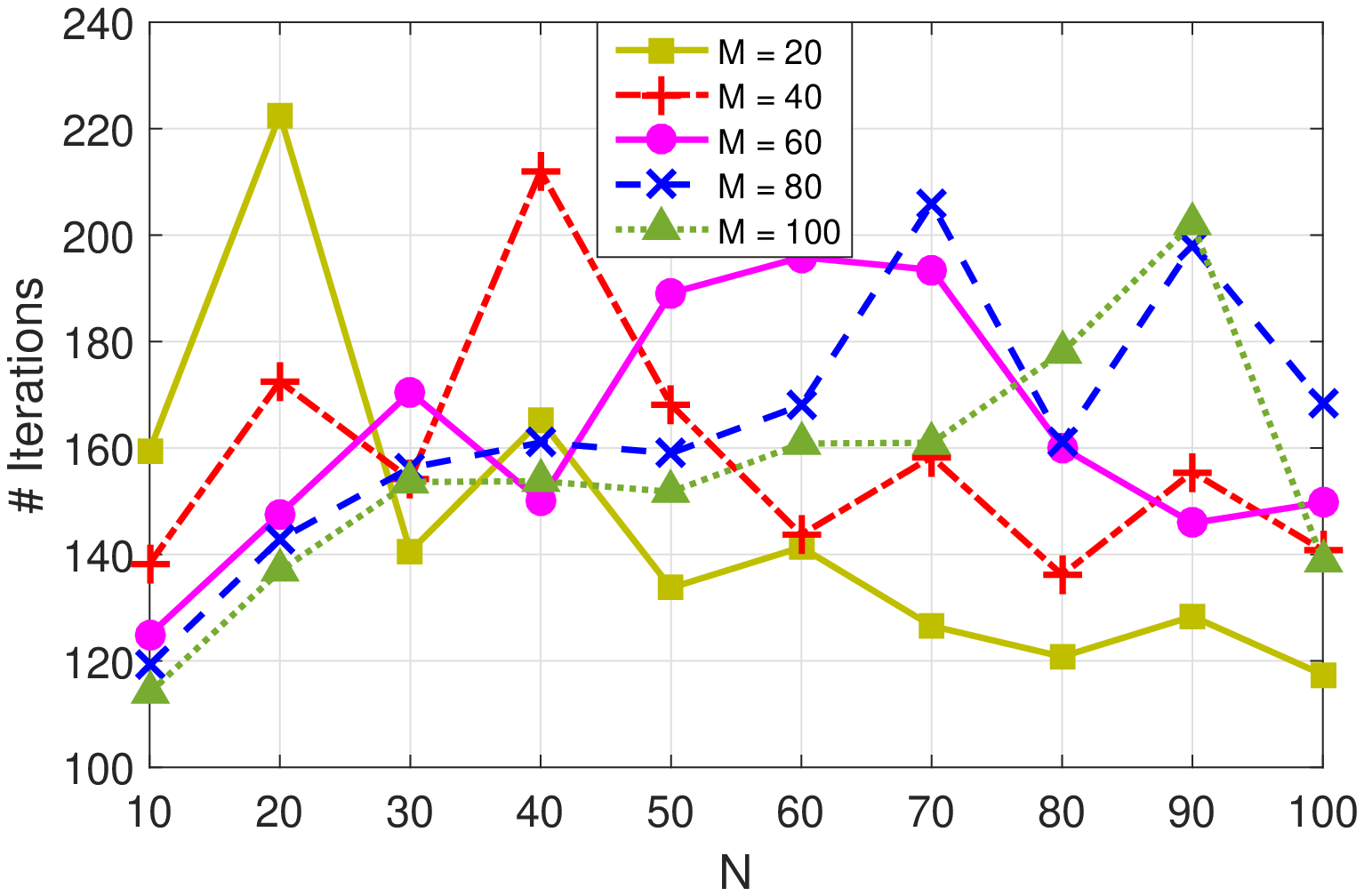}
	    \label{con1}
	}\hspace*{-1.9em}
		 \subfigure[Tolerance = $10^{-5}$]{
	     \includegraphics[width=0.245\textwidth,height=0.10\textheight]{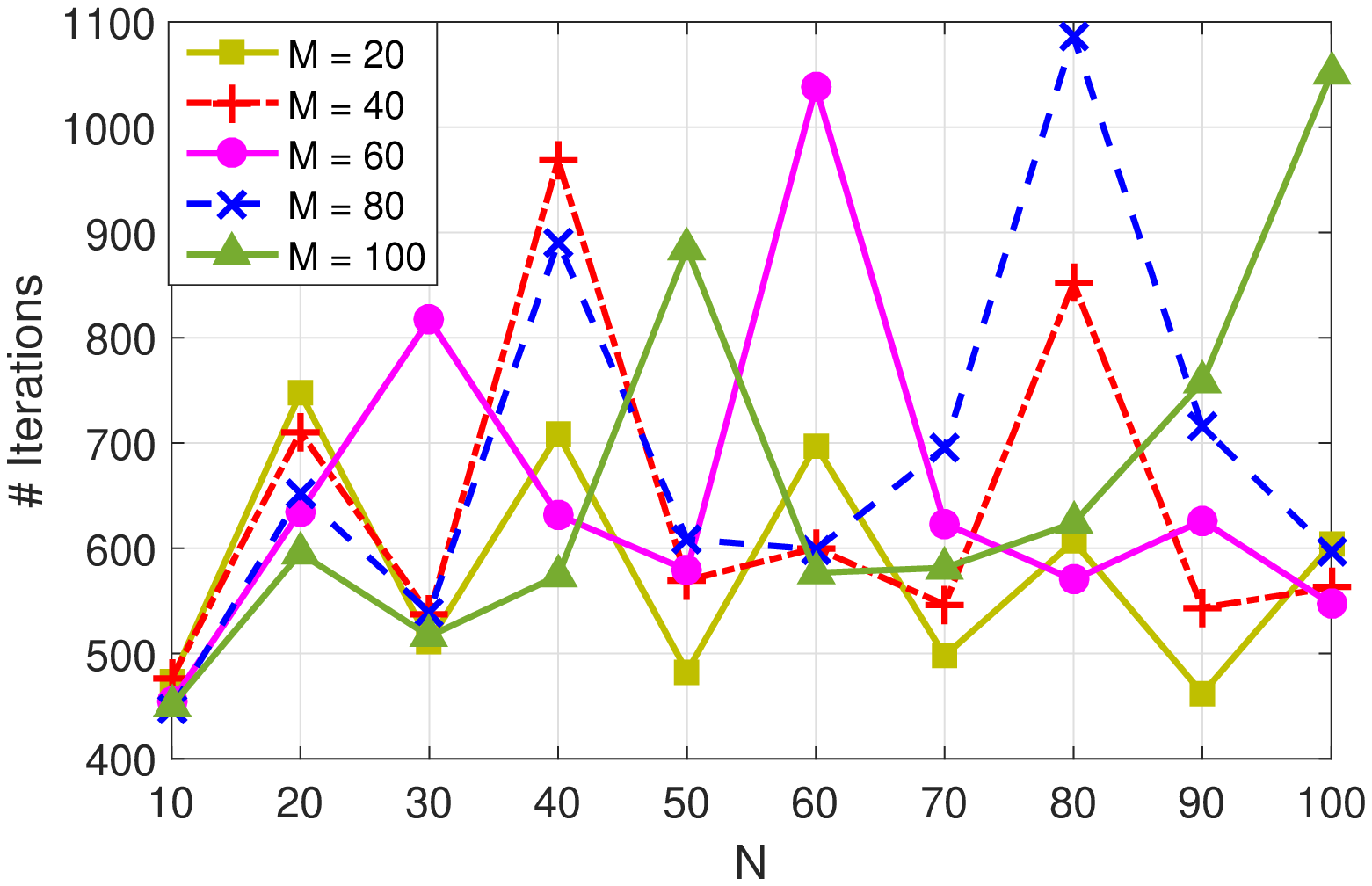}
	     \label{con2}
	}\vspace{-0.2cm} 
			\subfigure[Tolerance = $10^{-4}$]
		{
		  \includegraphics[width=0.245\textwidth,height=0.10\textheight]{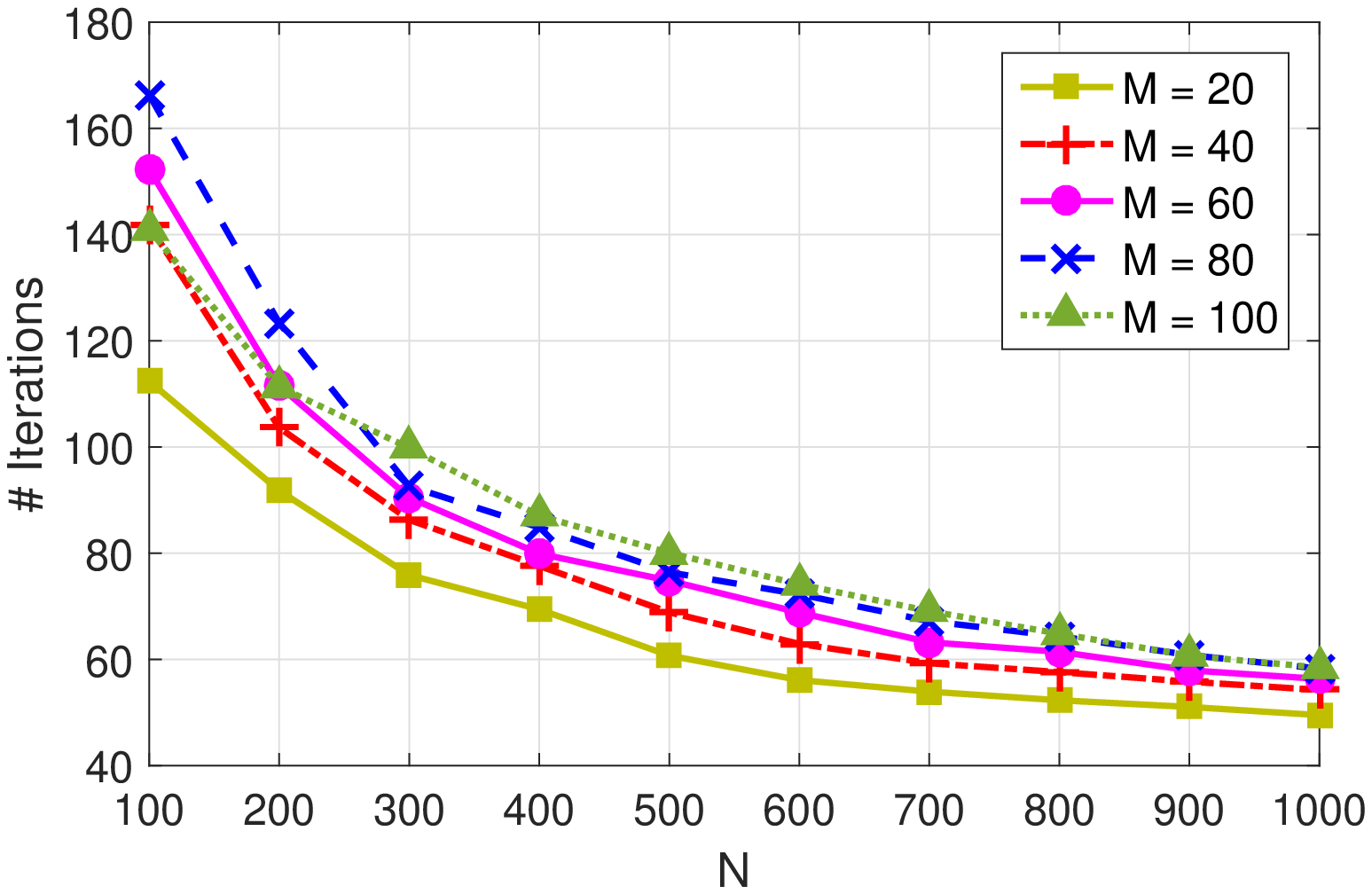}
	    \label{con3}
	} \hspace*{-1.9em}
		 \subfigure[Tolerance = $10^{-5}$]{
	     \includegraphics[width=0.245\textwidth,height=0.10\textheight]{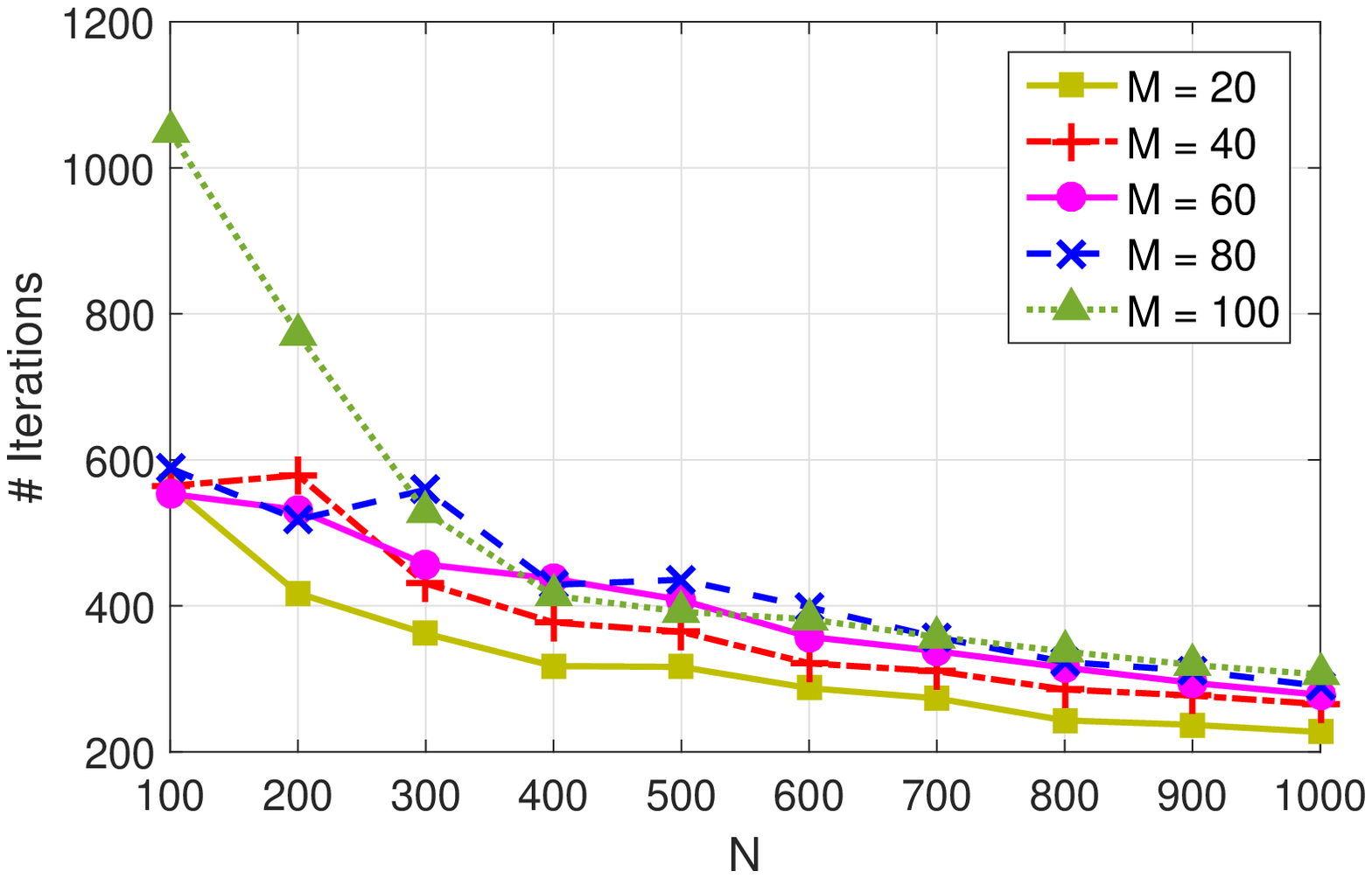}
	     \label{con4}
	}\vspace{-0.2cm}
	\caption{Number of iterations of PropDyn}
\end{figure}

The proportional dynamics mechanism (\textbf{PropDyn}) has low complexity and can be implemented in  a distributed manner. 
The convergence properties of this algorithm in the base case with 8 ENs and 4 services is shown in Figs.~\ref{fig:PDptrace}, \ref{fig:PDbtrace}. As we can see, the prices and the bids converge after a few tens of iterations. The running time of the algorithm is in order of milliseconds. 
Figs.~ \ref{con1}--\ref{con4} show the convergence properties of \textbf{PropDyn} as we change the number of ENs and the number of services. The  the stopping condition is defined according the relative change of the prices between iterations (i.e., absolute value of the price deviation over the price in the previous iteration). In other words, at iteration $t+1$, if $\max_j \Big(|p_j(t+1) - p_j(t)|/p_j(t) \Big)$ is sufficiently small (i.e., smaller than a tolerance threshold), the \textbf{PropDyn} algorithm terminates/converges. We have plotted the number of iterations of the \textbf{PropDyn} algorithm with different values of number of services (N) and number of ENs (M) as well as different tolerance thresholds in Figs. \ref{con1}--\ref{con4}. The goal of these figures is to give a rough idea about the number of iterations. 
For each of these figures, we generated 50 different datasets and took the average number of iterations over 50 runs. 
Theoretical results on the convergence rate of \textbf{PropDyn} can be found in \cite{fwu07,lzha09,lzha11}. Specifically, the number of iterations depends on the number of ENs, the tolerance threshold, and the system parameters $a_{i,j}$-which is indirectly related to the number of services. 


\begin{figure*}[ht]
		\subfigure[Utility in one instance]{
		  \includegraphics[width=0.35\linewidth,height=0.10\textheight]{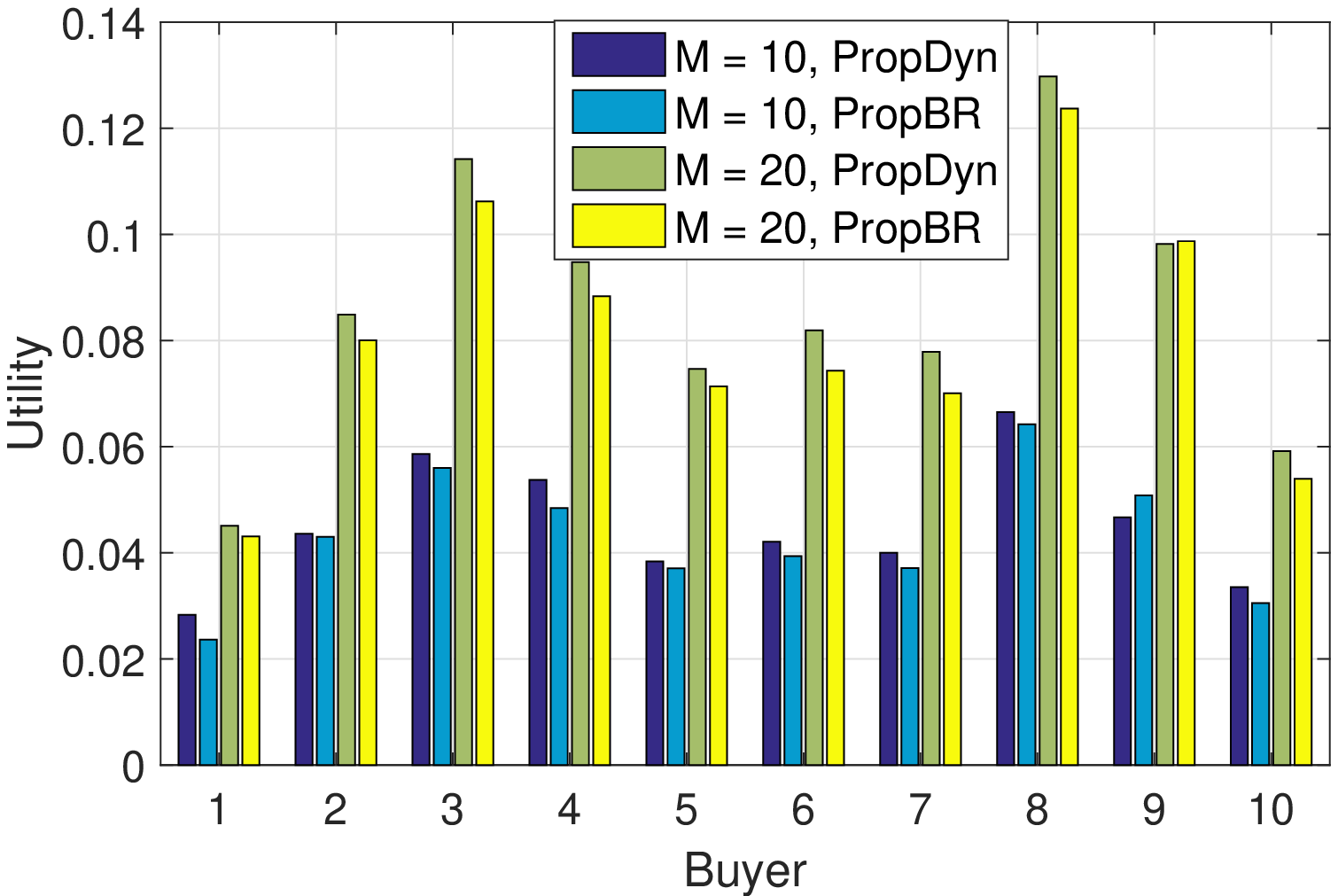}
	    \label{fig:PDite}
	} \hspace*{-1.9em}
		 \subfigure[Average utility in case 1]{
	     \includegraphics[width=0.35\linewidth,height=0.10\textheight]{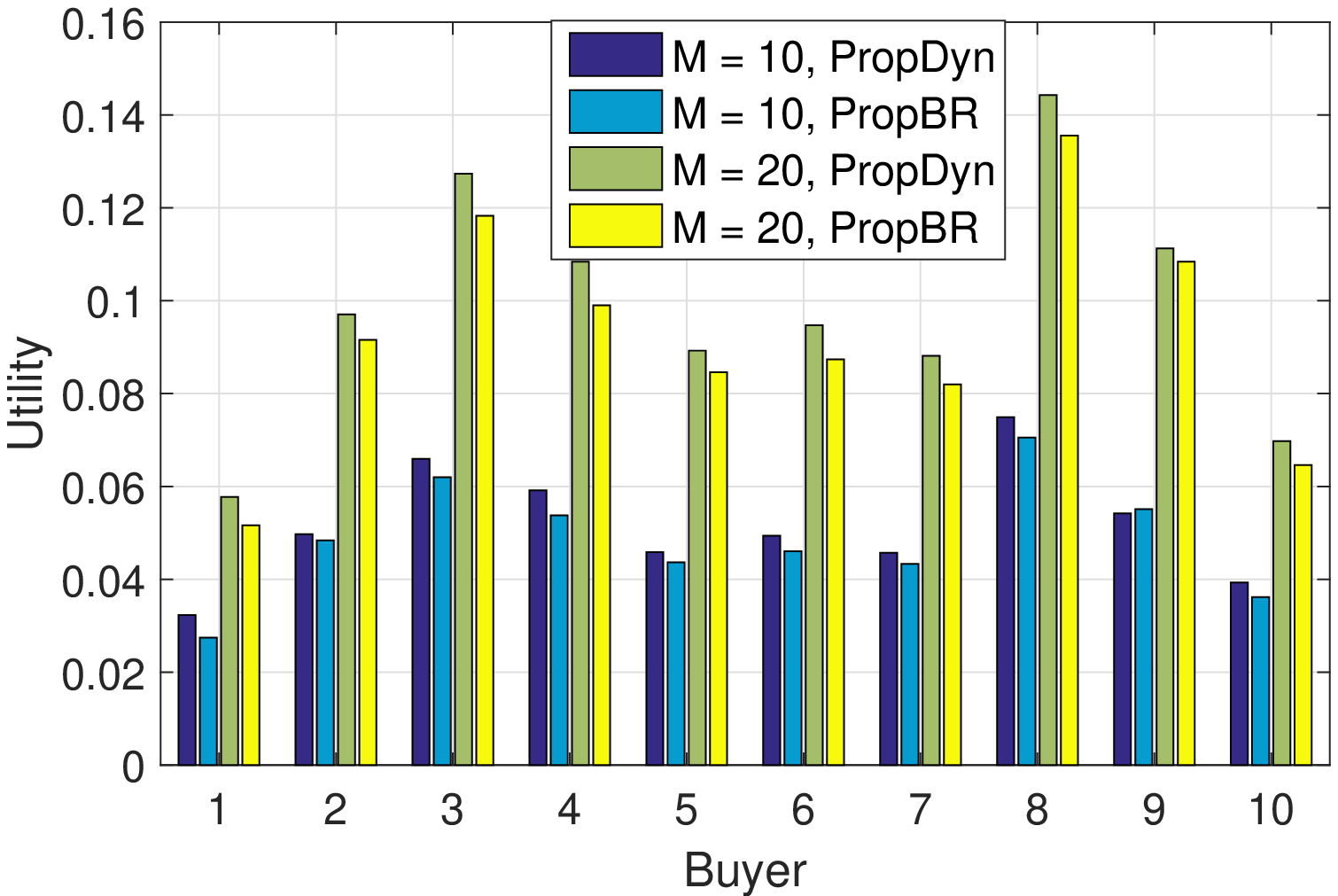}
	     \label{fig:ave1}
	} \hspace*{-1.9em} 
  \subfigure[Average utility in case 2]{
	     \includegraphics[width=0.35\linewidth,height=0.10\textheight]{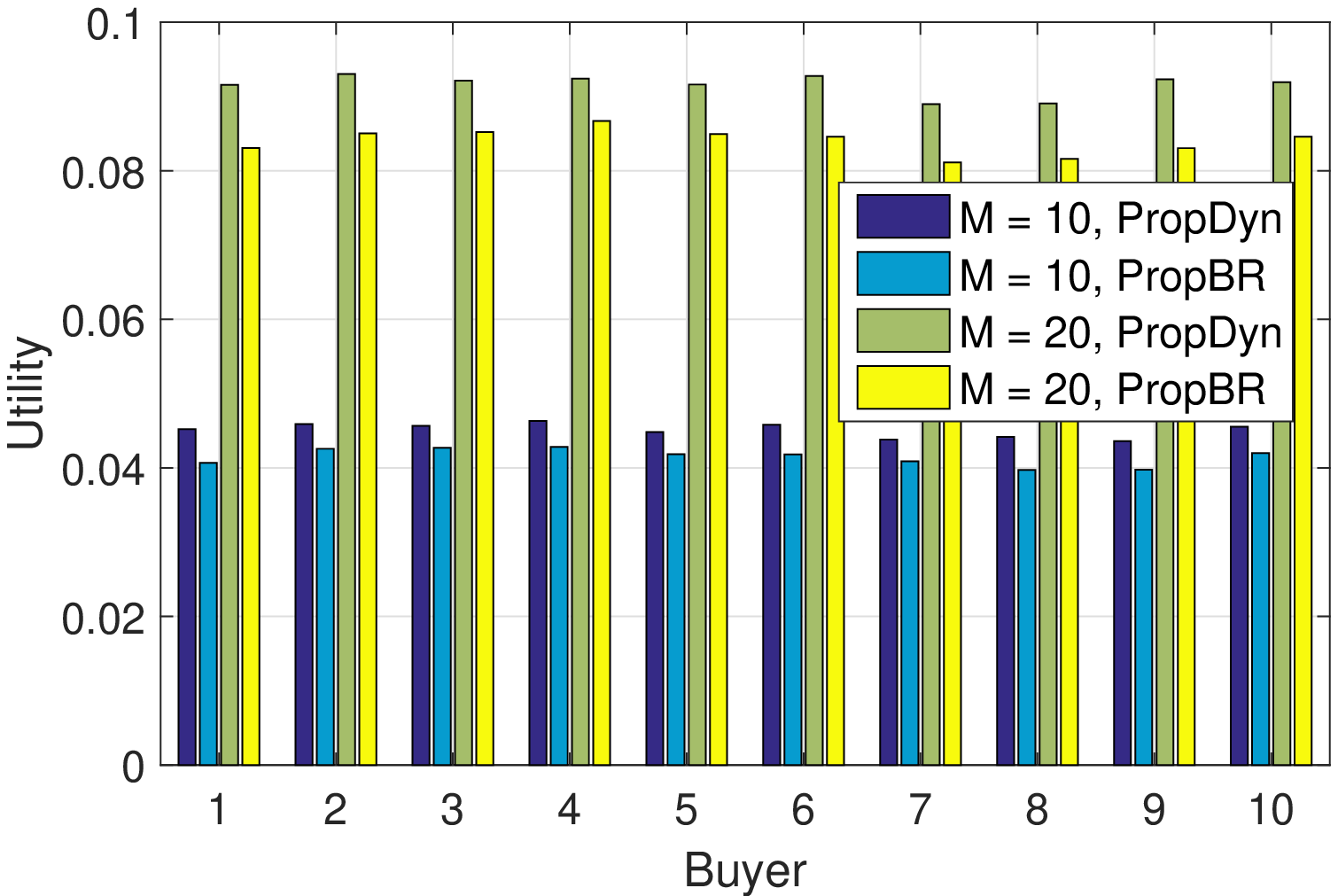}
	     \label{fig:ave2}
	} 
	\caption{Utility comparison between PropBR and PropDyn}
\end{figure*}

Figs.~\ref{fig:PDite},~\ref{fig:ave1},~\ref{fig:ave2} compare the buyers' utilities in the \textbf{PropDyn} and \textbf{PropBR} schemes. We select a particular instance with a set of 10 buyers and 20 ENs from the generated system data. Note that we have run simulation with numerous instances  and obtain similar trends. The utility of each buyer in the particular instance is presented in Fig.~\ref{fig:PDite}. As we can see, the utility values are higher for most of the buyers in the PropDyn scheme compared to the those in the PropBR scheme. In \ref{fig:ave1}, we add a random variable to each $a_{i,j}$ and run the schemes 100 times and take the average results. In \ref{fig:ave2}, we generate $a_{i,j}$ randomly in the range between 0.01 and 0.09. As we can observe, the buyers' utilities tend to be higher in the PropDyn scheme in comparison with the PropBR scheme. Furthermore, the PropBR requires  buyers to know more system information 
to play their BR actions in each round. 
The numerical results show that it brings almost no benefit to the buyers (no utility gain in most cases) to play PropBR scheme.. Hence, we can infer that \textit{the buyers should just follow the PropDyn scheme and obtain an ME allocation.}


\subsubsection{Function Approximation Algorithm}

\begin{figure}[ht]
		\subfigure[Number of iterations ($p_6$)]{
		  \includegraphics[width=0.245\textwidth,height=0.10\textheight]{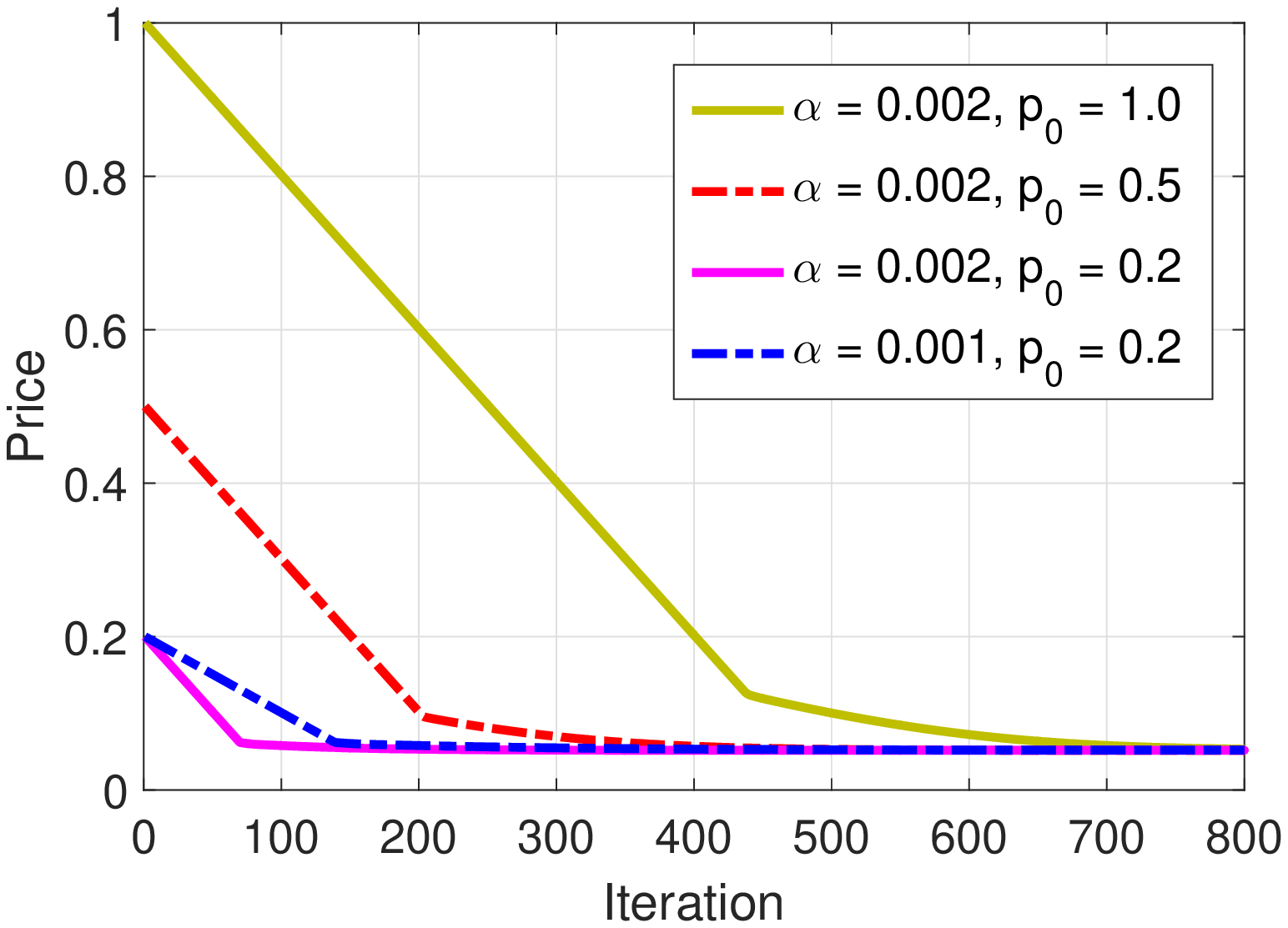}
	    \label{fig:cesite}
	} \hspace*{-1.9em} 
		 \subfigure[Price convergence]{
	     \includegraphics[width=0.245\textwidth,height=0.10\textheight]{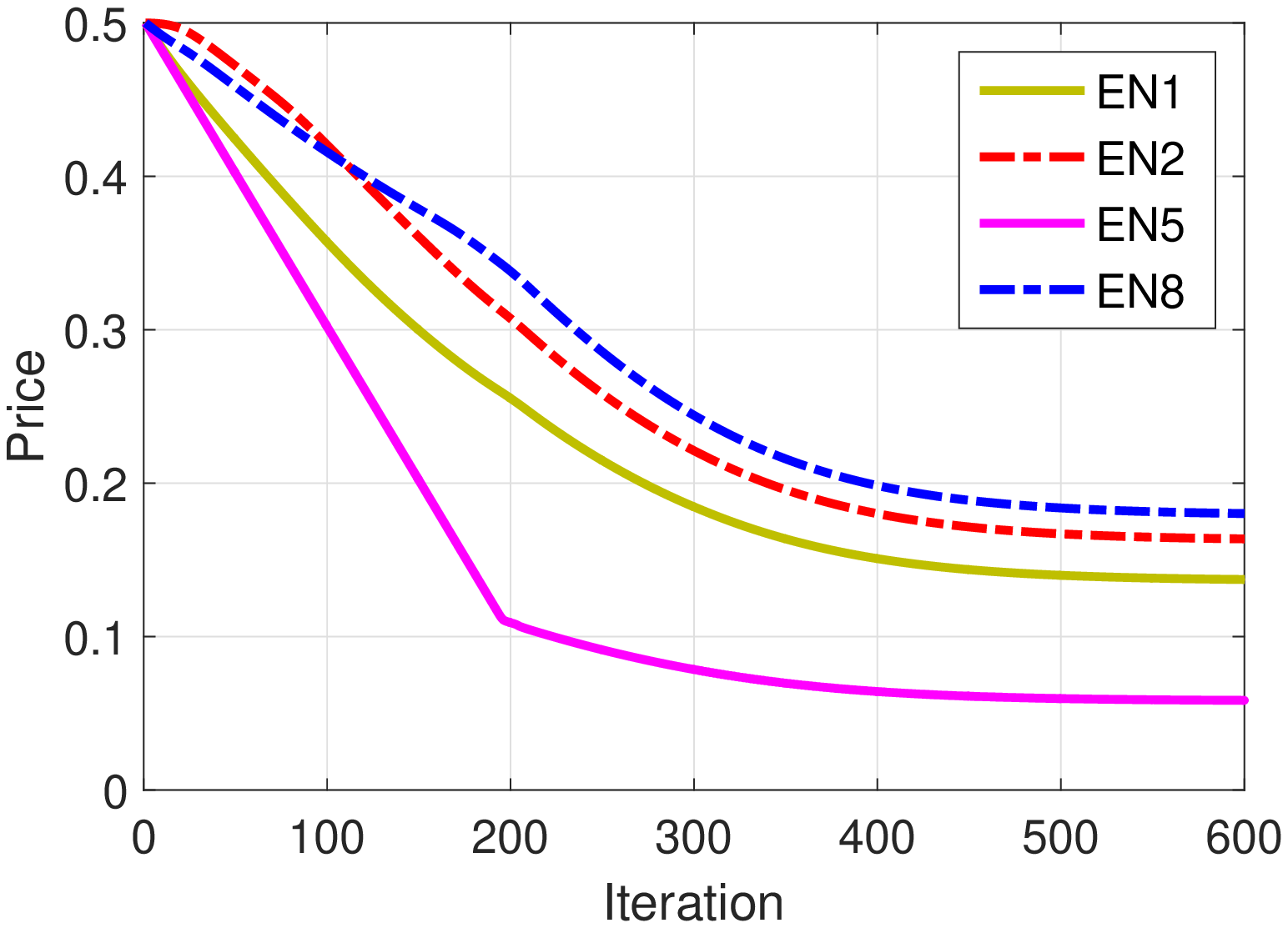}
	     \label{fig:cesprice}
	}\vspace{-0.2cm}
	\caption{Convergence of CES approximation}
\end{figure}

The convergence properties of the CES approximation scheme as well as its performance are reported. Thanks to the closed form expression of the optimal demand, the algorithm runs very fast even with high number of iterations. As expected, the number of iterations depends strongly on the step size and the initial prices. The convergence of EN6's price ($p_6$) is shown in Fig.~\ref{fig:cesite}. The number of iterations decreases as the initial prices are close to the final ME prices, which are unknown. The number of iterations decreases as the step size increases, but we cannot increase the step size $\gamma$ too much to ensure convergence. Fig.~\ref{fig:cesprice} presents the price traces of different ENs until convergence with $\alpha = 0.001 $ and  $p_0 = 0.2$. 

\begin{figure}[ht]
		\subfigure[Utility comparison]{
		  \includegraphics[width=0.245\textwidth,height=0.10\textheight]{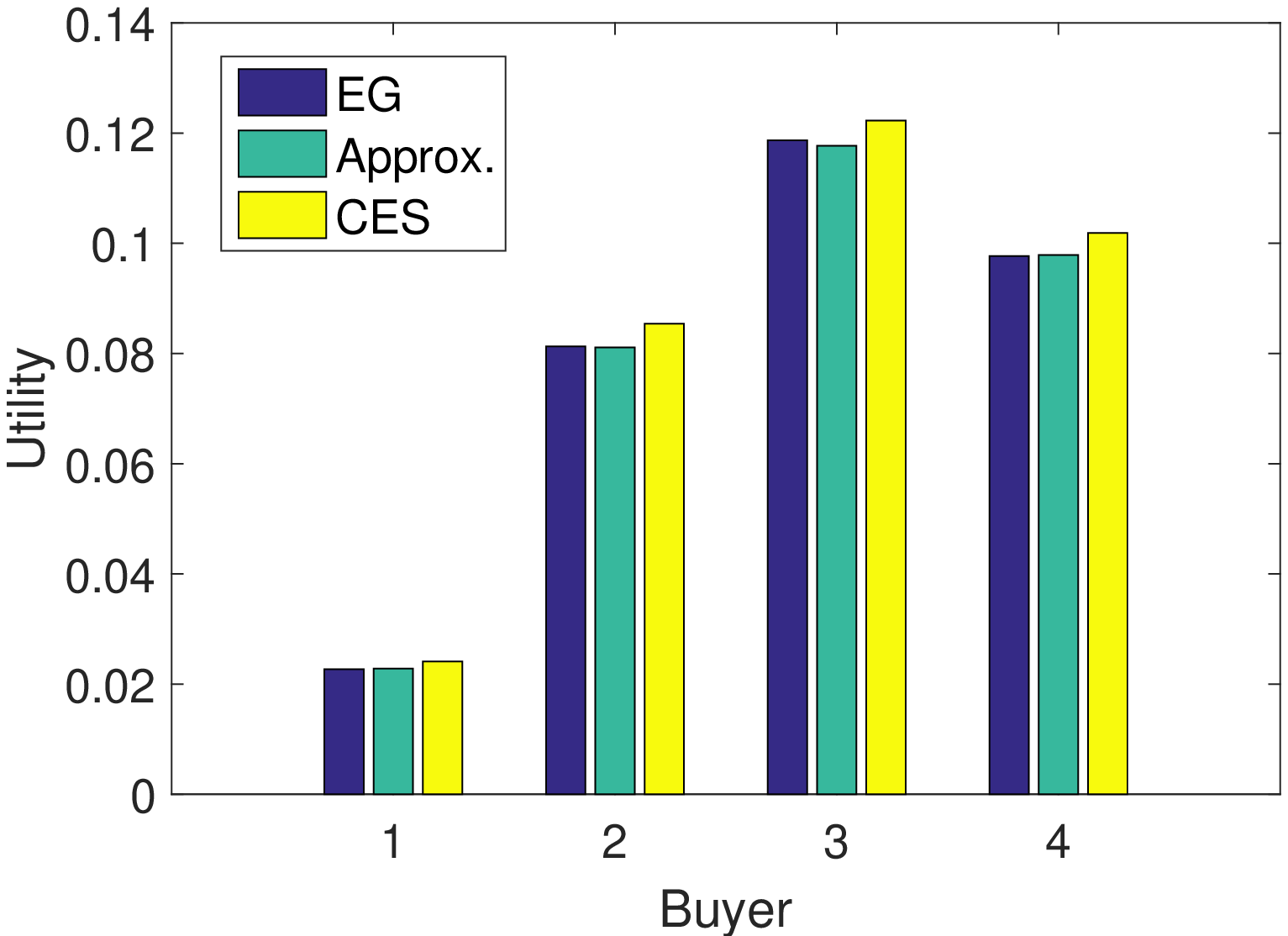}
	    \label{fig:cesu}
	} \hspace*{-1.9em}
		 \subfigure[Varying $\rho$]{
	     \includegraphics[width=0.245\textwidth,height=0.10\textheight]{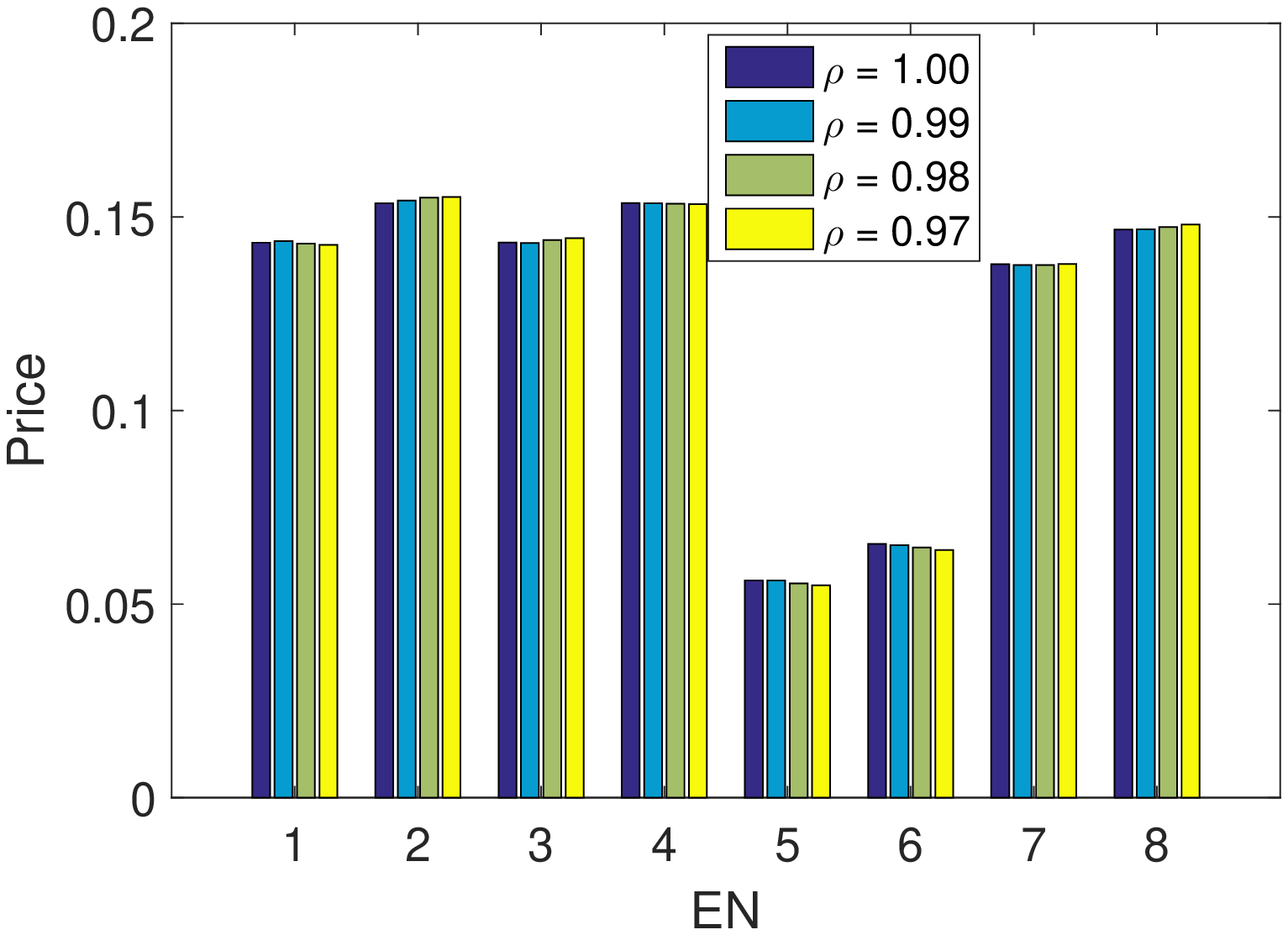}
	     \label{fig:cesp}
	}\vspace{-0.2cm}
	\caption{CES approximation utility comparison}
\end{figure}

In Fig.~\ref{fig:cesu}, we study the performance of the approximation scheme by comparing utility of the buyers under the centralized convex program (EG), the approximation CES utility (CES), and the approximation linear utility (\textbf{Approx.}). In the \textit{Approx.} scheme, the utility of buyer $i$ is $x_{i,j} a_{i,j}$ where $x_{i,j}$ is the solution of the optimization problem with CES approximation utilities. As we can observe, the values of the utilities are very similar, which confirms that the proposed approximation scheme performs well. In this figure, we set $\rho$ to be 0.99. Finally, the equilibrium prices with different values of $\rho$ is shown in Fig.~\ref{fig:cesp}. It is easy to see that the prices are almost equal for different values of $\rho$.

%
%

\begin{figure*}[ht]
	  \subfigure[Price]{
	     \includegraphics[width=0.35\textwidth,height=0.10\textheight]{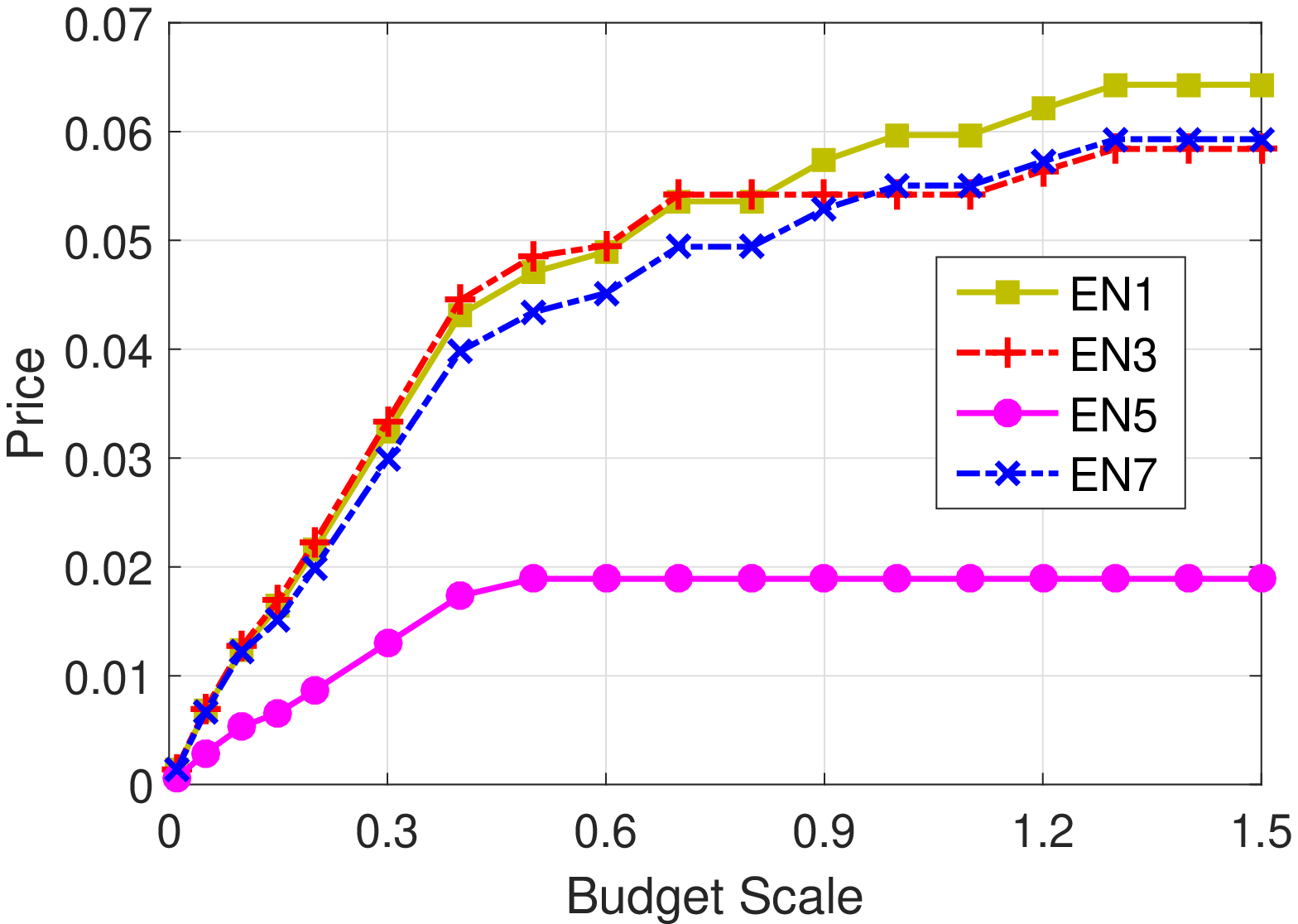}
	     \label{fig:quasiprice}
	} \hspace*{-1.9em}  
		\subfigure[Utility]{
		  \includegraphics[width=0.35\textwidth,height=0.10\textheight]{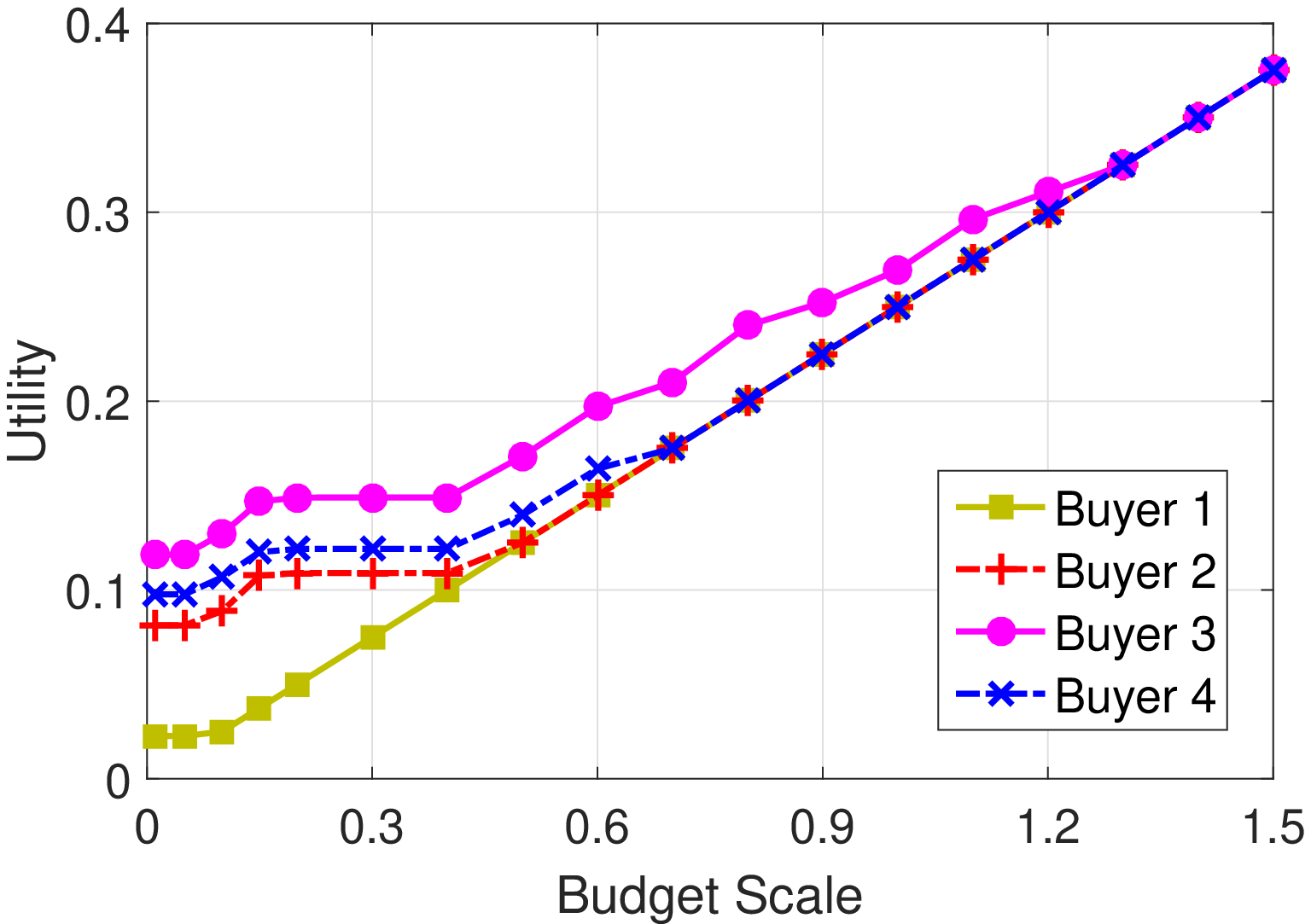}
	    \label{fig:quasiuti}
	} \hspace*{-1.9em}  
		 \subfigure[Surplus Ratio (surplus/budget)]{
	     \includegraphics[width=0.35\textwidth,height=0.10\textheight]{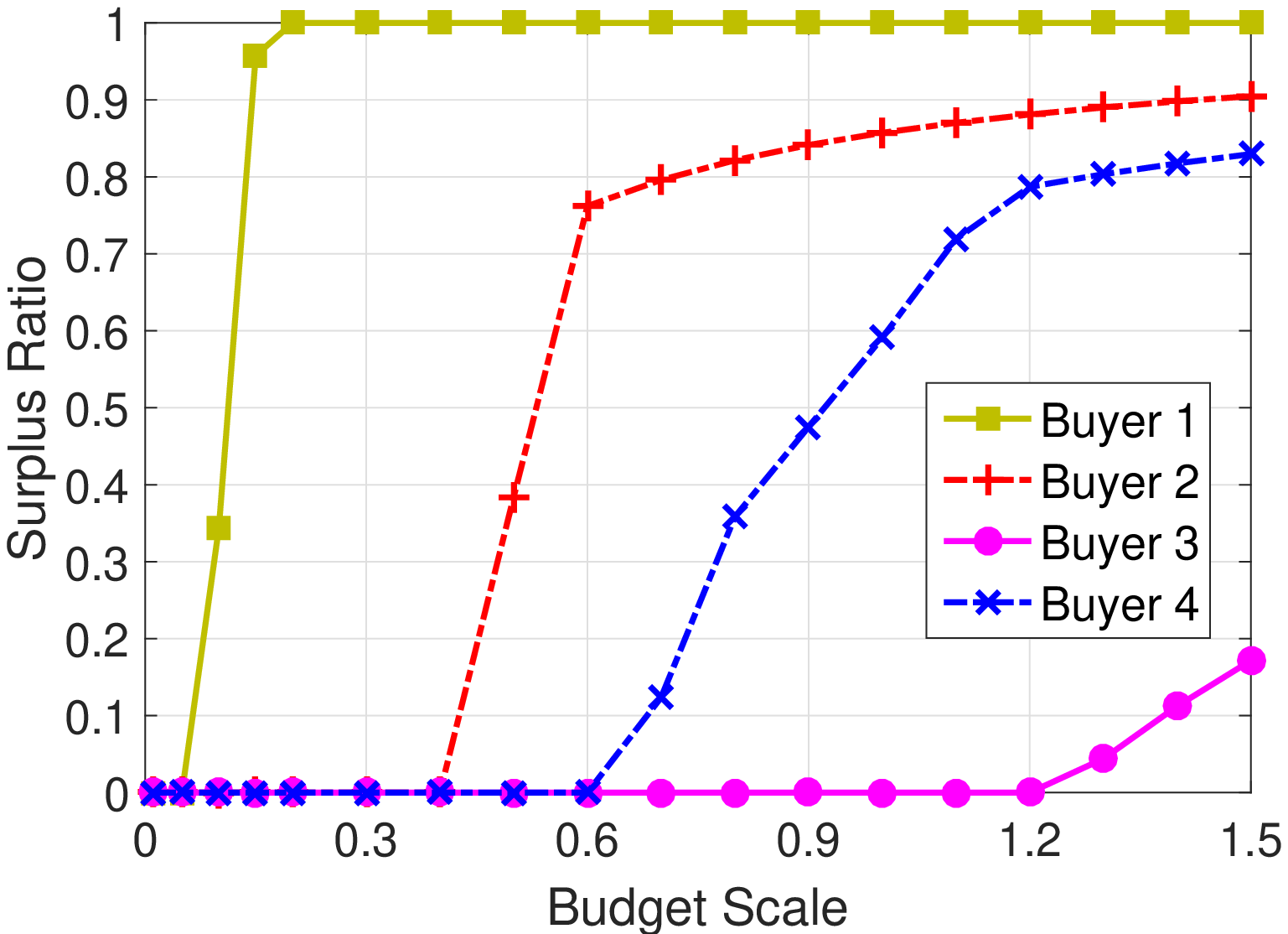}
	     \label{fig:quasisurplus}
	}\vspace{-0.1cm}

	\caption{ME in the net profit maximization model  }
\end{figure*}

\subsection{Net Profit Maximization Model}
\label{sim2}

We now evaluate the second model where 
the services aim to maximize their net profits.  We use the same system setting with 4 services and 8 ENs in  the \textbf{base case} as before. From the objective function of the buyer, we know that a buyer will buy resource from an EN only when the price of the EN is less than or equal to its utility gain from the EN. 
In the revenue maximization case as in the \textbf{basic model}, the equilibrium prices increase linearly at the same rate as the budget. However, as can be seen in Fig.~\ref{fig:quasiprice}, this property does not hold in the net profit maximization model. Budget scale is the scaling factor by which we multiply the original budget. The figure shows that the equilibrium prices increase then become saturated after certain values of the budgets. At these (saturated) prices, buying resources from the ENs or not does not change the utility for a buyer (i.e., $p_j = a_{i,j}$). When the budget is large enough, the utilities of the buyers become equal to their budgets. It means procuring resources or not does not bring any additional benefit to the buyers. These results are shown in Figs.~ \ref{fig:quasiuti},~\ref{fig:quasisurplus}. 

\begin{figure}[ht]
	\centering
		\subfigure[Prices]{
		  \includegraphics[width=0.245\textwidth,height=0.10\textheight]{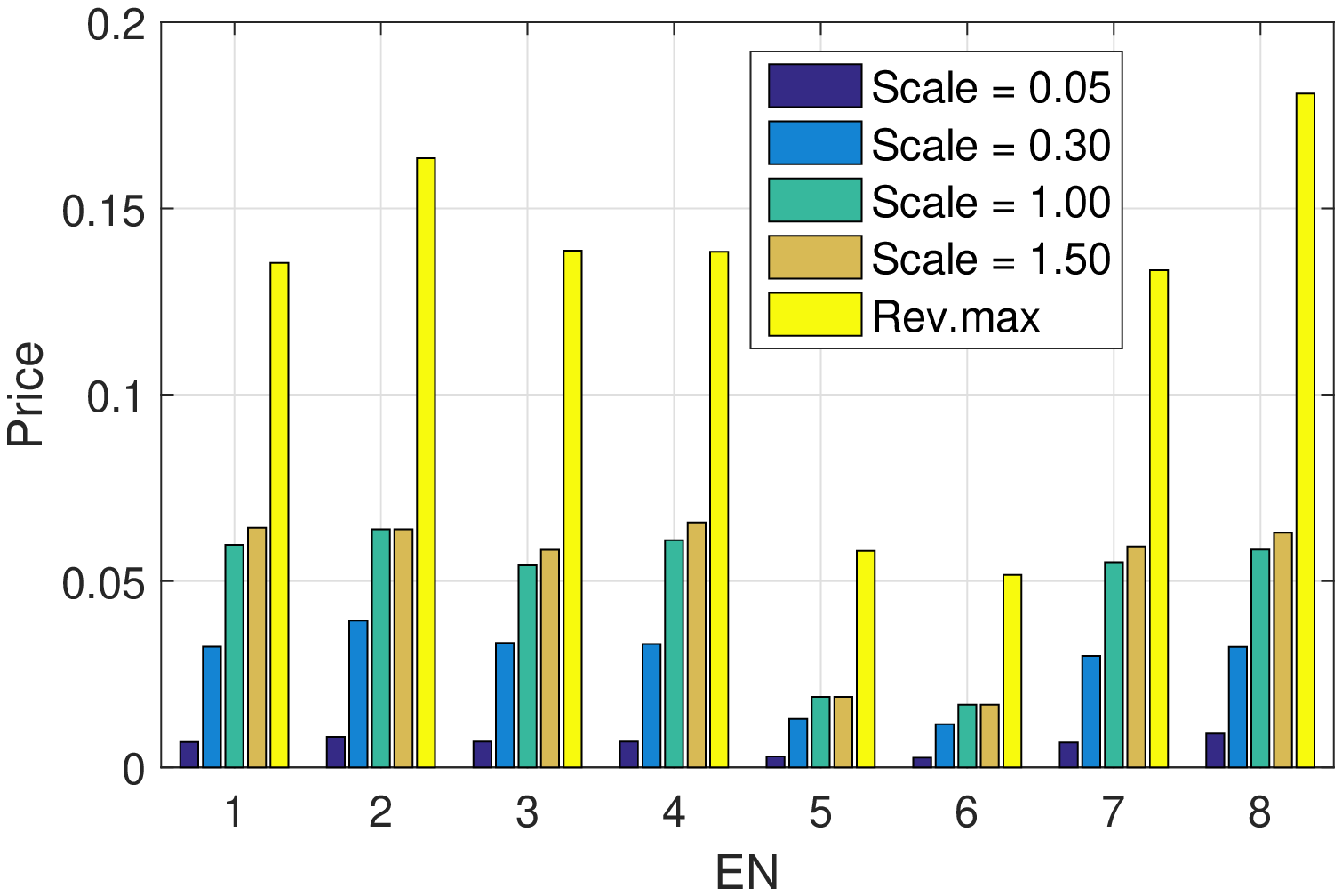}
	    \label{fig:quasip}
	} \hspace*{-1.9em}
		 \subfigure[Utilities]{
	     \includegraphics[width=0.245\textwidth,height=0.10\textheight]{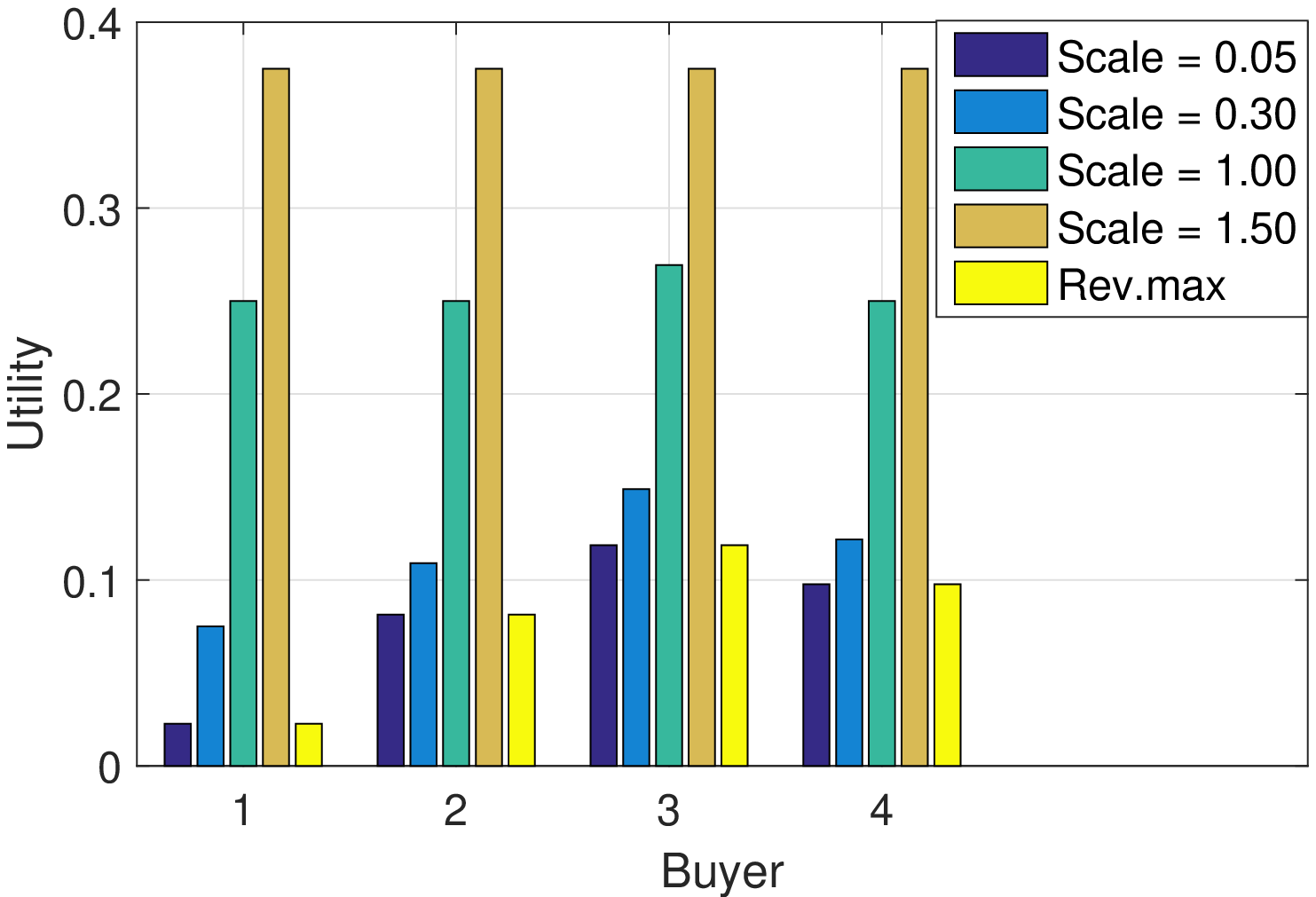}
	     \label{fig:quasiu}
	}\vspace{-0.2cm}
	\caption{Impact of budget on equilibrium prices and utilities}
\end{figure}

Figs.~\ref{fig:quasip}, \ref{fig:quasiu} present equilibrium prices and optimal utilities, respectively, as the budget varies.
\textbf{Rev.max} corresponds to the first model (i.e., revenue maximization) with scale equal to 1. As we can observe, for the same budget (i.e., scale = 1), equilibrium prices in the second model are smaller than equilibrium prices in the first model because in the net profit maximization model, a service only buys resource from an EN that gives it positive gain.
Also, the service utilities at the equilibrium in the second model is greater than those in the first model due to lower equilibrium prices and budget surplus is considered in the second model. Finally, in the second model, the equilibrium prices and optimal utilities increase as the budget increases. As explained above, equilibrium prices become saturated  in the second model at certain points. Hence, the equilibrium prices increase very little as budget scale increases from 1 to 1.5.

\section{Conclusion and Future Works}
\label{con}
 In this work, we consider the resource allocation for an EC system which consists  geographically distributed heterogeneous ENs with different configurations and a collection of services with different desires and buying power. 
Our main contribution is to suggest the famous concept of General Equilibrium in Economics as an effective solution for the underlying EC resource allocation problem. The proposed solution produces an ME that not only Pareto-efficient but also possesses many attractive fairness properties. The potential of this approach are well beyond EC applications. 
For example, it can be used to share storage space in edge caches to different service providers. We can also utilize the proposed framework to share resources (e.g., communication, wireless channels) to different users or groups of users (instead of services and service providers). 
Furthermore, the proposed model can extend to the multi-resource scenario where each buyer needs a combination of different resource types (e.g., storage, bandwidth, and compute) to run its service. We will formally report these cases (e.g., network slicing, NFV chaining applications) in our future work. 

The proposed framework could serve as a first step to understand new business models and  unlock the enormous potential of the future EC ecosystem.  There are several future research directions. For example, we will investigate the ME concept in the case when several edge networks cooperate with each other to form an edge/fog federation. Investigating the impacts of the strategic behavior on the efficiency of the ME is another interesting topic. Note that N. Chen {\em et. al.} \cite{nche16} have shown that the gains of buyers for strategic behavior in Fisher markets are small. Additionally, in this work, we implicitly assume the demand of every service is unlimited. It can be verified that we can add the maximum number of requests constraints to the EG program to capture the limited demand case, and the solution of this modified problem is indeed an ME. However, although the optimal utilities of the services in this case are unique, there can have infinite number of equilibrium prices. We are investigating this problem in our ongoing work. Also, integrating the operation cost of ENs into the proposed ME framework
 is a subject of our future work. Finally, how to compute market equilibria with more complex utility functions that capture practical aspects such as task moving expenses among ENs and data privacy is an interesting future research direction. 

'

\bibliographystyle{IEEEtran}


\appendices

\section{Proof of Proposition 6.1}

Consider two following problems
\beqn
\label{prob1}
 \textbf{(P1)} \quad && \underset{x_{i} \geq 0}{\text{maximize}} ~ u_i(x_i) \\
\label{61d1}
 && \text{subject to} ~  \sum_j p_j x_{i,j} \leq B_i \\ 
\label{subprob}
 \textbf{(P2)}  \quad && \underset{x_i \geq 0}{\text{maximize}} ~ B_i \ln u_i(x_i) - \sum_j p_j x_{i,j} 
\eeqn
where $u_i(x_i) = \sum_j \big(a_{i,j}x_{i,j} \big)^{\rho}$. We will show that the given a positive price vector $p$, each service $i$ can solve either \textbf{(P1)} or \textbf{(P2)}, which have the same optimal solution.
\beqn
\label{cesdemand}
x_{i,j} = \Big( \frac{a_{i,j}^{\rho}}{p_j}\Big)^{\frac{1}{1-\rho}} \frac{B_i}{ \sum_{j=1}^M \Big( \frac{a_{i,j}}{p_j} \Big)^{\frac{\rho}{1-\rho}}} , \quad \forall i,~j.
\eeqn
Consider \textbf{(P1)}. Let $\lambda_i$ is the dual variable associated with constraint (\ref{61d1}). The Lagrangian is
\beqn
L_i(x_i,\lambda_i) =  \sum_j \big(a_{i,j}x_{i,j} \big)^{\rho} + \lambda_i \Big( B_i - \sum_j p_j x_{i,j} \Big)
\eeqn
Take the first derivative and set it equal to zero, we have
\beqn
\frac{\partial L_i}{\partial x_{i,j}} = \rho a_{i,j} \big( a_{i,j} x_{i,j} \big)^{\rho - 1} - \lambda_i p_j &=& 0, \quad \forall j.\\ 
\label{temp1}
 \Rightarrow  ~\rho a_{i,j}^{\rho} x_{i,j}^{\rho - 1} &=& \lambda_i p_j,~ \forall j.
\eeqn
 If $\lambda_i = 0 \Rightarrow   a_{i,j} x_{i,j} = 0, ~ \forall j  \Rightarrow u_i(x_i) = 0.$ Clearly, optimal value of $u_i(x_i) >0$. Hence, $\lambda_i > 0.$ Also, as mentioned in the report, $p_j > 0, ~\forall j$. Thus, $x_{i,j} > 0,~\forall j.$
From (\ref{temp1}), we have 
\beqn
\label{temp2}
\rho a_{i,k}^{\rho}  x_{i,k}^{\rho - 1} &=& \lambda_i p_k, \quad \forall k \in \mathcal{M}.
\eeqn
Since at least one element of vector $a_i$ is positive, we assume $a_{i,j} > 0.$ From  (\ref{temp1}) and  (\ref{temp2}), we have
\beqn
\frac{\rho a_{i,k}^{\rho}  x_{i,k}^{\rho - 1}   }{ \rho a_{i,j}^{\rho} x_{i,j}^{\rho - 1}   } = \frac{p_k}{p_j}, \quad \forall k \in \mathcal{M} \\
\label{temp3}
\text{Hence,} ~ ~x_{i,k} = \Big(\frac{p_k}{p_j} \Big)^{\frac{1}{\rho - 1 }} \Big( \frac{a_{i,k}}{a_{i,j}} \Big)^{\frac{\rho}{1-{\rho}}} x_{i,j}, \quad \forall k \in  \mathcal{M}.
\eeqn
At the optimality, $\lambda_i > 0$, hence, $\sum_k p_k x_{i,k} = B_i.$ 
Substituting value of every $x_{i,k}$ from (\ref{temp3}), we can compute the value of $x_{i,j}$ and infer the value of every $x_{i,k}, ~\forall k \in  \mathcal{M}.$ We obtain the optimal solution as shown in (\ref{cesdemand}). Note that this optimal solution is  positive. 

Next, consider \textbf{(P2)}, which is the sub-problem to be solved by every service $i$. Define  $f_i(x_i) = B_i \ln u_i(x_i) - \sum_j p_j x_{i,j}$, where $u_i(x_i) = \sum_j \big(a_{i,j}x_{i,j} \big)^{\rho}.$ Take the first derivative and set it to be equal to zero, we have
\beqn
\frac{\partial f_i(x_i)}{\partial x_{i,j}} =  B_i \rho a_{i,j} \frac{ \Big( a_{i,j}x_{i,j}\Big)^{\rho - 1}}{ u_i(x_i)} - p_j = 0, \quad \forall j \\
\label{temp4}
\Rightarrow  B_i \rho a_{i,j}^{\rho} x_{i,j}^{\rho - 1} = p_j u_i(x_i),\quad \forall j.
\eeqn
Following similar steps as for \textbf{(P1)} above, we obtain the closed form expression of the optimal solution of \textbf{(P2)}.

\section{Proof of Proposition 7.1}
The EG convex program is
\beqn
\label{EGprogram1}
\vspace{-0.1in} 
\underset{\mathcal{X},u}{\text{maximize}} ~\sum_{i=1}^N B_i \ln u_i
\eeqn
subject to
\vspace{-0.3in} 
\beqn
\label{EQ11}
u_i =\sum_{j=1}^M a_{i,j} x_{i,j}, \quad \forall i  \\
\vspace{-0.5in}
\label{EQ12}
\sum_{i=1}^N x_{i,j} \leq  1, \quad \forall j \\
\label{EQ13}
x_{i,j} \geq 0, \quad \forall i,~j.
\eeqn

Let $\eta_i$, $p_j$, and $\nu_{i,j}$ be the dual variables associated with (\ref{EQ11}), (\ref{EQ12}), and (\ref{EQ13}), respectively. The Lagrangian function of the EG program (\ref{EGprogram1})-(\ref{EQ13}) is
\beqn
L(u,\mathcal{X},\eta,p,\nu) = \sum_i B_i \ln u_i + \sum_j p_j ( 1 - \sum_i x_{i,j})  \\ \nonumber
+ \sum_i \eta_i \Big( \sum_j a_{i,j}x_{i,j} - u_i \Big)  + \sum_i \sum_j \nu_{i,j} x_{i,j}. \\ \nonumber
\eeqn
Define $f_i(u_i) = - B_i \ln u_i$, we have
\beqn
L(u,\mathcal{X},\eta,p,\nu) =  \sum_i \Big( - \eta_i u_i - f_i(u_i) \Big)  + \sum_j p_j   \\ \nonumber
+ \sum_i \sum_j \Big( - p_j +  \eta_i a_{i,j} + \nu_{i,j}  \Big) x_{i,j}.
\eeqn
The dual function is 
\beqn
g(\eta,p,\nu) = \underset{u,\mathcal{X}}{\text{maximize}} ~~ L(u,\mathcal{X},\eta,p,\nu).
\eeqn
The dual problem is
\beqn
 \underset{\eta \geq 0,p \geq 0,\nu \geq 0}{\text{minimize}} ~~g(\eta,p,\nu).
\eeqn
According to the KKT conditions, we have
\beqn
\frac{\partial L(u,\mathcal{X},\eta,p,\nu)}{\partial x_{i,j}} = - p_j +  \eta_i a_{i,j} + \nu_{i,j}  = 0, \quad \forall i,~j
\eeqn
which gives us $ - p_j +  \eta_i a_{i,j} = - \nu_{i,j} \leq 0, ~ \forall i,~j$. Thus, $ \eta_i a_{i,j} \leq p_j, ~ \forall i,~j$.
Furthermore, the dual objective is equivalent to
\beqn
&& \underset{\eta,p,\nu}{\text{min}} \Big(\underset{u}{\text{max}} \sum_i \Big( - \eta_i u_i - f_i(u_i) \Big)  + \sum_j p_j   \Big) \\ \nonumber
&=& \underset{\eta,p,\nu}{\text{min}}  \Big( \sum_j p_j +  \sum_i \underset{u}{\text{max}} \big(-\eta_i u_i - f_i(u_i) \big) \Big) \\ \nonumber
&=&  \underset{\eta,p,\nu}{\text{min}}  \Big( \sum_j p_j +  \sum_i f_i^*(-\eta_i) \big) \Big) .
\eeqn
Note that the (Fenchel) conjugate function of a function $h :  \mathbb{R}^n \rightarrow \mathbb{R}  $  is defined as follows \cite{boyd}.
\beqn
h^*(\mu) :=\underset{x}{\text{sup}} \big\{ \mu^T x - h(x) \big\}.
\eeqn
We need to compute the conjugate of $f_i(u_i) = - B_i \ln u_i$. 
Consider function $h(x) = - A \ln x$ where A is a constant. First, we have the conjugate of $ h_1(x) = - \ln x$ is $h_1^*(\mu) = - 1 - \ln (-\mu)$. Also, the conjugate function has the following property \cite{boyd}: if $h_2(x) =  A h_1(x)$  for some constant A, then $h_2^* (\mu) = A h_1^* (\frac{\mu}{A})$. Thus, the   conjugate of  $h(x) = - A \ln x$ is $h^*(\mu) = - A - A \ln \big(\frac{-\mu}{A} \big) =  - A +  A \ln A - A \ln (-\mu)$.
Therefore, the conjugate of $f_i(u_i)$ is
\beqn
f_i^*(\mu_i) = - B_i + B_i \ln B_i - B_i \ln (-\mu_i).
\eeqn
By ignoring the constant terms $( - B_i + B_i \ln B_i) $, the dual objective is equivalent to
\beqn
\underset{\eta,p,\nu}{\text{min}}  \Big( \sum_j p_j  -  \sum_i B_i \ln \eta_i  \Big).
\eeqn
We  have shown that the convex program (\ref{dual1}) can be inferred directly from the EG convex program (\ref{EGprogram1})-(\ref{EQ13}). 
 
\beqn
\label{dual1}
\underset{p,\eta}{\text{minimize}} ~&&\sum_{j=1}^M p_j - \sum_{i=1}^N B_i \ln (\eta_i) \\ \nonumber
\label{dual1EQ1}
\text{subject to} ~ &&p_j \geq a_{i,j} \eta_i, ~\forall i,~j;  ~ p_j \geq 0,~\forall j.
\eeqn
Note that it can be verified that the KKT conditions of problems (\ref{dual1}) and (\ref{EGprogram1})-(\ref{EQ13}) are equivalent, where $x_{i,j}$ is the dual variable corresponding to the first constraint of  problem (\ref{dual1}). Hence, problem (\ref{dual1}) captures the market clearing prices (i.e. equilibrium prices).

\section{Proof of Theorem 7.2}
Consider problems \textbf{(P3)} and \textbf{(P4)} as follows
\beqn
\label{dual2}
\textbf{(P3)} \quad \underset{p,\eta}{\text{minimize}}~~ \sum_{j=1}^M p_j - \sum_{i=1}^N B_i \ln (\eta_i)
\eeqn
subject to
\beqn
\label{dual2EQ1}
p_j \geq a_{i,j} \eta_i, ~\forall i,~j;~ \eta_i \leq 1,~ \forall i;~\eta_i \geq 0, ~\forall i;~p_j \geq 0, ~\forall j. \nonumber
\eeqn
\beqn
\label{EGprogram2}
\textbf{(P4)}  \quad \underset{\mathcal{X},u,s}{ \text{maximize}} ~~\sum_{i=1}^N \Big( B_i \ln u_i - s_i \Big) 
\eeqn
subject to 
\vspace{-0.4in}
\beqn
\label{EQ21} 
\nonumber
&& u_i \leq \sum_{j=1}^M a_{i,j} x_{i,j} + s_i,\forall i \\ \nonumber
\vspace{-0.2in}
&& \sum_{i=1}^N x_{i,j} \leq 1, \forall j;x_{i,j} \geq 0, \forall i,j;~ s_i \geq 0, ~\forall i.
\eeqn

- \textbf{Dual program construction:} First we show that the convex program \textbf{(P4)} is the dual program of \textbf{(P3)}. This can be shown in a similar way to proof of  Proposition 7.1 as follows.
 Let $\lambda_{i,j}$, $\theta_i$, $\gamma_i$, and $\upsilon_j$ be the dual variables associated with constraints 
the first, second, third, and last constraints in problem \textbf{(P3)}, respectively. The Lagrangian of problem (\ref{dual2}) is
\beqn
 L(p,\eta,\lambda,\theta,\gamma,\upsilon) = \sum_j p_j - \sum_i B_i \ln \eta_i  - \sum_i \gamma_i \eta_i \\ \nonumber
 + \sum_i \sum_j \lambda_{i,j} \big( a_{i,j} \eta_i - p_j \big)  +  \sum_i \theta_i (\eta_i - 1) - \sum_j \upsilon_j p_j.
\eeqn
Define $f_i(\eta_i) = B_i \ln \eta_i$, we have
\beqn
 L(p,\eta,\lambda,\theta,\gamma,\upsilon) = \sum_j (1 - \sum_i \lambda_{i,j} - \upsilon_j) p_j   \\ \nonumber
 + \sum_i \big( ( \sum_j \lambda_{i,j}  a_{i,j} +\theta_i - \gamma_i) \eta_i - f_i({\eta_i}) \big)  - \sum_i \theta_i .
\eeqn

The dual function is 
\beqn
g(\lambda,\theta,\gamma,\upsilon) = \underset{p,\eta}{\text{minimize}} ~~ L(p,\eta,\lambda,\theta,\gamma,\upsilon) .
\eeqn

The dual problem is
\beqn
 \underset{\lambda \geq 0,\theta \geq 0,\gamma \geq 0, \upsilon \geq 0}{\text{maximize}} ~~g(\lambda,\theta,\gamma,\upsilon) .
\eeqn
The rest of this construction is similar to Proposition 7.1. By setting $x_{i,j} = \lambda_{i,j}$, $s_i = \theta_i $, and  $u_i = \sum_j \lambda_{i,j}  a_{i,j} +\theta_i - \gamma_i$, we can obtain problem \textbf{(P4)}.

- \textbf{Main proof of of theorem 7.2}. We need to show that the solution of \textbf{(P4)} is an exact ME. 
At the equilibrium, total money spent and surplus money of every service equals to its budget, i.e., $\sum_j p_j x_{i,j} + s_i = B_i, ~\forall i.$ Additionally, the optimal utility of every service is unique.  
Clearly, \textbf{(P4)} can be written in the following form.
\beqn
\label{EG2}
\quad \underset{\mathcal{X}}{ \text{maximize}} ~~\sum_{i=1}^N \Big( B_i \ln \big( s_i + \sum_{j=1}^M a_{i,j} x_{i,j} \big) - s_i \Big)
\eeqn
subject to 
\vspace{-0.2in}
\beqn
\label{tempp1}
 \sum_{i=1}^N x_{i,j} \leq 1, \quad \forall j \\
\label{tempp2}
x_{i,j} \geq 0, \quad \forall i,j \\
\label{tempp3}
s_i \geq 0, \quad \forall i.
\eeqn
Let $p_j$, $\nu_{i,j}$, and $\beta_i$ be the dual variables corresponding to (\ref{tempp1}), (\ref{tempp2}), and (\ref{tempp3}), respectively. The Lagrangian is
\beqn
L(\mathcal{X},s,p,\nu,\beta) = \sum_i \Big( B_i \ln \big( s_i + \sum_{j=1}^M a_{i,j} x_{i,j} \big) - s_i \Big)  \\ \nonumber
 + \sum_j p_j ( 1 - \sum_i x_{i,j}) +  \sum_i \sum_j \nu_{i,j} x_{i,j} + \sum_i s_i \beta_i. \\ \nonumber
\eeqn
Since the feasible region of (\ref{EG2})-(\ref{tempp2}) is closed, compact, and convex,  Slater's condition holds and the KKT conditions are necessary and sufficient for optimality. The KKT conditions render 
\beqn
\label{kkt1}
\frac{\partial L}{\partial s_i} = B_i \frac{1}{u_i(x_i,s_i)} - 1 + \beta_i = 0,~\forall i \\
\vspace{-0.1in}
\label{kkt2}
\frac{\partial L}{\partial x_{i,j}} = B_i \frac{a_{i,j}}{u_i(x_i,s_i)} - p_j + \nu_{i,j} = 0, \quad \forall i,~j \\
\vspace{-0.2in}
\label{kkt3}
p_j (1 - \sum_i x_{i,j}) = 0, ~ \forall j;~ \nu_{i,j} x_{i,j} = 0, ~\forall i,~j \\
\vspace{-0.1in}
\label{kkt4}
s_i \beta_i = 0,~\forall i; p_j \geq 0,~\forall j;~ \beta_i \geq 0,~\forall i;~\nu_{i,j} \geq 0,~\forall i,~j.
\eeqn
where $u_i(x_i,s_i) = s_i + \sum_j a_{i,j} x_{i,j}, ~\forall i $. For simplicity, we will write $u_i(x_i,s_i)$ as $u_i$.
It can be inferred from (\ref{kkt1})-(\ref{kkt4}) the following
\beqn
\label{con1}
\forall i,j: \frac{u_i}{B_i} \leq \frac{a_{i,j}}{p_j}\\
\label{con2}
\forall i,j: \text{if} ~x_{i,j} > 0 \Rightarrow \nu_{i,j} = 0 \Rightarrow \frac{u_i}{B_i} = \frac{a_{i,j}}{p_j}\\
\label{con3}
\forall j: p_j > 0 \Rightarrow \sum_i {x_{i,j}} = 1; ~ \sum_i {x_{i,j}} < 1   \Rightarrow p_j = 0.
\eeqn
Similar to the analysis of the EG program, conditions (\ref{con1}) and (\ref{con2}) imply that $x_{i,j} > 0 $ if and only if $j \in D_i(p)$, i.e., each service buys resources only from ENs giving it the MBB. This also maximizes $u_i$.
Next, we show that  total money spent and surplus money of every service equals to its budget. From (\ref{kkt1}), we have
\beqn
B_i \frac{s_i}{u_i} - s_i + s_i \beta_i = 0,~\forall i
\eeqn 
Since $s_i \beta_i = 0,~\forall i$, we have 
\beqn
\label{tem1}
B_i \frac{s_i}{u_i} =  s_i ,~\forall i
\eeqn
Similarly, since $ \nu_{i,j} x_{i,j} = 0$, by multiplying both sides of  (\ref{kkt2}) by $x_{i,j}$, we have
\beqn
\label{tem2}
B_i \frac{a_{i,j} x_{i,j}}{u_i} = p_j x_{i,j}, ~\forall i
\eeqn 
From (\ref{tem1}) and (\ref{tem2}), we have
\beqn
B_i \frac{s_i}{u_i} + \sum_j B_i \frac{a_{i,j} x_{i,j}}{u_i}  =  s_i + \sum_j p_j x_{i,j}, ~\forall i,~j
\eeqn
Hence, $B_i = s_i + \sum_j p_j x_{i,j},~\forall i,~j$.
From (\ref{kkt1}) and (\ref{kkt4}), if $s_i > 0 \Rightarrow \beta_i = 0,~\forall i \Rightarrow  u_i = B_i, ~\forall i.$  
Therefore, if a buyer has surplus, her utility is equal to her budget. It means buying resources or not does not change her utility since she can keep her money and buy nothing and still has a utility which equals to her budget. This happens when there is no ENs with prices strictly lower than the buyer's utilities gained from the ENs (i.e., $p_j \geq a_{i,j},~\forall j$). Finally, the utility of a buyer with no surplus money is always greater or equal to the her budget, i.e., $u_i \geq B_i$, which can be inferred  from (\ref{kkt1}) and $\beta_i \geq 0$.

\section{Concave Homogeneous Utility Functions}
\label{homo}
Indeed, the EG program not only captures the Fisher market with linear utilities \cite{EG} but also applies to a wider class of concave homogeneous (degree one) utility functions \cite{EG1}. The proof can be found in \cite{AGT}. We present the proof here for completeness. The EG program can be written as follows.
\beqn
\label{EG11}
\vspace{-0.1in} 
&& \underset{\mathcal{X},u}{\text{maximize}} ~\sum_{i=1}^N B_i \ln u_i(x_i) \\
&& \text{subject to}~\sum_{i=1}^N x_{i,j} \leq  1, \quad \forall j;~ x_{i,j} \geq 0, \quad \forall i,~j.
\eeqn
 Define $x_i^*$ as the optimal solution to the EG program. Let $p_j$ and $\nu_{i,j}$ be dual variables.
The KKT conditions give us 
\beqn
p_j \big(1-\sum_i x_{i,j}^* \big) = 0,\quad \forall ~j \\
\label{con2}
x_{i,j}^* \nu_{i,j} = 0, \quad \forall i,~j\\
\label{con3}
\frac{B_i}{u_i (x_i^*)} \frac{\partial u_i(x_i^*)}{\partial x_{i,j}} = p_j - \nu_{i,j}, \quad \forall ~i,j.
\eeqn
For every service $i$, multiply both sides of (\ref{con3}) by $x_{i,j}^*$ and sum up the resulting equalities, we have
\beqn
\frac{B_i}{u_i (x_i^*)} \sum_j x_{i,j}^* \frac{\partial u_i(x_i^*)}{\partial x_{i,j}} = \sum_{j} (p_j - \nu_{i,j}) x_{i,j}^*.
\eeqn 
From the Euler's theorem for homogeneous functions, we have $u_i(x_i) = \sum_j x_{i,j} \frac{\partial u_i}{\partial x_{i,j}}$ for homogeneous function $u_i$. 
Also, combined with (\ref{con2}), we have
\beqn
B_i = \sum_{j} p_j x_{i,j}^*.
\eeqn
Thus, $x_i^*$ exhausts the budget of buyer $i$ at price $p$.

We now show that $x_i^*$ maximizes $u_i(x_i)$. 
Let $x'_i \in \mathbb{R}_+^M$ be any allocation to buyer $i$ satisfying $\sum_{j} p_j x'_{i,j} \leq B_i$. From the concavity of $u_i$, we have.
\beqn
u_i(x'_i) - u_i(x_i^*) &\leq& \nabla u_i(x_i^*) (x'_i - x_i^*) \\ \nonumber
&=& \frac{u_i(x_i^*)}{B_i} \sum_j (p_j - \nu_{i,j})(x'_{i,j} - x_{i,j}^*) \\ \nonumber
&=& \frac{u_i(x_i^*)}{B_i}  \Big( \sum_j (p_j x'_{i,j} - \nu_{i,j} x'_{i,j}) - B_i  \Big) \\ \nonumber
&\leq& \frac{u_i(x_i^*)}{B_i} \Big(\sum_j p_j x'_{i,j} - B_i \Big) \leq 0. \\ \nonumber
\eeqn
Therefore, $x_i^*$ is an optimal demand bundle of buyer $i$ at price vector $p$.
Hence, every buyer is allocated its optimal resource bundle at the equilibrium prices. The market clearing condition proof is similar to that of the linear utility case in the manuscript.
We have shown that all the requirements of an ME are satisfied.

\section{Other Discussions}
- \textit{Net profit maximization and concave homogeneous revenue functions}: Instead of linear revenue function $u_i(x_i)$ as considered throughout this work, here, we consider the case where $u_i(x_i)$ is concave and homogeneous of degree one. Then, let $g_i(x_i,s_i) = u_i{x_i} + s_i$.
For any constant $d > 0$:
\beqn
g_i(d x_i, d s_i) &=& u_i{d x_i} + d s_i = d u_i(x_i) + d s_i\\ \nonumber
	&=& d \big( \sum_j a_{i,j} x_{i,j} + s_i \big) = d u_i(x_i, s_i),~\forall ~i.
\eeqn 
Thus, $g_i(x_i,s_i)$ is a concave homogeneous (of degree one) function. By following similar the proof above in Appendix D for the EG program, we can verify that our new convex optimization problem (as shown below) for the net profit maximization model also applies to the wider class of concave homogeneous functions.

\beqn
\label{probx}
  \quad \underset{\mathcal{X},u,s}{ \text{maximize}} ~~\sum_{i=1}^N \Big( B_i \ln (u_i(x_i) + s_i) - s_i \Big) 
\eeqn
subject to 
\beqn
\label{EQ29} 
\nonumber
&& \sum_{i=1}^N x_{i,j} \leq 1, \forall j;x_{i,j} \geq 0, \forall i,j;~ s_i \geq 0, ~\forall i.
\eeqn

- \textit{Multiple resource types:} Consider an example of the model with multiple resource types. For example, a service may need a combination of bandwidth, computing, and memory. Let $d_i = (d_{i,1},...,d_{i,K})$ be the base demand vector giving service $i$, in which $d_{i,k}$ is the amount of resource $k$ in the base demand vector of service $i$.
The utility of service $i$ is commonly defined as follows: $u_i(x_i) = \min_k  \{ \frac{x_{i,k}}{d_{i,k}} \}$.
It is easy to see that $u_i(x_i)$ is concave and homogeneous of degree one. If there are multiple ENs, we will take sum utility over all the ENs. Therefore, the proposed framework can apply to the multi-resource case. 


- \textit{Another decentralized implementation of the basic model (revenue maximization):} By carefully analyzing 
the centralized combinatorial algorithm proposed  for the linear Fisher market in the remarkable piece of work of
 by N.R. Devanur {\em et. al.} \cite{ndev08}, we observe that it can be modified to work in a distributed fashion as follows. In each round, given new prices, instead of reporting its optimal bundle(s), each service will report its maximum bang-per-buck (MBB) value and the ENs in its demand set $D_i$. The platform will collect these information to construct and compute a new equality graph as defined in \cite{ndev08}, and compute a balance flow (maxflow with minimum surplus). By this small modification, the combinatorial algorithm in \cite{ndev08} can be implemented in decentralized manner. The algorithm terminates when the max-flow is equal to the total budget of the buyers. Note that before the prices converge to equilibrium prices, the total budget of the buyers and the total prices of the ENs are different, which means the intermediate solution does not satisfy the market clearing condition. Interested readers can refer to \cite{ndev08} for more details.

\end{document}